\documentclass[11pt,openany]{book}
\usepackage[dvips,usenames,dvipsnames]{color}
\usepackage[dvips]{graphicx}
\usepackage{natbib}
\usepackage[dvips,plainpages=false,pagebackref,colorlinks,citecolor=blue,filecolor=PineGreen,urlcolor=PineGreen,linkcolor=Mulberry,pagecolor=black]{hyperref}

\newcommand{\pubname}{circular}        
\newcommand{\Pubname}{Circular}       

\newcommand{\degr}{\mbox{$^{\circ}$}}
\newcommand{\D}{\hphantom{9}}
\newcommand{\DD}{\hphantom{99}}
\newcommand{\DDD}{\hphantom{999}}

\newcommand{\E}[1]{\mbox{$\times 10^{#1}$}}

\newcommand{\uas}{\mbox{$\mu$as}}

\newcommand{\AsA}{{\it The Astronomical Almanac}}
\newcommand{\UTC}{\mbox{UTC}}
\newcommand{\TAI}{\mbox{TAI}}
\newcommand{\TT}{\mbox{TT}}
\newcommand{\TCG}{\mbox{TCG}}
\newcommand{\TCB}{\mbox{TCB}}
\newcommand{\TDB}{\mbox{TDB}}
\newcommand{\UTI}{\mbox{UT1}}
\newcommand{\GMST}{\mbox{GMST}}
\newcommand{\GAST}{\mbox{GAST}}
\newcommand{\ssec}{\mbox{$\!^{\rm s}$}}
\newcommand{\ssun}{\mbox{\tiny$_{\bigcirc\hspace{-0.61em}\cdot}$}} 
\newcommand{\bm}[1]{\mbox{\boldmath${#1}$\unboldmath}}
\newcommand{\vcip}{\mbox{${\bf n}_{_{\rm GCRS}}$}}
\newcommand{\vcio}{\mbox{\boldmath$\sigma$\unboldmath$_{_{\rm GCRS}}$}}
\newcommand{\veqx}{\mbox{\boldmath$\Upsilon$\unboldmath$_{_{\rm GCRS}}$}}
\newcommand{\Ec}{\mbox{E$_{\sigma}$}}
\newcommand{\Et}{\mbox{E$_{\varpi}$}}
\newcommand{\Eq}{\mbox{E$_{_\Upsilon}$}}
\newcommand{\w}[1]{\hphantom{#1}}
\newcommand{\NOPRINT}[1]{\null}
\newcommand{\refsep}{\qquad\textcolor{black}{\dots}}

\bibpunct{(}{)}{;}{a}{\ }{,}

\begin{document}

\begin{titlepage}
  \begin{center}
    \textsf{ \rule{0in}{0.5in} \\
         {\Large UNITED STATES NAVAL OBSERVATORY \\ \rule{0in}{0in} \\
          CIRCULAR NO. 179 \\ \rule{0in}{0in} \\}
         {\small \rule[1pt]{6.2in}{0.01in} \\
          U.S. Naval Observatory, Washington, D.C. 20392 \hspace{2.2in} 2005 Oct 20\\
         \rule[8pt]{6.2in}{0.01in} }
         \\ \rule{0in}{1.0in} \\
        {\LARGE The IAU Resolutions \\ on Astronomical Reference Systems, \\
        Time Scales, and Earth Rotation Models \\ \rule{0in}{0in} \\}
        {\Large Explanation and Implementation}
        \\ \rule{0in}{1.0in} \\ by \\ \rule{0in}{0in} \\ {\large George H. Kaplan}
          }
  \end{center}
\end{titlepage}

\frontmatter
\markboth{}{}
\tableofcontents
\chapter*{Introduction}\label{intro}
\addcontentsline{toc}{chapter}{Introduction}
\markboth{INTRODUCTION}{INTRODUCTION}

The series of resolutions passed by the International
Astronomical Union at its General Assemblies in 1997 and 2000 are
the most significant set of international agreements in
positional astronomy in several decades and arguably since the
Paris conference of 1896.  The approval of these resolutions
culminated a process --- not without controversy --- that began
with the formation of an inter-commission Working Group on
Reference Systems at the 1985 IAU General Assembly in Delhi.

The resolutions came at the end of a remarkable decade for
astrometry, geodesy, and dynamical astronomy.  That decade
witnessed the successes of the Hipparcos satellite and the Hubble
Space Telescope (in both cases, after apparently fatal initial
problems), the completion of the Global Positioning System,
25-year milestones in the use of very long baseline
interferometry
(VLBI) and lunar laser ranging (LLR) for astrometric and
geodetic
measurements, the discovery of Kuiper Belt objects and
extra-solar planets, and the impact of comet Shoemaker-Levy 9
onto Jupiter. At the end of the decade, interest in near-Earth
asteroids and advances in sensor design were motivating plans for
rapid and deep all-sky surveys. Significant advances in theory
also took place, facilitated by inexpensive computer power and
the Internet. Positional and dynamical astronomy were enriched by
a deeper understanding of chaos and resonances in the solar system,
advances in the theory of the rotational dynamics of the Earth,
and increasingly sophisticated models of how planetary and stellar systems 
form and evolve.  It is not too much of an exaggeration to say that
as a result of these and similar developments, the old idea that
astrometry is an essential tool of astrophysics was rediscovered.
The IAU resolutions thus came at a fortuitous time, providing a
solid framework for interpreting the modern high-precision
measurements that are revitalizing so many areas of astronomy.

This \pubname\ is an attempt to explain these resolutions and
provide guidance on their implementation.  This publication is
the successor to USNO Circular~163 (1982), which had a similar
purpose for the IAU resolutions passed in 1976, 1979, and 1982.
Both the 1976--1982 resolutions and those of 1997--2000 provide
the specification of the fundamental astronomical reference
system, the definition of time scales to be used in astronomy,
and the designation of conventional models for Earth orientation
calculations (involving precession, nutation, and Universal
Time).  It will certainly not go unnoticed by readers familiar
with Circular~163 that the current publication is considerably
thicker. This reflects both the increased complexity of the
subject matter and the wider audience that is addressed.

Of course, the IAU resolutions of 1997--2000 did not arise in a
vacuum. Many people participated in various IAU working groups,
colloquia, and symposia in the 1980s and 1990s on these topics,
and some important resolutions were in fact passed by the IAU in
the early 1990s.  Furthermore, any set of international standards
dealing with such fundamental matters as space and time must to
some extent be based on, and provide continuity with, existing
practice.  Therefore, many of the new resolutions carry
``baggage'' from the past, and there is always the question of
how much of this history (some of it quite convoluted) is
important for those who simply wish to implement the latest
recommendations. Material in this \pubname\ generally avoids
detailed history in an effort to present the most succinct and
least confusing picture possible. 
However, many readers will be involved with modifying
existing software systems, and some mention of previous practice
is necessary simply to indicate what needs to be changed.  A
limited amount of background material also sometimes aids in
understanding and provides a context for the new recommendations.
The reader should be aware that the presentation of such material
is selective and no attempt at historical completeness is
attempted.

It must be emphasized that the resolutions described here affect 
astronomical quantities only at the level of some tens of milliarcseconds
or less at the present epoch.  And, despite misinformation to
the contrary, familiar concepts such as the equinox and sidereal time have not
been discarded.  The largest systematic change is
due to the new rate of precession, which is 0.3~arcsecond per century less
than the previous (1976) rate;  the change affects some types of
astronomical coordinates and sidereal time.   Astronomical software applications that work
acceptably well now at the arcsecond or 0.1-arcsecond level (which would include
most telescope control systems) will continue to work at that level, even
when used with new sources of reference data, such as
the Hipparcos, Tycho-2, or 2MASS star catalogs or the VCS3 radio source catalog.
Applications that are independent of the rotation of the Earth, such as those for differential
(small-field) astrometry, are largely unaffected.   For these kinds
of systems, changes to computer code that implement the new resolutions are
recommended as a long-term goal, to maintain standardization of
algorithms throughout the astronomical community, but are not an
immediate practical necessity.  (Perhaps some readers will 
stop right here and file this \pubname\ on the shelf!) 

\section*{Overview of the Resolutions}\label{intro.overview}
\addcontentsline{toc}{section}{Overview of the Resolutions}

The IAU resolutions described in this \pubname\ cover a range
of fundamental topics in positional astronomy:

\begin{itemize}
\item {\bf Relativity}\quad Resolutions passed
in 2000 provide the relativistic metric tensors for reference
systems with origins at the solar system barycenter and the
geocenter, and the transformation between the two systems. While
these are mostly of use to theorists --- for example, in the
formulation of accurate models of observations ---
they provide a proper relativistic framework for current and future
developments in precise astrometry, geodesy, and dynamical
astronomy.  (See Chapter~\ref{rel}.)
\item {\bf Time Scales}\quad Resolutions passed in 1991 and
2000 provide the definitions of
various kinds of astronomical time and the relationships between
them.  Included are time scales based on the Syst\`{e}me
International (SI) second (``atomic'' time scales) as well as
those based on the rotation of the Earth.   (See Chapter~\ref{time}.)
\item {\bf The Fundamental Astronomical Reference System}\quad
A resolution passed in 1997 established the International
Celestial Reference System (ICRS), a high precision coordinate
system with its origin at the solar system barycenter and ``space
fixed'' (kinematically non-rotating) axes.  The resolution
included the specification of two sets of benchmark objects and
their coordinates, one for radio observations (VLBI-measured
positions of pointlike extragalactic sources) and one for optical
observations (Hipparcos-measured positions of stars).  These two
sets of reference objects provide the practical implementation of
the system and allow new observations to be related to it. (See
Chapter~\ref{refsys}.)
\item {\bf Precession and Nutation}\quad
Resolutions passed in 2000 provided a new precise definition of
the celestial pole and endorsed a specific theoretical
development for computing its instantaneous motion.  The
celestial pole to which these developments refer is called the Celestial Intermediate
Pole (CIP); the instantaneous equatorial plane is orthogonal to
the CIP.  There are now new precise algorithms for computing the
pole's position on the celestial sphere at any time, in the form
of new expressions for precession and nutation.  (See Chapter~\ref{prenut}.)
\item {\bf Earth Rotation}\quad A resolution passed in 2000
establishes new reference points, one on the celestial sphere and
one on the surface of the Earth, for the measurement of the
rotation of the Earth about its axis. The new points are called,
respectively, the Celestial Intermediate Origin (CIO) and the
Terrestrial Intermediate Origin (TIO).  Both lie in the instantaneous equatorial
plane.  The rotation of the Earth is simply the geocentric angle,
$\theta$, between these two points, a linear function of
Universal Time (UT1).  The CIO is analogous to the equinox, the
reference point on the celestial sphere for sidereal time. 
Unlike the equinox, however, the CIO has no motion along the
instantaneous equator, and unlike sidereal time, $\theta$ is not
``contaminated'' by precession or nutation.  The new
CIO-TIO-based Earth rotation paradigm thus allows a clean
separation of Earth rotation, precession, and nutation in the
transformation between terrestrial and celestial reference
systems.  (See Chapter~\ref{erot}.)
\end{itemize}

This \pubname\ also includes a brief description of the de facto
standard solar system model, produced and distributed by the Jet
Propulsion Laboratory (see Chapter~\ref{eph}).  This model, labeled
DE405/LE405, provides the positions and velocities of the nine
major planets and the Moon with respect to the solar system
barycenter for any date and time between 1600 and 2200. The
positions and velocities are given in rectangular coordinates,
referred to the ICRS axes.  This ephemeris is not the subject of
any IAU resolutions but has become widely adopted
internationally; for example, it is the basis for the tabulations
in \AsA\ and it underlies some of the
other algorithms presented in this \pubname.

The 1997 and 2000 IAU resolutions form an interrelated and
coherent set of standards for positional astronomy.  For example,
the definitions of the SI-based time scales rely on the
relativity resolutions, and the position of the Celestial
Intermediate Pole and the Celestial Intermediate Origin can only be
properly computed using the new precession and nutation
expressions.  Many other links between the resolutions exist.  In
fact, attempting to apply the resolutions selectively can lead to
quite incorrect (or impossible to interpret) results.  This
\pubname\ is meant to provide an explanatory and computational
framework for a holistic approach to implementing these
resolutions in various astronomical applications.  The author
hopes that what is presented here does justice to the efforts of
the many people who worked very hard over the last decade to take
some important scientific ideas and work out their practical
implications for positional astronomy, to the benefit of the
entire scientific community.

\section*{About this \Pubname}\label{intro.about}
\addcontentsline{toc}{section}{About this \Pubname}

The chapters in this \pubname\ reflect the six main subject areas
described above.  Each of the chapters contains a list of the
relevant IAU resolutions, a summary of the recommendations, an
explanatory narrative, and, in most chapters, a collection of
formulas used in implementing the recommendations.  The
references for all chapters are collected in one list
at the end of the \pubname\ (p.~\pageref{refstart} ff.).  The reference
list is in the usual style for astronomy journal articles.
At the end of the references, a list of Uniform Resource Locators (URLs)
is given (p.~\pageref{urls}) for documents and data
that are available on the World Wide Web.  These URLs are numbered
and they are referred to in the text by number --- for example, a PDF
version of this \pubname\ can be found at \citet{url-thispub}. 

It is assumed that readers have a basic knowledge of positional astronomy;
that the terms right ascension, declination, sidereal time, precesssion, nutation,
equinox, ecliptic, and ephemeris are familiar.  Some experience in computing
some type of positional astronomy data is useful,
because the ultimate purpose of the \pubname\ is to
enable such computations to be
carried out in accordance with the 1997 and 2000 IAU resolutions.  The
explanatory narratives deal primarily with new or unfamiliar concepts
introduced by the resolutions --- concepts that would not generally be described
in most introductory textbooks on positional astronomy.  This \pubname\ is not
a substitute for such textbooks.

IAU resolutions are referred to in the text in the form ``res.~N
of year'', for example, ``res.~B1.2 of 2000''.   The year refers
to the year of the IAU General Assembly that passed the
resolution.  The proceedings of each General Assembly, including
the text of the resolutions, are usually published the following
year.  The References section of this \pubname\ lists the various
proceedings volumes under ``IAU''.  An
online reference for the text of
IAU resolutions (beginning with those passed at the 1994
General Assembly) is the IAU Information Bulletin (IB) series, at
\citet{url-iaubull}.  Resolutions are printed in the January IB following a General
Assembly, i.e., IB numbers 74, 81, 88, 94, etc.   This
\pubname\ contains two appendices containing the complete text of
the resolutions passed by the 1997 and 2000 General Assemblies,
which are the main focus of attention here.

Errata in this \pubname\ and updates to it are provided at \citet{url-thispub}.

\section*{Other Resources}\label{intro.resources}
\addcontentsline{toc}{section}{Other Resources}

An increasing number of publications, data, and software related
to the recent IAU resolutions are becoming available. 

A major online resource for implementing the IAU resolutions
involving Earth rotation and time (Chapters~\ref{time}, \ref{prenut},
and \ref{erot} here) is the document of conventions used by the International
Earth Rotation and Reference Systems Service (IERS): \citet{iers03},
IERS Technical Note No.~32, edited by
D.~D.~McCarthy and G.~Petit.  It is available in printed form 
from the IERS and also on the web at \citet{url-iersconv}.   The online
document contains links to Fortran subroutines that implement the
recommended models.  The document also contains algorithms
specific to geodetic applications, such as tidal and 
geopotential models, that have not been the subject of IAU
action and are not discussed in this \pubname.  The IERS also
maintains an online list of FAQs on the IAU resolutions
\citep{url-iersfaq}.
 
The IAU Working Group on Nomenclature for Fundamental 
Astronomy (2003--2006) has a website \citep{url-wgnfa} with
many helpful documents, including a list of definitions
(some of which are used in this \pubname) and other
educational material. 
 
In addition to the IERS software, two other packages of
computer subroutines are available for
implementing the IAU resolutions: the Standards of Fundamental
Astronomy (SOFA), at \citet{url-sofa}, and
the Naval Observatory Vector Astrometry Subroutines (NOVAS), at 
\citet{url-novas}.   SOFA is a collection of routines managed by an international
panel, the SOFA Reviewing Board, that works under the auspices of
IAU Division~1 and is chaired by P.~Wallace.  The board has
adopted a set of Fortran coding
standards for algorithm implementations (C versions are
contemplated for the future) and is soliciting code from the
astrometric and geodetic communities that implements IAU models.
Subroutines are adapted to the coding standards and validated for
accuracy before being added to the SOFA collection. NOVAS is an
integrated package of subroutines, available in Fortran and C,
for the computation of a wide variety of common astrometric
quantities and transformations.  NOVAS dates back to the 1970s
but has been continually updated to adhere to subsequent IAU
resolutions.

\AsA, beginning with the 2006 edition, is also a resource for implementing
the IAU resolutions.  Not only does it list various algorithms arising
from or consistent with the resolutions,
but its tabular data serve as numerical checks for independent developments.
Both SOFA and NOVAS subroutines are used in preparing the tabulations
in \AsA, and various checks have been made to ensure the consistency of the 
output of the two software packages.

\section*{Acknowledgments}\label{intro.acknow}
\addcontentsline{toc}{section}{Acknowledgments}

Many people have contributed in some way to this \pubname.  I have had many
interesting and enlightening discussions with James Hilton, Victor Slabinski, and
Patrick Wallace.  I wish to thank Bill Tangren for
setting me straight on some of the basic concepts of general relativity, and 
Sergei Klioner and Sergei Kopeikin for patiently trying to
explain to me the relativistic aspects of the 2000 IAU resolutions.  Sergei Klioner
provided some text that I have used verbatim.  I hope that the final product
properly represents their input.  The e-mail discussions within the IAU Working
Group on Nomenclature for Fundamental Astronomy have also been quite 
valuable and have contributed to what is presented here.  The many scientific
papers written by Nicole Capitaine
and her collaborators have been essential references.  Victor Slabinski and Dennis
McCarthy of the U.S. Naval Observatory, Ken Seidelmann of the University of Virginia,
Catherine Hohenkerk of Her Majesty's Nautical Almanac Office, and Myles Standish
of the Jet Propulsion Laboratory carefully reviewed drafts of this \pubname\
and made many substantive suggestions for improvement.   The remaining defects
in the \pubname\ are, of course, my sole responsibility.\\[0.1in]   

\hfill --- George Kaplan\hspace{1in}

\chapter*{A Few Words about Constants}\label{const}
\addcontentsline{toc}{chapter}{A Few Words about Constants}
\markboth{CONSTANTS}{CONSTANTS}

This \pubname\ does not contain a list of adopted fundamental astronomical
constants, because the IAU is no longer maintaining such a list.
The last set of officially adopted constant values was the IAU (1976)
System of Astronomical Constants.  That list is almost entirely obsolete.
For a while, an IAU working group maintained a list of ``best estimates'' of various
constant values, but the IAU General Assembly of 2003 did not renew
that mandate.  It can be argued that a list of fundamental astronomical
constants is no longer possible, given the complexity of the models now
used and the many free parameters that must be adjusted in each model to
fit observations.  That is, there are more constants now to consider, and their
values are theory dependent.  In many cases, it would be incorrect to attempt to use a constant value, obtained from the fit of one theory to observations, with another theory.


We are left with three {\it defining constants} with IAU-sanctioned values
that are intended to be fixed:

\begin{enumerate}
\item The Gaussian gravitational constant: $k = 0.01720209895$.  The dimensions of $k^2$
are \linebreak[4]AU$^3$M\ssun$^{-1}$d$^{-2}$ where AU is the astronomical unit, M\ssun\ is the
solar mass, and d is the day of 86400 seconds.
\item The speed of light:  $c = 299\,792\,458\;\mbox{m}\,\mbox{s}^{-1}$.
\item The fractional difference in rate between the time scales TT and TCG:
$\mbox{L}_G = 6.969290134\E{-10}$.  Specifically, the derivative $d\TT/d\TCG =
1 - \mbox{L}_G$.  (See Chapter~\ref{refsys}.)
\end{enumerate}

The \citet{iers03} includes a list of constants as its Table~1.1.  Several useful ones from this list that are not highly theory dependent (for astronomical use, at least) are: 

\begin{enumerate}
\item Equatorial radius of the Earth:  $a_E = 6\,378\,136.6\;\mbox{m}$.
\item Flattening factor of the Earth:  $f = 1/298.25642$.
\item Dynamical form factor of the Earth: $J_2 = 1.0826359\E{-3}$.
\item Nominal mean angular velocity of Earth rotation: $\omega = 7.292115\E{-5}\; \mbox{rad}\,\mbox{s}^{-1}$.
\item Constant of gravitation: $G = 6.673\E{-11}\;\mbox{m}^3\mbox{kg}^{-1}\mbox{s}^{-2}$
    \quad(CODATA 2002 recommended value: $6.6742\E{-11}\;\mbox{m}^3\mbox{kg}^{-1}\mbox{s}^{-2}$).
\end{enumerate}

The first four values above were recommended by Special Commission~3 of the International Association of Geodesy; the first three are so-called ``zero tide'' values.   (The need to introduce the concept of ``zero tide'' values indicates how theory creeps into even such basic constants as the radius of the Earth as the precision of measurement increases.  See section 1.1 of the \citet{iers03}.)  Planetary masses, the length of the astronomical unit,  and related constants used in or obtained from the Jet Propulsion Laboratory DE405/LE405 ephemeris are listed with its description in Chapter~\ref{eph}.  The rate of general precession in longitude (the ``constant of precession'') is given in Chapter~\ref{prenut} on the precession and nutation theories.

The World Geodetic System 1984 (WGS 84), which is the basis for coordinates obtained from GPS,
uses an Earth ellipsoid with $a_E = 6378137$~m and $f = 1/298.257223563$.

Some astronomical ``constants'' (along with reference data such as star positions) actually represent quantities that slowly vary, and the values given must therefore be associated with a specific epoch.  That epoch is now almost always 2000 January~1, $12^{\mbox{h}}$ (JD 2451545.0), which can be expressed in any of the usual time scales.   If, however, that epoch is considered an event at the geocenter and given in the TT time scale, the epoch is designated J2000.0.  See Chapter~\ref{time}.

\chapter*{Abbreviations and Symbols Frequently Used}\label{abbrev}
\addcontentsline{toc}{chapter}{Abbreviations and Symbols Frequently Used}
\markboth{ABBREVIATIONS \& SYMBOLS}{ABBREVIATIONS \& SYMBOLS}

{\bf and index to most relevant sections}\\[0.1in]

\begin{tabular}{p{0.7in}p{4.8in}l}
$\alpha$ & right ascension & \ref{refsys.bg} \\
$\delta$ & declination & \ref{refsys.bg} \\
$\Delta\psi$ & nutation in [ecliptic] longitude (usually expressed in arcseconds) &
           \ref{prenut.models}, \ref{prenut.formulas.nut} \\
$\Delta\epsilon$ & nutation in obliquity (usually expressed in arcseconds) &
           \ref{prenut.models}, \ref{prenut.formulas.nut} \\
$\epsilon$ & mean obliquity of date & \ref{prenut.formulas.pre} \\
$\epsilon'$ & true obliquity of date (= $\epsilon + \Delta\epsilon$) & \ref{prenut.formulas.nut} \\
$\epsilon_0$ & mean obliquity of J2000.0 & \ref{prenut.models}, \ref{prenut.formulas.pre} \\
$\theta$ & Earth Rotation Angle & \ref{time.erot}, \ref{erot.nro} \\
$\uas$ & microarcecond (= $10^{-6}$ arcsecond $\approx$ 4.8\E{-12} radian) \\
$\sigma$ & a non-rotating origin or, specifically, the CIO & \ref{erot.nro}  \\
\boldmath$\sigma$\unboldmath & unit vector toward a non-rotating origin or, specifically, the CIO &
          \ref{erot.formulas.cardinal}\\
$\Upsilon$ & the equinox & \ref{refsys.bg}, \ref{erot.nro} \\
\boldmath$\Upsilon$\unboldmath & unit vector toward the equinox & \ref{erot.formulas.cardinal} \\
as\hspace{0.6em}{\it or}\hspace{0.6em}$''$ & arcsecond (= 1/3600 degree $\approx$ 4.8\E{-6} radian) \\
AU & astronomical unit(s) \\
{\bf B} & frame bias matrix & \ref{refsys.formulas} \\
BCRS & Barycentric$^1$ Celestial Reference System & \ref{rel.rs} \\
BIPM & Bureau International des Poids et Mesures \\
{\bf C} & matrix for transformation from GCRS to \Ec & \ref{prenut.formulas.alt}, \ref{erot.formulas.tercel} \\
cen & century, specifically, the Julian century of 36525 days of 86400 seconds \\
CIO & Celestial Intermediate Origin$^{2\;3}$ & \ref{erot.nro}, \ref{erot.formulas.cardinal} \\
CIP & Celestial Intermediate Pole & \ref{prenut.pole}, \ref{erot.formulas.cardinal} \\
CIRS & (See \Ec) \\
\Eq & instantaneous (true) equator and equinox of date & \ref{erot.transform}, \ref{erot.formulas.tercel} \\
\Ec & Celestial Intermediate Reference System (CIRS) & \ref{erot.transform}, \ref{erot.formulas.tercel} \\
\Et & Terrestrial Intermediate Reference System (TIRS) & \ref{erot.transform}, \ref{erot.formulas.tercel} \\
$\mathcal{E}_\Upsilon$ & equation of the equinoxes & \ref{time.erot}, \ref{time.formulas.erot} \\
$\mathcal{E}_o$ & equation of the origins & \ref{erot.formulas.cardinal.cio1}, \ref{erot.formulas.ha} \\
ESA & European Space Agency \\
FK$n$ & $n^{\rm th}$ Fundamental Catalog (Astronomisches Rechen-Institut, Heidelberg)
         & \ref{refsys.recent} \\
GAST & Greenwich apparent sidereal time & \ref{time.erot}, \ref{time.formulas.erot} \\
GCRS & Geocentric Celestial Reference System & \ref{rel.rs} \\
\end{tabular}

\begin{tabular}{p{0.7in}p{4.8in}l}
GMST & Greenwich mean sidereal time & \ref{time.erot}, \ref{time.formulas.erot} \\
GPS & Global Positioning System \\
HCRF & Hipparcos Celestial Reference Frame & \ref{refsys.rs}, \ref{refsys.icrs} \\
IAG & International Association of Geodesy \\
IAU & International Astronomical Union \\
ICRF & International Celestial Reference Frame & \ref{refsys.rs}, \ref{refsys.icrs} \\
ICRS & International Celestial Reference System & \ref{refsys.rs}, \ref{refsys.icrs}\\
IERS & International Earth Rotation and Reference System Service \\
ITRF & International Terrestrial Reference Frame & \ref{erot.transform} \\
ITRS & International Terrestrial Reference System & \ref{erot.transform}, \ref{erot.formulas.geod} \\
IUGG & International Union of Geodesy and Geophysics \\
J2000.0 & the epoch 2000 January 1, 12$^{\rm h}$ TT (JD 2451545.0 TT) at the geocenter
            & \ref{time.si} \\
 \quad  & (``J2000.0 system'' is shorthand for the celestial reference system defined by the mean dynamical equator and equinox of J2000.0.) & \ref{refsys.bg} \\
 JD & Julian date (time scale used should be specified) \\
JPL & Jet Propulsion Laboratory \\
mas & milliarcsecond (= $10^{-3}$ arcsecond $\approx$ 4.8\E{-9} radian) \\
{\bf N} & nutation matrix (for transformation from mean to true system of date)
        & \ref{prenut.formulas}, \ref{prenut.formulas.nut} \\
{\bf n} & unit vector toward the CIP (celestial pole) & \ref{prenut.formulas},\ref{erot.formulas.cardinal} \\
NOVAS & Naval Observatory Vector Astrometry Subroutines (software) & \ref{intro.resources} \\
{\bf P} & precession matrix (for transformation from J2000.0
     system to mean system of date)
        & \ref{prenut.formulas}, \ref{prenut.formulas.nut} \\ \\
${\bf R}_1(\phi)$ & rotation matrix to transform column 3-vectors from one cartesian coordinate
system to another.  Final system is formed by rotating original system about its own x-axis by
angle $\phi$
(counterclockwise as viewed from the +x direction):
    \begin{displaymath}
    {\bf R}_1(\phi) = \left( \begin{array}{ccc}
                     1  &              0         &             0        \\
                     0  &  \D\cos\phi  &  \D\sin\phi \\
                     0  &     -\sin\phi   &  \D\cos\phi
               \end{array} \right)
    \end{displaymath}                                                           \\
${\bf R}_2(\phi)$ & rotation matrix to transform column 3-vectors from one cartesian coordinate
system to another.  Final system is formed by rotating original system about its own y-axis by
angle $\phi$
(counterclockwise as viewed from the +y direction):
    \begin{displaymath}
    {\bf R}_2(\phi) = \left( \begin{array}{ccc}
                     \D\cos\phi  &  0 &   -\sin\phi \\
                                  0        &  1 &             0      \\
                      \D\sin\phi  &  0 & \D\cos\phi
               \end{array} \right)
    \end{displaymath}                                                          \\
${\bf R}_3(\phi)$ & rotation matrix to transform column 3-vectors from one cartesian coordinate
system to another.  Final system is formed by rotating original system about its own z-axis by
angle $\phi$
(counterclockwise as viewed from the +z direction):
    \begin{displaymath}
    {\bf R}_3(\phi) = \left( \begin{array}{ccc}
                     \D\cos\phi  &  \D\sin\phi   &  0 \\
                         -\sin\phi  &  \D\cos\phi  &  0 \\
                                   0        &               0        & 1
               \end{array} \right)
    \end{displaymath}                                                          \\  \\
\end{tabular}

\begin{tabular}{p{0.7in}p{4.8in}l}
$s$ & CIO locator: the difference between two arcs on the celestial sphere, providing the
     direction toward the CIO & \ref{erot.formulas.cardinal.cio3} \\
SI & Syst\`{e}me International d'Unit\'{e}s (International System of Units) \\
SOFA & Standards of Fundamental Astronomy (software) & \ref{intro.resources} \\
$T$ & unless otherwise specified, time in Julian centuries (36525
days of 86400 seconds)
       from  JD 2451545.0 (2000 Jan 1.5) \\
\quad   &  (The time scale used should be specified, otherwise TT is understood.) \\
T$_{\rm eph}$ & time argument of JPL planetary and lunar ephemerides & \ref{time.si}, \ref{eph.de405} \\
TAI & International Atomic Time & \ref{time.si} \\
TCB & Barycentric$^1$ Coordinate Time & \ref{rel.rs}, \ref{time.si} \\
TCG & Geocentric Coordinate Time & \ref{rel.rs}, \ref{time.si}\\
TDB & Barycentric$^1$ Dynamical Time & \ref{time.si} \\
TIO & Terrestrial Intermediate Origin$^2$ & \ref{erot.nro}, \ref{erot.formulas.geod} \\
TIRS & (See \Et) \\
TT & Terrestrial Time & \ref{time.si} \\
UCAC & USNO CCD Astrographic Catalog & \ref{refsys.icrs.data} \\
USNO & U.S. Naval Observatory \\
UT1 & Universal Time (affected by variations in length of day) & \ref{time.erot}, \ref{time.formulas.erot} \\
UTC & Coordinated Universal Time (an atomic time scale) & \ref{time.utc} \\
VLBI & very long baseline [radio] interferometry \\
{\bf W} & ``wobble'' (polar motion) matrix (for transformation from ITRS to \Et) & \ref{erot.formulas.geod} \\
WGS 84 & World Geodetic System 1984 & \ref{erot.transform} \\
$ \left. \begin{array}{l}X\\Y\\Z\end{array}\right\} $ &  components of  \vcip, unit vector toward the CIP
    with respect to the GCRS & \ref{prenut.formulas}, \ref{erot.formulas.cardinal}\\ \\
$ \left. \begin{array}{l}x_p\\y_p\end{array}\right\} $ &  standard
polar motion parameters, defining location
         of the CIP in the ITRS & \ref{erot.formulas.geod} \\
\end{tabular}
\\*[0.7in]

$^1$\quad  ``Barycentric'' always refers to the solar system barycenter, the center of mass of all bodies in the solar system.\\

$^2$\quad  The fundamental reference points referred to here as the Celestial Intermediate Origin (CIO) and the Terrestrial Intermediate Origin (TIO) were called, respectively, the Celestial Ephemeris Origin (CEO) and the Terrestrial Ephemeris Origin (TEO) in the IAU resolutions of 2000.  The IAU Working Group on Nomenclature for Fundamental Astronomy \citep{url-wgnfa} has recommended the change of nomenclature with no change in the definitions.  The new terminology is already in use in \AsA\ and in IERS documents, and will undoubtedly be adopted by the IAU General Assembly in 2006.  It is used throughout this \pubname, except in the verbatim text of the IAU resolutions.\\

$^3$\quad The abbreviation CIO was used throughout much of the 20th century to designate the
Conventional International Origin, the reference point for the measurement of polar motion.

\mainmatter
\chapter{Relativity}
\markboth{RELATIVITY}{RELATIVITY}\label{rel}
\fbox{\parbox{6.5in}{
Relevant IAU resolutions:\quad A4.I, A4.II, A4.III, A4.IV of 1991; B1.3, B1.4, B1.5 of 2000}}
\\
\addcontentsline{toc}{section}{Summary}
\begin{quotation}
\noindent{\bf Summary}\quad
In 2000, the IAU defined a system of space-time coordinates for
(1) the solar system, and (2) the Earth, within the framework of
General Relativity, by specifying the form of the metric tensors
for each and the 4-dimensional space-time transformation between
them.  The former is called the Barycentric Celestial Reference
System (BCRS) and the latter is called the Geocentric Celestial Reference
System (GCRS).  The BCRS is the system appropriate for the basic
ephemerides of solar system objects and astrometric reference data
on galactic and extragalactic objects.  The GCRS is the system
appropriate for describing the rotation of the Earth, the orbits of
Earth satellites, and geodetic quantities such as instrument
locations and baselines. The analysis of precise observations
inevitably involves quantities expressed in both systems and the
transformations between them. 
\end{quotation}

\section{Background}\label{rel.bg}

Although the theory of relativity has been with us for a century
(Einstein's first papers on special relativity were published in
1905), it has only been within the last few decades that it has become
a routine consideration in positional astronomy.  The reason is simply
that the observational effects of both special and general relativity
are small.  In the solar system, deviations from Newtonian physics did
not need to be taken into account --- except for the advance
of the perihelion of Mercury --- until the advent of highly precise
``space techniques'' in the 1960s and 1970s:  radar ranging,
spacecraft ranging, very long baseline interferometry (VLBI), pulsar
timing, and lunar laser ranging (LLR). More recently, even optical
astrometry has joined the list, with wide-angle satellite measurements
(Hipparcos) at the milliarcsecond level.  Currently, the effects of
relativity are often treated as small corrections added to
basically Newtonian developments.  But it has become evident that the
next generation of instrumentation and theory will require a more
comprehensive approach, one that encompasses definitions of such basic
concepts as coordinate systems, time scales, and units of measurement
in a relativistically consistent way.  It may remain the case
that, for many applications, relativistic effects can either be ignored
or handled as second-order corrections to Newtonian formulas.  
However, even in such simple cases, the establishment of a
self-consistent relativistic framework has benefits --- it at least allows
the physical assumptions and the errors involved to be more clearly
understood.

In 1991, the IAU made a series of recommendations concerning how 
the theory of relativity could best be incorporated into positional astronomy.
These recommendations and their implications were studied by several
working groups in the 1990s and some deficiencies were noted.  As
a result, a series of new recommendations was proposed and discussed at
IAU Colloquium 180 \citep{johnston}.  The new recommendations
were passed by the IAU General Assembly in 2000.  It is these
recommendations that are described briefly in this chapter.

In special relativity, the Newtonian idea of absolute time in all inertial
reference systems is replaced by the concept that time runs differently
in different inertial systems, in such a way that the speed of light 
has the same measured value in all of them.  In both Newtonian physics
and special relativity, inertial reference systems are preferred:
physical laws are simple when written in terms of inertial coordinates.  In
general relativity, however, time (and even space-time) is influenced
not only by velocity but also by gravitational fields, and there are no
preferred reference systems.  One can use, in principle, any reference system
to model physical processes.  For an infinitely small space-time region around an
observer (considered to be a massless point), one can introduce so-called locally
inertial reference systems where, according to the Einstein's
equivalence principle, all physical laws have the same form as in an
inertial reference system in special relativity.  Such locally inertial
reference systems are used to describe observations taken by the
point-like observer.  In general-relativistic reference systems of
finite spatial extent, the geometry of space-time is
defined by a {\it metric tensor}, a 4$\times$4 matrix of
mathematical expressions, that serves as an operator on two 4-vectors.
 In its simplest application, the metric tensor directly yields the
generalized (4-dimensional) distance between two neighboring
space-time events.  The metric tensor effectively determines the equations
through which physics is described in the reference system.  

Time in general relativity can be understood as follows.  \label{reltimes}As a particle
moves through space-time, each point (a space-time event) on the path that
it follows can be characterized by a set of four numbers. These four
numbers are the values of the four coordinates in four-dimensional
space-time for a given coordinate system.  For the same path in a different
coordinate system, the numbers will, in general, be different.  {\it Proper time}
is simply the time kept by a clock
co-moving with the particle, in whatever trajectory and gravity field it finds itself.
Proper time is always measurable if a clock is available that can travel with the
particle.  {\it Coordinate time} is one of the four independent variables used to
characterize a space-time event.  Coordinate time is {\it not} measurable.  The 
coordinate time of a reference system is the independent argument of the
equations of motion of bodies in that reference system.
The IAU resolutions on relativity passed in 2000 are concerned
with two coordinate frames, one barycentric and one
geocentric, and the coordinate times used in each one.

\section{The BCRS and the GCRS}\label{rel.rs}

In res.~B1.3 of 2000, the IAU defined two coordinate frames for use in
astronomy, one with its origin at the solar system barycenter and one
with its origin at the geocenter.  In current astronomical usage these
are referred to as {\it reference systems}.  (The astronomical
distinction between {\it reference systems}\/ and {\it reference frames}\/
is discussed in Chapter~\ref{refsys}.)  The two systems are the Barycentric
Celestial Reference System (BCRS) and the Geocentric Celestial
Reference System (GCRS).  Harmonic coordinates are recommended for
both systems (i.e., the harmonic gauge is used).  The resolution provides the
specific forms of the metric tensors for the two coordinate systems and the
4-dimensional transformation between them.  (The latter would reduce to a Lorentz
transformation for a fictitious Earth moving with constant velocity in
the absence of gravitational fields.)  The general forms of the gravitational
potentials, which appear in the metric tensors, are also presented. In
res.~B1.4, specific expansions of the Earth's gravitational potential 
in the GCRS are recommended.  In res.~B1.5, the relationship between the
coordinate time scales for the two reference systems, Barycentric
Coordinate Time (TCB), and Geocentric Coordinate Time (TCG), is given.
Each of the resolutions is mathematically detailed, and the
formulas may be found in the text of the resolutions at the end of
this \pubname. For interested readers, the paper titled ``The IAU
2000 Resolutions for Astrometry, Celestial Mechanics, and Metrology in
the Relativistic Framework: Explanatory Supplement'' \citep{soffel}, is
highly recommended as a narrative on the background, meaning, and
application of the relativity resolutions.  Here we will make only
very general comments on the BCRS and GCRS, although the time scales
TCB and TCG are described in a bit more detail in Chapter~\ref{time}.

The BCRS is a ``global'' reference system in which the positions and
motions of bodies outside the immediate environment of the Earth are
to be expressed.  It is the reference system appropriate for the
solution of the equations of motion of solar system bodies (that is,
the development of solar system ephemerides) and within which the
positions and motions of galactic and extragalactic objects are most
simply expressed.  It is the system to be used for most
positional-astronomy reference data, e.g., star catalogs.   The GCRS
is a ``local'' reference system for Earth-based measurements and the
solution of the equations of motion of bodies in the near-Earth
environment, e.g., artificial satellites.  The time-varying position of
the Earth's celestial pole is defined within the GCRS (res.~B1.7 of
2000).  Precise astronomical observations involve both systems:  the
instrumental coordinates, boresights, baselines, etc., may be
expressed in the GCRS, but in general we want the astronomical results
expressed in the BCRS where they are easier to interpret.  Thus it is
unavoidable that data analysis procedures for precise techniques will
involve both GCRS and BCRS quantities and the transformation between
them.  For example, the basic equation for VLBI delay (the time
difference between wavefront arrivals at two antennas) explicitly
involves vectors expressed in both systems --- antenna-antenna
baselines are given in the GCRS, while solar system coordinates and
velocities and quasar directions are expressed in the BCRS.  Various
relativistic factors connect the two kinds of vectors.

In the 2000 resolutions, the coordinate axes of the two
reference systems do not have a defined orientation.  They are
described as {\it kinematically nonrotating}, which means that the
axes have no systematic rotation with respect to distant objects in
the universe (and specifically the radio sources that make up the ICRF
--- see Chapter~\ref{refsys}).  Since the axis directions are not specified, one
interpretation of the 2000 resolutions is that the BCRS and GCRS in
effect define families of coordinate systems, the members of which
differ only in overall orientation.  The IAU Working Group on
Nomenclature for Fundamental Astronomy has recommended that the
directions of the coordinate axes of the BCRS be understood to
be those of the International Celestial Reference System (ICRS)
described in Chapter~\ref{refsys}.  And, since the transformation
between the BCRS and GCRS is specified in the resolutions, the
directions of the GCRS axes are also implicitly defined by this
understanding.  Here are the definitions of the two
systems recommended by the working group:

\begin{quote} 
{\bf Barycentric Celestial Reference System
(BCRS):}\quad A system of barycentric space-time coordinates for the
solar system within the framework of General Relativity with metric
tensor specified by the IAU 2000 Resolution B1.3. Formally, the metric
tensor of the BCRS does not fix the coordinates completely, leaving the 
final orientation of the spatial axes undefined. However, for all practical
applications, unless otherwise stated, the BCRS is assumed to be
oriented according to the ICRS axes.\pagebreak

{\bf Geocentric Celestial Reference System (GCRS):}\quad A system of
geocentric space-time coordinates within the framework of General
Relativity with metric tensor specified by the IAU 2000 Resolution
B1.3. The GCRS is defined such that the transformation between BCRS and GCRS
spatial coordinates contains no rotation component, so that GCRS is
kinematically non-rotating with respect to BCRS. The
equations of motion of, for example, an Earth satellite with respect
to the GCRS will contain relativistic Coriolis forces that come mainly
from geodesic precession. The spatial orientation of the GCRS is
derived from that of the BCRS, that is, unless otherwise stated, by the
orientation of the ICRS.
\end{quote}

Because, according to the last sentence of the GCRS definition, the orientation
of the GCRS is determined by that of the BCRS, and therefore the ICRS, in this
\pubname\ the GCRS will often be described as the ``geocentric ICRS''.
However, this sentence does not imply that the spatial orientation
of the GCRS is the {\it same} as that of the BCRS (ICRS).  The relative
orientation of these two systems is embodied in the 4-dimensional
transformation given in res.~B1.3 of 2000, which, we will see in the
next section, is itself embodied in the algorithms used to compute
observable quantities from BCRS (ICRS) reference data.  From another
perspective, the GCRS is just a rotation (or series of rotations) of the
international geodetic system (discussed in Chapter~\ref{erot}).  The geodetic
system rotates with the crust of the Earth, while the GCRS has no
systematic rotation relative to extragalactic objects.

The above definition of the GCRS also indicates some of the subtleties
involved in defining the spatial orientation of its axes.  Without the
kinematically non-rotating constraint, the GCRS would have a slow
rotation with respect to the BCRS, the largest component of which is
called geodesic (or de Sitter-Fokker) precession.  This rotation,
approximately 1.9 arcseconds per century, would be inherent in the
GCRS if its axes had been defined as {\it dynamically}\/ non-rotating
rather than {\it kinematically}\/ non-rotating.  By imposing the
latter condition, Coriolis terms must be added (via the inertial parts of the
potentials in the metric; see notes to res.~B1.3 of 2000) to the equations of motion of bodies
expressed in the GCRS.  For example, as mentioned above, the motion of the
celestial pole is defined within the GCRS, and geodesic precession
appears in the precession-nutation theory rather than in the
transformation between the GCRS and BCRS.  Other
barycentric-geocentric transformation terms that affect the equations
of motion of bodies in the GCRS because of the axis-orientation
constraint are described in \citet[section~3.3]{soffel} and
\citet[section~6]{kopeikin}.

\section{Computing Observables}\label{rel.obs}

Ultimately, the goal of these theoretical formulations is to facilitate the
accurate computation of the values of observable astrometric quantities (transit
times, zenith distances, focal plane coordinates, interferometric delays, etc.) at the
time and place of observation, that is, in the {\it proper reference system}\/
of the observer.  There are some subtleties involved because in Newtonian physics
and special relativity, observables are directly related to some inertial
coordinate, while according to the rules of general relativity, observables must
be computed in a coordinate-independent manner. 

In any event, to obtain observables, there are a number of calculations that must be
performed.  These begin with astrometric reference data: a precomputed
solar system ephemeris and, if a star is involved, a star catalog with
positions and proper motions listed for a specified epoch.   The
computations account for the space motion of the object (star or
planet), parallax (for a star) or light-time (for a planet),
gravitational deflection of light, and the aberration of light due to the
Earth's motions.  Collectively, these calculations will be referred to in this
\pubname\ as the algorithms for {\it proper place}.   For Earth-based
observing systems, we must also account for  precession, nutation, Earth rotation,
and polar motion. There are classical expressions for all these
effects (except gravitational deflection), and relativity {\it
explicitly}\/ enters the procedure in only a few places, usually as
added terms to the classical expressions and in the formulas that link
the various time scales used.  It has become common, then, to view
this ensemble of calculations as being carried out entirely in a
single reference system;  or, two reference systems, barycentric and
geocentric, that have parallel axes and differ only in the origin of
coordinates (that is, they are connected by a Galilean
transformation).  For example, the coordinate system defined by the ``equator and
equinox of J2000.0'', can be thought of as either
barycentric or geocentric.  The relativistic effects then are interpreted simply
as ``corrections'' to the classical result.

While such a viewpoint may be aesthetically tidy, it breaks down at
high levels of accuracy and for some types of observations.  Relativity theory leads
to a more correct, albeit more subtle, interpretation for the same set of
calculations.  It is represented by the BCRS-GCRS paradigm wherein some of the
quantities are expressed relative to the BCRS and others are relative to the
GCRS.  The two systems are quite different in a number of ways,
as described in the previous section.  The situation is easiest to describe
if we restrict the discussion to a fictitious observer at the center of
the Earth, that is, to observations referred to the geocenter.  The transformation between
the two systems is not explicit in the normal algorithms, but is embodied in the
relativistic terms in the expressions used for aberration or VLBI
delay.  The distinction between the two systems is most obvious in the
formulation for angular variables.  There, the algorithms for space
motion, parallax, light-time, and gravitational deflection\footnote{In
the case of the observer at the geocenter, we neglect the gravity field of
the Earth itself in computing gravitational deflection.} all use vectors expressed
in the BCRS (star catalogs and solar system ephemerides are inherently
BCRS), while the series of rotations for precession, nutation, Earth
rotation, and polar motion (if applied in that order) starts with vectors expressed
in the GCRS.  In essence, the aberration calculation connects the two systems because it
contains the transformation between them:  its input is a pair of vectors in the
BCRS and its output is a vector in the GCRS.  In the VLBI case, aberration does
not appear explicitly, but the conventional algorithm for the delay observable
incorporates vectors expressed in both systems, with appropriate
conversion factors obtained from the BCRS-GCRS transformation.\footnote{Part of
the expression for VLBI delay, in the time domain, accounts for what would be
called aberration in the angular domain;
it is possible to compute aberration from the VLBI delay algorithm.  See
\citet{kaplan98}.}  

For an observer on or near the Earth's surface the calculations have
to include the position and velocity of the observer relative to the
geocenter.  These are naturally expressed in the GCRS but for some of the
calculations (parallax, light-time, light deflection, and aberration) they must be
added to the position and velocity of the geocenter relative to the
solar system barycenter, which are expressed in the BCRS.  Thus
another GCRS-BCRS transformation is indicated, although the velocity
is sufficiently small that a Galilean transformation (simple vector
addition) suffices for current observational accuracy \citep{klioner}.
Correct use of the resulting vectors results in the
values of the observables expressed, not in the GCRS, but in the proper
reference system of the observer.

\section{Concluding Remarks}\label{rel.remarks}

The 2000 IAU resolutions on relativity define a framework for future
dynamical developments within the context of general relativity. For
example, \citet{klioner} has described how to use the framework to
compute the directions of stars as they would be seen by a precise
observing system in Earth orbit.  However, there is much unfinished
business. The apparently familiar concept of the ecliptic plane has
not yet been defined in the context of relativity resolutions.  A
consistent relativistic theory of Earth rotation is still some years
away; the algorithms described in Chapter~\ref{prenut} are not such a theory,
although they contain all the main relativistic effects and are quite
adequate for the current observational precision.

A local reference system similar to the GCRS can be easily constructed
for any body of an N-body system in exactly the same way as the GCRS, simply
by changing the notation so that the subscript $E$ denotes a body other
than the Earth.  In particular, a celenocentric reference system
for the Moon plays an important role in lunar laser ranging.  

It is also worth noting that the 2000 resolutions do not describe the
proper reference system of the observer --- the local, or topocentric,
system in which most measurements are actually taken.  (VLBI observations are
unique in that they exist only after data from various individual antennas are
combined; therefore they are referred to the GCRS {\it ab initio}.)   A
kinematically non-rotating version of the proper reference system of
the observer is just a simplified version of the GCRS:  $x^i_E$ should
be understood to be the BCRS position of the observer ($v^i_E$ and $a^i_E$ are
then the observer's velocity and acceleration) and one should neglect
the internal potentials.  See \citet{kv93,kopeikin91,kopeikin,klioner04}.

One final point: the 2000 IAU resolutions as adopted apply
specifically to Einstein's theory of gravity, i.e., the general theory
of relativity.  The Parameterized Post-Newtonian (PPN) formalism (see, e.g., 
\citet{will}) is more general, and the 2000 resolutions have been discussed in the PPN
context by \citet{ks00} and \citet{kopeikin}.  In the 2000 resolutions,
it is assumed that the PPN parameters $\beta$ and $\gamma$ are both 1.

\chapter{Time Scales}\label{time}
\markboth{TIME SCALES}{TIME SCALES}

\fbox{\parbox{6.5in}{
Relevant IAU resolutions:\quad A4.III, A4.IV, A4.V, A4.VI of 1991;  C7 of 1994;  B1.3, B1.5, B1.7, B1.8, B1.9, and B2 of 2000}}
\\  
\addcontentsline{toc}{section}{Summary} \begin{quotation} \noindent{\bf Summary}\quad The IAU has not established any new time scales since 1991, but more recent IAU resolutions have redefined or clarified those already in use, with no loss of continuity.  There are two classes of time scales used in astronomy, one based on the SI (atomic) second, the other based on the rotation of the Earth.   The SI second has a simple definition that allows it to be used (in practice or in theory) in any reference system.  Time scales based on the SI second include TAI and TT for practical applications, and TCG and TCB for theoretical developments.  The latter are to be used for relativistically correct dynamical theories in the geocentric and barycentric reference systems, respectively.  Closely related to these are two time scales, TDB and T$_{\rm eph}$, used in the current generation of ephemerides.  Time scales based on the rotation of the Earth include mean and apparent sidereal time and UT1.   Because of irregularities in the Earth's rotation, and its tidal deceleration, Earth-rotation-based time scales do not advance at a uniform rate, and they increasingly lag behind the SI-second-based time scales.  UT1 is now defined to be a linear function of a quantity called the {\it Earth Rotation Angle}, $\theta$.  In the formula for mean sidereal time,  $\theta$ now constitutes the ``fast term''.  The widely disseminated time scale UTC is a hybrid:  it advances by SI seconds but is subject to one-second corrections ({\it leap seconds\/}) to keep it within 0.\ssec9 of UT1.  That procedure is now the subject of debate and there is a movement to eliminate leap seconds from UTC.\end{quotation}

\section{Different Flavors of Time}\label{time.flavors}

The phrase {\it time scale} is used quite freely in astronomical contexts, but there is sufficient confusion surrounding astronomical times scales that it is worthwhile revisiting the basic concept. A time scale is simply a well defined way of measuring time based on a specific periodic natural phenomenon.  The definition of a time scale must provide a description of the phenomenon to be used (what defines a period, and under what conditions), the rate of advance (how many time units correspond to the natural period), and an initial epoch (the time reading at some identifiable event).  For example, we could define a time scale where the swing of a certain kind of pendulum, in vacuum at sea level, defines one second, and where the time 00:00:00 corresponds to the transit of a specified star across a certain geographic meridian on an agreed-upon date.

As used in astronomy, a time scale is an idealization, a set of specifications written on a piece of paper.  The instruments we call clocks, no matter how sophisticated or accurate, provide some imperfect approximation to the time scale they are meant to represent.  In this sense, time scales are similar to spatial reference systems (see Chapter~\ref{refsys}), which have precise definitions but various imperfect realizations.  The parallels are not coincidental, since for modern high-precision applications we actually use space-time reference systems (see Chapter~\ref{rel}).  All time scales are therefore associated with specific reference systems.

Two fundamentally different groups of time scales are used in astronomy.  The first group of time scales is based on the second that is defined as part of the the Syst\`{e}me International (SI) --- the ``atomic'' second --- and the second group is based on the rotation of the Earth.  The SI second is defined as $9,192,631,770$ cycles of the radiation corresponding to the ground state hyperfine transition of Cesium~133 \citep{si}, and provides a very precise and constant rate of time measurement, at least for observers local to the apparatus in which such seconds are counted.  The rotation of the Earth (length of day) is quite a different basis for time, since it is variable and has unpredictable components.  It must be continuously monitored through astronomical observations, now done primarily with very long baseline [radio] interferometry (VLBI).  The SI-based time scales are relatively new in the history of timekeeping, since they rely on atomic clocks first put into regular use in the 1950s.  Before that, all time scales were tied to the rotation of the Earth.  (Crystal oscillator clocks in the 1930s were the first artificial timekeeping mechanisms to exceed the accuracy of the Earth itself.)  As we shall see, the ubiquitous use of SI-based time for modern applications has led to a conundrum about what the relationship between the two kinds of time should be in the future.  Both kinds of time scales can be further subdivided into those that are represented by actual clock systems and those that are simply theoretical constructs.

General reviews of astronomical time scales are given in \citet{pkstf} and Chapter~2 of the \citet{seid92}.

\section{Time Scales Based on the SI Second}\label{time.si}

Let us first consider the times scales based on the SI second.  As a simple count of cycles of microwave radiation from a specific atomic transition, the SI second can be implemented, at least in principle, by an observer anywhere.  Thus, SI-based time scales can be constructed or hypothesized on the surface of the Earth, on other celestial bodies, on spacecraft, or at theoretically interesting locations in space, such as the solar system barycenter.  According to relativity theory, clocks advancing by SI seconds may not appear to advance by SI seconds by an observer on another space-time trajectory.  In general, there will be an observed difference in rate and possibly higher-order or periodic differences, depending on the relative trajectory of the clock and the observer and the gravitational fields involved.  The precise conversion formulas can be mathematically complex, involving the positions and velocities not just
of the clock and observer but also those of an ensemble of massive bodies (Earth, Sun, Moon, planets).  These considerations also apply to coordinate time scales established for specific reference systems.  The time-scale conversions are taken from the general 4-dimensional space-time transformation between the reference systems given by relativity theory (see Chapter~\ref{rel}).  While the rate differences among these time scales may seem inconvenient, the universal use of SI units, including ``local'' SI seconds, means that the values of fundamental physical constants determined in one reference system can be used in another reference system {\it without} scaling factors.

Two SI-second-based times have already been mentioned in Chapter~\ref{rel}: these are {\it coordinate time} scales (in the terminology of General Relativity) for theoretical developments based on the Barycentric Celestial Reference System (BCRS) or the Geocentric Celestial Reference System (GCRS). These time scales are called, respectively, Barycentric Coordinate Time (TCB) and Geocentric Coordinate Time (TCG).  With respect to a time scale based on SI seconds on the surface of the Earth, TCG advances at a rate 6.97\E{-10} faster, while TCB advances at a rate 1.55\E{-8} faster. TCB and TCG are not likely to come into common use for practical applications, but they are beginning to appear as the independent argument for some theoretical developments in dynamical astronomy (e.g., \citet{moisson}).   However, none of the current IAU recommended models used in the analysis of astrometric data use TCB or TCG as a basis, and neither time scale appears in the main pages of \AsA.  This simply reflects the fact that there has not been enough time or motivation for a new generation of dynamical models to be fully developed within the IAU-recommended relativistic paradigm.

For practical applications, International Atomic Time (TAI) is a commonly used time scale based on the SI second on the Earth's surface at sea level (specifically, the rotating geoid). TAI is the most precisely determined time scale that is now available for astronomical use.  This scale results from analyses by the Bureau International des Poids et Mesures (BIPM) in S\`{e}vres, France, of data from atomic time standards of many countries, according to an agreed-upon algorithm.  Although TAI was not officially introduced until 1972, atomic time scales have been available since 1956, and TAI may be extrapolated backwards to the period 1956--1971 (See \citet{nelson} for a history of TAI). An interesting discussion of whether TAI should be considered a {\it coordinate time} or a kind of modified {\it proper time}\footnote{These terms are described in Chapter~\ref{rel}, p.~\pageref{reltimes}.} in the context of general relativity has been given by \citet{guinot86}.   In any event, TAI is readily available as an integral number of seconds offset from UTC, which is extensively disseminated; UTC is discussed at the end of this chapter.  The TAI offset from UTC is designated $\Delta$AT = TAI--UTC.  (For example, from 1999 through 2005, $\Delta$AT = 32~s.)  $\Delta$AT increases by 1~s whenever a positive {\it leap second\/} is introduced into UTC (see below).  The history of $\Delta$AT values can be found on page~K9 of \AsA\ and the current value can be found at the beginning of each issue of \citet{IERSBullA} \citep{url-ierseop}.

The astronomical time scale called Terrestrial Time (TT), used widely for geocentric and topocentric ephemerides (such as in \AsA), is defined to run at a rate of $(1-\mbox{L}_G)$ times that of TCG, where   $\mbox{L}_G = 6.969290134\E{-10}$.  The rate factor applied to TCG to create TT means that TT runs at the same rate as a time scale based on SI seconds on the surface of the Earth.  $\mbox{L}_G$ is now considered a defining constant, not subject to further revision.  Since TCG is a theoretical time scale that is not kept by any real clock, for practical purposes, TT can be considered an idealized form of TAI with an epoch offset:  TT = TAI + $32.\ssec184$.  This expressssion for TT preserves continuity with previously-used (now obsolete) ``dynamical'' time scales, Terrestrial Dynamical Time (TDT) and Ephemeris Time (ET).   That is, ET $\rightarrow$ TDT $\rightarrow$ TT can be considered a single
continuous time scale.\\

{\bf Important Note:}\quad The ``standard epoch'' for modern astrometric reference data, designated J2000.0, is expressed as a TT instant:  J2000.0 is 2000 January 1, $12^{\rm h}$ TT (JD 2451545.0 TT) at the geocenter.\\

The fundamental solar system ephemerides from the Jet Propulsion Laboratory (JPL) that are the basis for many of the tabulations in \AsA\ and other national almanacs were computed in a barycentric reference system and are distributed with the independent argument being a coordinate time scale called T$_{\rm eph}$ (Chapter~\ref{eph} describes the JPL ephemerides). T$_{\rm eph}$ differs in rate 
 from that of TCB, the IAU recommended time scale for barycentric developments; the rate of
T$_{\rm eph}$ matches that of TT, on average, over the time span of the ephemerides.  One may treat T$_{\rm eph}$ as functionally equivalent to Barycentric Dynamical Time (TDB), defined by the IAU in 1976 and 1979.  Both are meant to be ``time scales for equations of motion referred to the barycenter of the solar system'' yet (loosely speaking) match TT in average rate.  The original IAU definition of TDB specified that ``there be only periodic variations''  with respect to what we now call TT (the largest variation is 0.0016~s with a period of one year).  It is now clear that this condition cannot be rigorously fulfilled in practice; see \citet{standish98} for a discussion of the issue and the distinction between TDB and T$_{\rm eph}$.  Nevertheless, space coordinates obtained from the JPL ephemerides are consistent with TDB, and it has been said that ``T$_{\rm eph}$ is what TDB was meant to be.''  Therefore, barycentric and heliocentric data derived from the JPL ephemerides are often tabulated with
TDB shown as the time argument (as in \AsA), and TDB is the specified time argument for many of the equations presented in this \pubname.\footnote{The IAU Working Group on Nomenclature for Fundamental Astronomy
is considering a recommendation to correct the definition of TDB so that a distinction between TDB and T$_{\rm eph}$ would no longer be necessary; TDB would be a linear function of TCB with a rate as close to that of TT as possible over the time span of the ephemeris to which it applies.}   Because T$_{\rm eph}$ ($\approx$TDB) is not based on the SI second, as is TCB, the values of parameters determined from or consistent with the JPL ephemerides will, in general, require scaling to convert them to SI-based units.  This includes the length of the astronomical unit.  Dimensionless quantities such as mass ratios are unaffected. 

The problem of defining relativistic time scales in the solar system has been treated by \citet{brumberg};
the paper is quite general but pre-dates the current terminology.  

\section{Time Scales Based on the Rotation of the Earth}\label{time.erot}

Time scales that are based on the rotation of the Earth are also used in astronomical applications, such as telescope pointing, that depend on the geographic location of the observer.  Greenwich sidereal time is the hour angle of the equinox measured with respect to the Greenwich meridian. Local sidereal time is the local hour angle of the equinox, or the Greenwich sidereal time plus the longitude (east positive) of the observer, expressed in time units.  Sidereal time appears in two forms, mean and apparent, 
depending on whether the mean or true equinox is the reference point.  The position of the mean equinox is affected only by precession while the true equinox is affected by both precession and nutation.  The difference between true and mean sidereal time is the {\it equation of the equinoxes\/},
which is a complex periodic function with a maximum amplitude of about 1~s.
Of the two forms, apparent sidereal time is more relevant to actual observations, since it
includes the effect of nutation.  Greenwich (or local) apparent sidereal time can be observationally obtained from the right ascensions of celestial objects transiting the Greenwich (or local) meridian.

Universal Time (UT) is also widely used in astronomy, and now almost always refers to the specific time scale UT1.  Historically, Universal Time (formerly, Greenwich Mean Time) has been obtained from Greenwich sidereal time using a standard expression.  In 2000, the IAU redefined UT1 to be a linear function of the {\it Earth Rotation Angle}, $\theta$, which is the geocentric angle between two
directions in the equatorial plane called, respectively, \pagebreak                          
the Celestial Intermediate Origin (CIO)
and the Terrestrial Intermediate Origin (TIO) (res.~B1.8 of 2000\footnote{In the resolution, these points are called the Celestial Ephemeris Origin (CEO) and the Terrestrial Ephemeris Origin (TEO).  The change in terminology has been recommended by the IAU Working Group on Nomenclature for Fundamental Astronomy and will probably be adopted at the 2006 IAU General Assembly.}). The TIO rotates with the Earth, while the CIO has no instantaneous rotation around the Earth's axis, so that $\theta$ is a direct measure of the Earth's rotational motion:
$\dot\theta = \omega$, the Earth's average angular velocity of rotation.   See Chapter~\ref{erot} for a more complete description of these new reference points.  The definition of UT1 based on sidereal time is still widely used, but the definition based on $\theta$ is becoming more common for precise applications.  In fact, the two definitions are equivalent, since the expression for sidereal time as a function of UT1 is itself now based on $\theta$.

Since they are mathematically linked, sidereal time, $\theta$, and UT1 are all affected by variations in the Earth's rate of rotation (length of day), which are unpredictable and must be routinely measured through astronomical observations. The lengths of the sidereal and UT1 seconds, and the value of $\dot\theta$, are therefore not precisely constant when expressed in a uniform time scale such as TT.  The accumulated difference in time measured by a clock keeping SI seconds on the geoid from that measured by the rotation of the Earth is $\Delta T$ = TT--UT1.  A table of observed and extrapolated values of $\Delta T$ is given in \AsA\ on page~K9.   The long-term trend is for $\Delta T$ to increase gradually because of the tidal deceleration of the Earth's rotation, which causes UT1 to lag increasingly behind TT.

In predicting the precise times of topocentric phenomena, like solar eclipse contacts, both TT and UT1 come into play.  Therefore, assumptions have to be made about the value of $\Delta T$ at the time of the phenomenon.  Alternatively, the circumstances of such phenomena can be expressed in terms of an imaginary system of geographic meridians that rotate uniformly about the Earth's axis ($\Delta T$ is assumed zero, so that UT1=TT) rather than with the real Earth; the real value of $\Delta T$ then does not need to be known when the predictions are made.  The zero-longitude meridian of the uniformly rotating system is called the {\it ephemeris meridian}. As the time of the phenomenon approaches and the value of $\Delta T$ can be estimated with some confidence, the predictions can be related to the real Earth: the uniformly rotating system is 1.002738 $\Delta T$ east of the real system of geographic meridians. (The 1.002738 factor converts a UT1 interval to the equivalent Earth Rotation Angle --- i.e., the sidereal/solar time ratio.)

\section{Coordinated Universal Time (UTC)}\label{time.utc} 

The worldwide system of civil time is based on Coordinated Universal Time (UTC), which is now ubiquitous and tightly synchronized.  (This is the de facto situation; most nations' legal codes, including that of the U.S., do not mention UTC specifically.)  UTC is a hybrid time scale, using the SI second on the geoid as its fundamental unit, but subject to occasional 1-second adjustments to keep it within 0.\ssec9 of UT1.  Such adjustments, called {\it leap seconds}, are normally introduced at the end of June or December, when necessary, by international agreement.  Tables of the remaining difference, UT1--UTC, for various dates are published by the International Earth Rotation and Reference System Service (IERS), at \citet{url-ierseop}.  Both past observations and predictions are available.  DUT1, an approximation to UT1--UTC, is transmitted in code with some radio time signals, such as those from WWV.  As previously discussed in the context of TAI,  the difference $\Delta$AT = TAI--UTC is an integral number of seconds,  a number that increases by 1 whenever a (positive) leap second is introduced into UTC.  That is, UTC and TAI share the same seconds ticks, they are just labeled differently.

Clearly UT1--UTC and $\Delta T$ must be related, since they are both measures of the natural ``error'' in the Earth's angle of rotation at some date.  In fact, $\Delta T$ = 32.\ssec184 + $\Delta$AT -- (UT1--UTC).

For the user, then, UTC, which is widely available from GPS, radio broadcast services, and the Internet, is the practical starting point for computing any of the other time scales described above.  For the SI-based time scales, we simply add the current value of $\Delta$AT to UTC to obtain TAI.  TT is then just 32.\ssec184 seconds ahead of TAI.  The theoretical time scales TCG, TCB, TDB, and T$_{\rm{eph}}$ can be obtained from TT using the appropriate mathematical formulas.  For the time scales based on the rotation of the Earth, we again start with UTC and add the current value of UT1--UTC to obtain UT1.  The various kinds of sidereal time can then be computed from UT1 using standard formulas.    

\begin{center}
\includegraphics[width=5.5in,trim=0.0in 0.3in 0.0in 0.1in,clip=true]{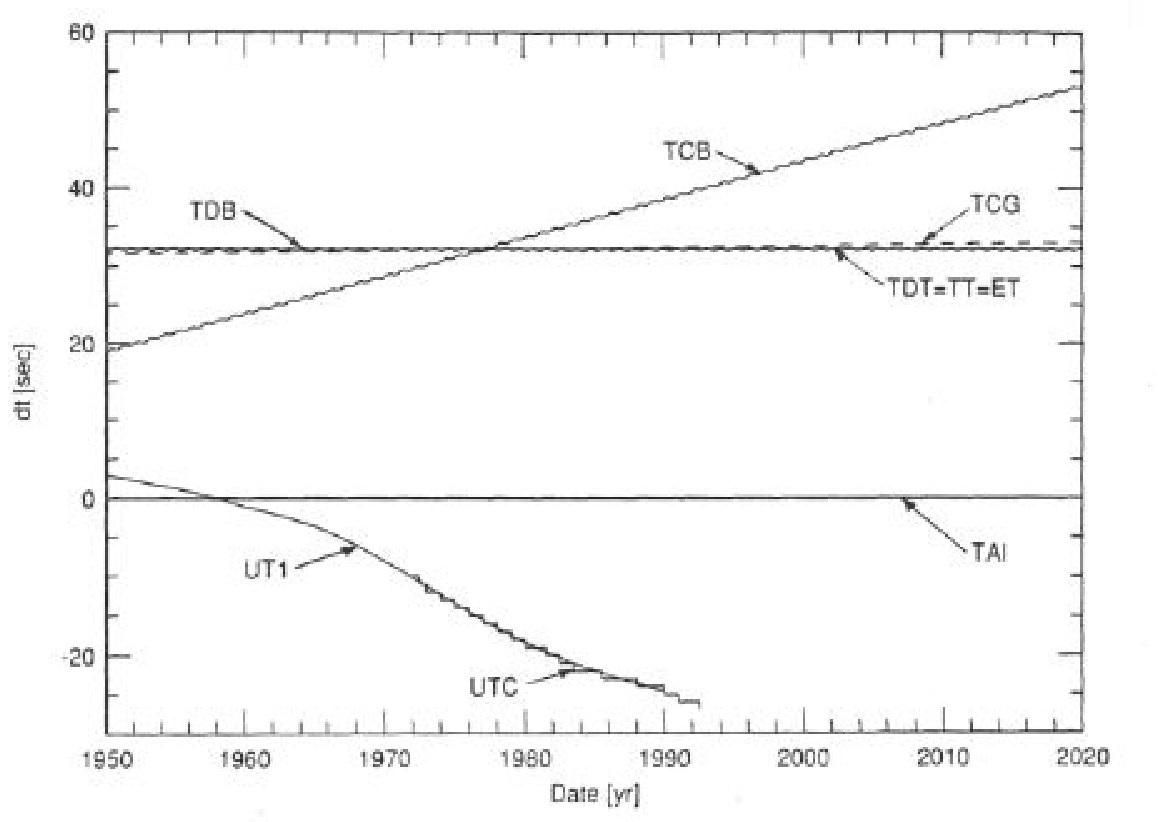}
\parbox{5.0in}{\small\vspace{0.2in}{\bf Figure 2.1}\quad Differences in readings of various time
scales compared to International Atomic Time (TAI).  TT and its predecessors, TDT and ET, are all
shown as TAI+32.184~s. The periodic terms of TCB and TDB are exaggerated by a factor of 100.  The
``stair-step'' appearance of UTC is due to the leap seconds inserted into that time scale so that it tracks UT1.  TT and the ``steps'' of UTC are parallel to the TAI line because they are all based on the SI second on the geoid.  TDB (or T$_{\rm eph}$) tracks TT on average over the time span of the specific ephemeris to which it applies.  Note the instant at the beginning of 1977 when TT, TCB, and TCG all had the same value.  The figure is from \citet{pkstf}.}\\[6ex]
\end{center}

\section{To Leap or Not to Leap}\label{time.leap}

Because of the widespread and increasing use of UTC for applications not considered three decades ago --- such as precisely time-tagging electronic fund transfers and other networked business transactions --- the addition of leap seconds to UTC at unpredictable intervals creates technical problems and legal issues for service providers.  There is now a movement to relax the requirement that UTC remain within 0.9 seconds of UT1.  The issue is compounded by the unavoidable scientific fact that the Earth's rotation is slowing due to tidal friction, so that the rate of addition of leap seconds to UTC must inevitably increase.  Aside from monthly, annual, and decadal variations, the Earth's angular velocity of rotation is decreasing linearly, which means that the accumulated lag in UT1 increases quadratically;  viewed over many centuries, the $\Delta T$ curve is roughly a parabola.  The formulas for sidereal time, and length of the old {\it ephemeris second} to which the SI second was originally calibrated, are based on the average (assumed fixed) rate of Earth rotation of the mid-1800s  \citep{nelson}.  All of our modern timekeeping systems are ultimately based on what the Earth was doing a century and a half ago.  

An IAU Working Goup on the Redefinition of Universal Time Coordinated (UTC) was established to consider the leap second problem and recommend a solution, working with the IERS, the International Union of Radio Science (URSI), the Radiocommunication Sector of the International Telecommunications Union (ITU-R), the International Bureau for Weights and Measures (BIPM), and the relevant navigational agencies (res.~B2  of 2000).  Possibilities include: using TAI for technical applications instead of UTC; allowing UT1 and UTC to diverge by a larger amount (e.g., 10 or 100 seconds) before a multi-second correction to UTC is made;  making a variable correction to UTC at regularly scheduled dates; eliminating the corrections to UTC entirely and allowing UTC and UT1 to drift apart; or changing the definition of the SI second.  No solution is ideal (including the status quo) and each of these possibilities has its own problems. For example, if we keep leap seconds, or a less frequent multi-second correction, can current systems properly time-tag the date and time of an event that occurs {\it during} the correction?  Does a time scale that diverges from UT1 provide a legally acceptable representation of civil time?  If corrections are made less frequently, will the possibility of technical blunders increase?  If leap seconds are eliminated, won't natural phenomena such as sunrise and sunset eventually fall out of sync with civil time?  How do we find all the existing computer code that assumes $|\mbox{UT1--UTC}|\le 0.9\,\mbox{s}$?  The matter is now being considered by the ITU-R, where a working group has proposed eliminating leap seconds from UTC entirely.  Contact Dr.\ Dennis McCarthy, U.S. Naval Observatory, dmc@maia.usno.navy.mil, for a copy of a report or if you wish to comment.  In any event, it would take a number of years for any proposed change to take place because of the many institutions and international bodies that would have to be involved.  

For scientific instrumentation, the use of TAI --- which is free of leap seconds --- has much to recommend it.  Its seconds can be easily synchronized to those of UTC (only the labels of the seconds are different).  It is straightforward to convert from TAI to any of the other time scales.  Use of TAI provides an internationally recognized time standard and avoids the need to establish an instrument-specific time scale when continuity of time tags is a requirement.   

\section{Formulas}\label{time.formulas}

\subsection{Formulas for Time Scales Based on the SI Second}\label{time.formulas.si}

For the SI-based time scales, the event tagged 1977 Jan\-uary~1, 00:00:00 TAI (JD 2443144.5 TAI) at the geocenter is special.  At that event, the time scales TT, TCG, and TCB all read 1977 January~1, 00:00:32.184 (JD 2443144.5003725).  (The 32.\ssec184 offset is the estimated difference between TAI and the old Ephemeris Time scale.)  This event will be designated $t_0$ in the following; it can be represented in any of the time scales, and the context will dictate which time scale is appropriate. 

From the perspective of a user, the starting point for computing all the time scales is Coordinated Universal Time (UTC).   From UTC, we can immediately get International Atomic Time (TAI):
\begin{equation}
                \TAI = \UTC + \Delta\mbox{AT}
\end{equation}
where $\Delta$AT, an integral number of seconds, is the accumulated number of leap seconds applied to UTC.  

The astronomical time scale Terrestrial Time (TT) is defined by the epoch $t_0$ and its IAU-specified rate with respect to Geocentric Coordinate Time (TCG):
\begin{equation}
   \frac{d\TT}{d\TCG} = 1-\mbox{L}_G \quad\mbox{where}\quad \mbox{L}_G = 6.969290134\E{-10}\;           \mbox{(exactly)}
\end{equation}
from which we obtain
\begin {equation}
    \TT = \TCG-\mbox{L}_G\,(\TCG-t_0)
\end{equation}
However, TCG is a theoretical time scale, not kept by any real clock system, so in practice,
\begin{equation}
    \TT = \TAI + 32.\ssec184
\end{equation}
and we obtain TCG from TT.

The relationship between TCG and Barycentric Coodinate Time (TCB) is more complex.  TCG and TCB are both coordinate time scales, to be used with the geocentric and barycentric reference systems (the GCRS and BCRS), respectively.  The exact formula for the relationship between TCG and TCB is given in res.~B1.5 of 2000, recom\-menda\-tion~2.   For a given TCB epoch, we have
\begin{equation}
\TCG =  \TCB - \frac{1}{c^2} \int_{t_0}^{\rm TCB} \Big( \frac{v_e^2}{2} + U_{ext}({\bf x}_e) \Big) dt 
      - \frac{{\bf v}_e}{c^2} \cdot ({\bf x} - {\bf x}_e) + \cdots
\end{equation}
where $c$ is the speed of light, ${\bf x}_e$ and ${\bf v}_e$ are the position and velocity vectors of the Earth's center with respect to the solar system barycenter, and $U_{ext}$ is the Newtonian potential of all solar system bodies apart from the Earth.  The integral is carried out in TCB since the positions and motions of the Earth and other solar system bodies are represented (ideally) as functions of TCB.   The last term on the right contains the barycentric position vector of the point of interest, ${\bf x}$, and will be zero for the geocenter, as would normally be the case.  The omitted terms are of maximum order $c^{-4}$.  Note that the transformation is ephemeris-dependent, since it is a function of the time series of ${\bf x}_e$ and ${\bf v}_e$ values.  The result is a ``time ephemeris'' associated with every spatial ephemeris of solar system bodies expressed in TCB.   It is to be expected that ephemeris developers will supply appropriate time conversion algorithms (software) to allow the positions and motions of solar system bodies to be retrieved for epochs in conventional time scales such as TT or TAI.  It is unlikely that ordinary ephemeris users will have to compute eq.~(2.5) on their own. 

The functional form of the above expressions may seem backwards for practical applications; that is, they provide TCG from TCB and TT from TCG.  These forms make sense, however, when one considers how an ephemeris of a solar system body (or bodies) or a spacecraft is developed.  The equations of motion for the body (or bodies) of interest are expressed in either the barycentric or geocentric system as a function of some independent coordinate time argument.  For barycentric equations of motion, expressed in SI units, we would be tempted immediately to identify this time argument  with TCB.  Actually, however, the association of the time argument with TCB is not automatic; it comes about only when the solution of the equations of motion is made to satisfy the boundary conditions set by the ensemble of real observations of various kinds.  Generally, these observations will be time-tagged in UTC, TAI, or TT (all of which are based on the SI second on the geoid) and these time tags must be associated with the time argument of the ephemeris.  The above formulas can be used to make that association, which then allows the ephemeris to be fit to the observations.  (More precisely, the space-time coordinates of the observation events must be transformed to the BCRS.)  As a consequence, the time argument of the ephemeris becomes a realization of TCB.   The fit of the computed ephemeris to observations usually proceeds iteratively, and every iteration of the spatial ephemeris produces a new time ephemeris.  With each iteration, the spatial coordinates of the ephemeris become better grounded in reality (as represented by the observations) and the time coordinate becomes a better approximation to TCB.  Viewed from this computational perspective, the ephemeris and its time argument are the starting point of the process and the sequence TCB~$\rightarrow$ TCG~$\rightarrow$ TT makes sense.

One can compute an ephemeris and fit it to observations using other formulas for the time scale conversions.  A completely valid and precise ephemeris can be constructed in this way, but its independent time argument could not be called TCB.   The values of various constants used in, or derived from, such an ephemeris would also not be SI-based and a conversion factor would have to be applied to convert them to or from SI units.   Such is the case with the solar system Development Ephemeris (DE) series from the Jet Propulsion Laboratory.  DE405 is now the consensus standard for solar system ephemerides and is described in Chapter~\ref{eph}.  The DE series dates back to the 1960s, long before TCB and TCG were defined, and its independent time argument is now called T$_{\rm eph}$.  T$_{\rm eph}$ can be considered to be TCB with a rate factor applied.  Or, as mentioned above, T$_{\rm eph}$ can be considered to be functionally equivalent to the time scale called TDB.  Both T$_{\rm eph}$ and TDB advance, on average, at the same rate as TT.  This arrangement makes accessing the DE ephemerides straightforward, since for most purposes, TT can be used as the input argument with little error.  The total error in time in using TT as the input argument is $<$2~ms, which for the geocentric position of the Moon would correspond to an angular error of $<$1~mas.   When more precision is required, the following formula can be used:
\begin{eqnarray}
           \mbox{T}_{\rm eph} \approx \TDB \approx \TT
            & + & 0.001657\, \sin \,(  628.3076\,T + 6.2401)                 \nonumber\\
            & + & 0.000022\, \sin \,(  575.3385\,T + 4.2970)                 \nonumber\\
            & + & 0.000014\, \sin \,(1256.6152\,T + 6.1969)                 \nonumber\\ 
            & + & 0.000005\, \sin \,(  606.9777\,T + 4.0212)                                     \\ 
            & + & 0.000005\, \sin \,(    52.9691\,T + 0.4444)                 \nonumber\\  
            & + & 0.000002\, \sin \,(    21.3299\,T + 5.5431)                 \nonumber\\
            & + & 0.000010\,T \sin \,( 628.3076\,T + 4.2490) \; + \cdots  \nonumber
\end{eqnarray}
where the coefficients are in seconds, the angular arguments are in radians, and $T$ is the number of Julian centuries of TT from J2000.0: $T = (\mbox{JD(TT)} - 2451545.0) / 36525$.    The above is a truncated form of a much longer and more precise series given by \citet{fairhead}.  The maximum error in using the above formula is about 10 $\mu s$ from 1600 to 2200;  that is, its precision is more than two orders of magnitude better than the approximation T$_{\rm eph}$$\approx$TDB$\approx$TT.  For even more precise applications, the series expansion by \citet{harada} is recommended.

A word of caution:  The idea that ``T$_{\rm eph}$ and TDB advance, on average, at the same rate as TT'' is problematic.  The independent time argument of a barycentric ephemeris (whether considered to be T$_{\rm eph}$, TDB, or TCB) has a large number of periodic components with respect to TT.  Some of the periods are quite long, and may extend beyond the time period of the ephemeris.  Thus, the ``average rate'' of the time argument of the ephemeris, with respect to TT, depends on the averaging method and the time span considered.  Differences in rate of some tens of microseconds per century are possible \citep{fairhead}.  These rate ambiguities are probably unimportant (amounting to fractional errors of only $\sim$$10^{-14}$) for retrieving positions and velocities from the ephemeris but do affect pulsar timing that is reduced to the barycentric time scale.

\subsection{Formulas for Time Scales Based on the Rotation of the Earth}\label{time.formulas.erot}

For the time scales that are based on the rotation of the Earth, we again start with UTC.
\begin{eqnarray}
   \UTI &     =        & \UTC + \mbox{(UT1--UTC)}  \\
           & \approx & \UTC + \mbox{DUT1}
\end{eqnarray}
where DUT1 is a broadcast approximation to UT1--UTC (precision $\pm$0.\ssec1).  We also have
\begin{equation}
     \UTI = \TT - \Delta T
\end{equation}
where $\Delta T$ = 32.\ssec184 + $\Delta$AT -- (UT1--UTC).    The most recent values of UT1--UTC and $\Delta$AT are listed in \citet{IERSBullA} \citep{url-ierseop}.  Values of $\Delta T$ are listed in \AsA\ on page K9. 

The Earth Rotation Angle, $\theta$, is 
\begin{equation}
    \theta =  0.7790572732640 + 1.00273781191135448 \; D_U
\end{equation}
where $D_U$ is the number of UT1 days from 2000 January~1, 12$^{\rm h}$ UT1:  $D_U$ = JD(UT1) -- 2451545.0.  The angle $\theta$ is given in terms of rotations (units of 2$\pi$ radians or 360\degr).  The above rate coefficient gives an Earth rotation period of 86164.0989036903511 seconds of UT1.  If we consider this to be the adopted average rotation period of the Earth in SI seconds, it is consistent with the nominal mean angular velocity of Earth rotation, $\omega$ = 7.292115\E{-5} radian s$^{-1}$, used by the International Association of Geodesy.  The above expression is taken directly from note 3 to res.~B1.8 of 2000.  An equivalent form of this expression (if the integral number of rotations is neglected) that is usually more numerically precise is
\begin{equation}
    \theta =  0.7790572732640 + 0.00273781191135448 \; D_U + \mbox{frac(JD(UT1))}
\end{equation}
where frac(JD(UT1)) is the fractional part of the UT1 Julian date, i.e., JD(UT1) modulus 1.0.  Then Greenwich mean sidereal time in seconds is
\begin{eqnarray}
      \GMST = & 86400\cdot\theta\quad  + &(0.014506 + 4612.156534\,T + 1.3915817\,T^2 \nonumber\\
                   &  & - 0.00000044\,T^3 - 0.000029956\,T^4 - 0.0000000368\,T^5) / 15  
\end{eqnarray}
where $T$ is the number of centuries of TDB (equivalently for this purpose, TT) from J2000.0:
$T$ = (JD(TDB) -- 2451545.0)/36525.  The polynomial in parentheses is the accumulated precession of the equinox in right ascension, in arcseconds, as given for the P03 solution (eq.~42) in \citet{cap03}.  Note that two time scales are now required to compute sidereal time: 
in the ``fast term'', $\theta$ is a function of UT1, while in the remaining terms, $T$ is
expressed in TDB (or TT).

To obtain Greenwich apparent sidereal time in seconds, we must add the {\it equation of the equinoxes}:
\begin{equation}
             \GAST = \GMST + \mathcal{E}_{\Upsilon} / 15
\end{equation}
which accounts for the motion of the equinox due to nutation.  An extended series is now used for the equation of the equinoxes.  The new series includes so-called {\it complementary terms} and more fully accounts for the accumulated effect of combined precession and nutation on the position of the equinox.
The equation of the equinoxes in arcseconds is
\begin{eqnarray}
       \mathcal{E}_{\Upsilon} = & \D & \Delta\psi \, \cos\epsilon          \nonumber\\
             & + &  0.00264096 \,\sin \,( \Omega )                                       \nonumber\\
             & + &  0.00006352 \,\sin \,( 2 \Omega )                                   \nonumber\\
             & + & 0.00001175 \,\sin \,( 2 F - 2 D + 3 \Omega )                \nonumber\\
             & + & 0.00001121 \,\sin \,( 2 F - 2 D + \Omega )                    \nonumber\\
             & - &  0.00000455 \,\sin \,( 2 F - 2 D + 2 \Omega )                                    \\
             & + & 0.00000202 \,\sin \,( 2 F            + 3 \Omega )               \nonumber\\
             & + & 0.00000198 \,\sin \,( 2 F            + \Omega )                  \nonumber\\
             & -  & 0.00000172 \,\sin \,( 3 \Omega )                                    \nonumber\\
             & - &  0.00000087 \, T \,\sin \,( \Omega ) + \cdots                  \nonumber
\end{eqnarray}
where $\Delta\psi$ is the nutation in longitude, in arcseconds; $\epsilon$ is the mean obliquity of the ecliptic; and $F$, $D$, and $\Omega$ are fundamental luni-solar arguments.  All of these quantities are functions of TDB (or TT);  see Chapter~\ref{prenut} for expressions (esp. eqs.~5.12, 5.15, \& 5.19).  The above series is a truncated form of a longer series given in the \citet{iers03}, but should be adequate for almost all practical applications.

Local mean sidereal time (LMST) and local apparent sidereal time (LAST) in seconds can then be computed respectively from
\begin{equation}
         \mbox{LMST} = \GMST + \left(\frac{3600}{15}\right)\,\lambda    \qquad\mbox{and}\qquad
         \mbox{LAST} = \GAST + \left(\frac{3600}{15}\right)\,\lambda
\end{equation}
where $\lambda$ is the longitude of the place of interest, in degrees, positive for places east of Greenwich.

In the above, ``Greenwich'' actually refers to a plane containing the geocenter, the Celestial Intermediate Pole (CIP), and the point called the Terrestrial Intermediate Origin (TIO).  These concepts are described in Chapters~\ref{prenut} and \ref{erot}.  Loosely, the CIP is the rotational pole, defined by the precession and nutation theories.  For astronomical purposes, the TIO can be considered to be a point on the rotational equator (the plane orthogonal to the CIP) essentially fixed at geodetic longitude 0.  Strictly, then, the longitude $\lambda$ should be measured around the axis of the CIP from the TIO to the location of interest. Because of polar motion, the pole of the conventional system of geodetic coordinates is not at the CIP so the longitude we need is not quite the same as the geodetic longitude.  The longitude, in degrees, to be used in eq.~2.15 is
\begin{equation}
                \lambda = \lambda_G + ( x_p \sin\lambda_G + y_p \cos\lambda_G) \, \tan \phi_G \;/3600
\end{equation}
where $\lambda_G$ and $\phi_G$ are the usual geodetic longitude and latitude of the place, with $\lambda_G$ in degrees (north latitudes and east longitudes are positive); and $x_p$ and $y_p$ are the coordinates of the pole (CIP) with respect to the geodetic system, in arcseconds  ($x_p$ and $y_p$ can be a few tenths of an arcsecond).  The geodetic system is formally the International Terrestrial Reference System (ITRS), which matches WGS-84 (available from GPS) to several centimeters.  The local meridian assumed by the formula for LAST, using the longitude $\lambda$, passes through the local zenith (orthogonal to the local surface of the WGS-84 ellipsoid) and the north and south celestial poles --- close to but not through the local geodetic north and south points.  This is the meridian that all stars with apparent topocentric right ascension equal to LAST will pass over at time UT1.  More information can be found in sections~\ref{erot.transform}, \ref{erot.formulas.geod}, and \ref{erot.formulas.ha}. 

   The above formulas are entirely geometric.  Not described are astronomical latitude and longitude, which are based on the local direction of gravity.  Astronomical latitude and longitude are affected by the deflection of the vertical caused by permanent gravitational anomalies and, at a much lower level, semidiurnal tides.  Astronomical latitude and longitude must be corrected for such effects to obtain geodetic latitude and longitude.

\chapter{The Fundamental Celestial Reference System}\label{refsys} 
\markboth{CELESTIAL REFERENCE SYSTEM}{CELESTIAL REFERENCE SYSTEM}

\fbox{\parbox{6.5in}{
Relevant IAU resolutions:\quad A4.VI, A4.VII of 1991; B5 of 1994; B2 of 1997; B1.2 of 2000}}
\\
\addcontentsline{toc}{section}{Summary}
\begin{quotation}
\noindent{\bf Summary}\quad Reference data for positional
astronomy, such as the data in astrometric star catalogs or
barycentric planetary ephemerides, are now specified within the
International Celestial Reference System (ICRS).  The ICRS is a
coordinate system whose origin is at the solar system barycenter
and whose axis directions are effectively defined by the adopted
coordinates of 212 extragalactic radio sources observed by
VLBI.  These radio sources (quasars and active galactic nuclei)
are assumed to have no observable intrinsic angular motions. Thus,
the ICRS is a ``space-fixed'' system (more precisely, a
kinematically non-rotating system) without an associated epoch.
However, the ICRS closely matches the conventional dynamical
system defined by the Earth's mean equator and equinox of J2000.0;
the alignment difference is at the 0.02 arcsecond level,
negligible for many applications.

Strictly speaking, the ICRS is somewhat of an abstraction, a
coordinate system that perfectly satisfies a list of criteria. The
list of radio source positions that define it for practical
purposes is called the International Celestial Reference Frame
(ICRF).  In the terminology that is now commonly used,
 a {\it reference system} like the ICRS is
``realized'' by a {\it reference frame} like the ICRF, and there
can be more than one such realization.  In the case of the ICRS,
there is, in fact, a second, lower-accuracy realization for work
at optical wavelengths, called the Hipparcos Celestial Reference
Frame (HCRF).  The HCRF is composed of the positions and proper
motions of the astrometrically well-behaved stars in the Hipparcos
catalog.

Astrometric data referred to the ICRS is becoming more common, with 
new catalogs now available in the optical, infrared, and radio.  

The ICRS is itself a specific example of a Barycentric Celestial
Reference System, incorporating the relativistic
metric specified in res.~B1.3 of 2000 for solar system barycentric
coordinate systems.  In other words, the ICRS provides the orientation
of the BCRS axes. 
\end{quotation}

\section{The ICRS, the ICRF, and the HCRF}\label{refsys.rs}

The fundamental celestial reference system for astronomical
applications is now the International Celestial Reference System
(ICRS), as provided in res.~B2 of 1997.   The ICRS is a coordinate system 
with its origin at the solar system barycenter and axis directions that are fixed
with respect to distant objects in the universe;  it is to be used to express
the positions and motions of stars, planets, and other celestial objects.  Its
relativistic basis is defined by res.~B1.3 of 2000;  in the words of
that resolution, it is a Barycentric Celestial Reference System (BCRS), and
as such its axes are {\it kinematically non-rotating} (see Chapter~\ref{rel}).
To establish the ICRS as a practical system, the IAU specified
a set of distant benchmark objects, observable at radio wavelengths, whose
adopted coordinates effectively define the directions of the ICRS axes.
This ``realization'' of the ICRS, called the
International Celestial Reference Frame (ICRF), is a set of high
accuracy positions of extragalactic radio sources measured by very
long baseline interferometry \citep{ma97,ma98}.   The
ICRS is realized at optical wavelengths --- but
at lower accuracy --- by the Hipparcos Celestial Reference Frame
(HCRF), consisting of the \textit{Hipparcos Catalogue}
\citep{hip} of star positions and motions, with certain exclusions
(res.~B1.2 of 2000).  The coordinates of the ICRF radio sources
and HCRF stars are given relative to the ICRS origin at the solar system
barycenter, and a number of transformations are required to obtain the
coordinates that would be observed from a given location on Earth
at a specific date and time.

Although the directions of the ICRS coordinate axes are not
defined by the kinematics of the Earth, the ICRS axes (as
implemented by the ICRF and HCRF) closely approximate the axes
that would be defined by the mean Earth equator and equinox of
J2000.0 (to within about 0.02 arcsecond), if the latter is considered to be
a barycentric system.  Because the ICRS axes are meant to be
``space fixed'', i.e., kinematically non-rotating, there is no date
associated with the ICRS.  Furthermore, since the defining radio sources
are assumed to be so distant that their angular motions, seen
from Earth, are negligible, there is no epoch associated with the ICRF. It
is technically incorrect, then, to say that the ICRS is a
``J2000.0 system'', even though for many current data sources, the
directions in space defined by the equator and equinox of J2000.0
and the ICRS axes are the same to within the errors of the data.

The ICRS, with its origin at the solar system barycenter and 
``space fixed'' axis directions, is meant to represent the most
appropriate coordinate system currently available for expressing
reference data on the positions and motions of celestial objects.

The IAU Working Group on Nomenclature for Fundamental Astronomy
has recommended the following definitions for the ICRS and ICRF:

\begin{quote}
{\bf International Celestial Reference System (ICRS):}\quad The idealized
barycentric coordinate system to which celestial positions are referred.
It is kinematically non-rotating with respect to the ensemble of distant extragalactic objects. It has no intrinsic orientation but was aligned close to the mean equator and dynamical equinox of J2000.0 for 
continuity with previous fundamental reference systems. Its orientation 
is independent of epoch, ecliptic or equator and is realized by a list 
of adopted coordinates of extragalactic sources.

{\bf International Celestial Reference Frame (ICRF):}\quad A set of
extragalactic objects whose adopted positions and uncertainties realize 
the ICRS axes and give the uncertainties of the axes.  It is also the name of 
the radio catalogue whose 212 defining sources are currently the most 
accurate realization of the ICRS. Note that the orientation of the ICRF 
catalogue was carried over from earlier IERS radio catalogs and was 
within the errors of the standard stellar and dynamic frames at the time 
of adoption. Successive revisions of the ICRF are intended to minimize 
rotation from its original orientation. Other realizations of the ICRS 
have specific names (e.g. Hipparcos Celestial Reference Frame).
\end{quote}

\section{Background: Reference Systems and Reference Frames}\label{refsys.bg}

The terminology that has become standard over the past decade or
so distinguishes between a {\it reference system\/} and a {\it
reference frame\/}.  A {\it reference system\/} is the complete
specification of how a celestial coordinate system is to be
formed.  Both the origin and the orientation of the 
fundamental planes (or axes) are defined.  A reference system
also incorporates a specification of the fundamental models
needed to construct the system; that is, the basis for the algorithms used to
transform between observable quantities and reference data in the system.
A {\it reference frame\/}, on the other hand, consists of a set of identifiable fiducial
points on the sky along with their coordinates, which serves as
the practical realization of a reference system.

For example, the fundamental plane of an astronomical reference
system has conventionally been the extension of the Earth's
equatorial plane, at some date, to infinity.  
Declination is the angular distance
north or south of this plane, and right ascension
is the angular distance measured eastward along the equator
from some defined reference point.  This reference point, the
right ascension origin, has traditionally been the {\it
equinox\/}: the point at which the Sun, in its yearly circuit of
the celestial sphere, crosses the equatorial plane moving from
south to north. The Sun's apparent yearly motion lies in the {\it
ecliptic}, the plane of the Earth's orbit.  The equinox,
therefore, is a direction in space along the nodal line defined by
the intersection of the ecliptic and equatorial planes;
equivalently, on the celestial sphere, the equinox is at one of
the two intersections of the great circles representing these
planes. Because both of these planes are moving, the coordinate
systems that they define must have a date associated with them;
such a reference system must therefore be specified as ``the
equator and equinox of (some date)''.

Of course, such a reference system is an idealization, because the
theories of motion of the Earth that define how the two planes move are
imperfect.  In fact, the very definitions of these planes are problematic
for high precision work.  Even if the fundamental planes of a reference
system are defined without
any reference to the motions of the Earth, there is no way magically
to paint them on the celestial sphere at any particular time.  Therefore,
in practice, we use a specific reference frame --- a set of fiducial
objects with assigned coordinates --- as the practical representation of an
astronomical reference system.  The scheme is completely analogous to
how terrestrial reference systems are established using survey 
control stations (geodetic reference points) on the Earth's surface.

Most commonly, a reference frame consists of a catalog of precise
positions (and motions, if measurable) of stars or extragalactic
objects as seen from the solar system barycenter at a specific
epoch (now usually ``J2000.0'', which is 12$^{\rm h}$ TT on 1~January
2000).  Each object's instantaneous position, expressed as right
ascension and declination, indicates the object's angular
distance from the catalog's equator and origin of right
ascension.  Any two such objects in the catalog (if they are not coincident or antipodal) 
therefore uniquely orient a spherical coordinate system on the sky --- a
reference frame.

A modern astrometric catalog contains data on a large number of
objects ($N$), so the coordinate system is vastly overdetermined.
The quality of the reference frame defined by a catalog depends on
the extent to which the coordinates of all possible pairs of
objects ($\approx N^2/2$) serve to define the identical equator
and right ascension origin, within the expected random errors.
Typically, every catalog contains {\it systematic errors}, that
is, errors in position that are similar 
for objects that are in the same area of the sky, or are of the
same magnitude (flux) or color (spectral index).  Systematic
errors mean that the reference frame is warped, or is effectively
different for different classes of objects. Obviously, minimizing
systematic errors when a catalog is constructed is at least as important
as minimizing the random errors.

To be useful, a reference frame must be implemented at the time of
actual observations, and this requires the computation of the
apparent coordinates of the catalog objects at arbitrary dates
and times.  The accuracy with which we know the motions of the
objects across the sky is an essential factor in this
computation.  Astrometric star catalogs list {\it proper motions},
which are the projection of each star's space motion onto the
celestial sphere, expressed as an angular rate in right ascension
and declination per unit time.  Because the tabulated proper
motions are never perfect, any celestial
reference frame deteriorates with time.  Moreover, systematic
errors in the proper motions can produce time-dependent warpings
and spurious rotations of the frame.  Therefore, the accuracy and
consistency of the proper motions are critical to the overall
quality, utility, and longevity of reference frames defined by
stars.   Even reference frames defined by extragalactic objects, which
are usually considered to have zero proper motion, may deteriorate,
because many of these objects show small apparent motions
that are artifacts of their emission mechanisms. 

The positions of solar system objects can also be used to define a
reference frame.  For each solar system body involved, an {\it
ephemeris\/} (pl.\ {\it ephemerides\/}) is used, which is simply a
table of the celestial coordinates of the body as a function of
time (or an algorithm that yields such a table).  A reference
frame defined by the ephemerides of one or more solar system
bodies is called a {\it dynamical reference frame\/}.  Because the
ephemerides used incorporate the motion of the Earth
as well as that of the other solar system bodies, dynamical
reference frames embody in a very fundamental way the moving
equator and ecliptic, hence the equinox.  They have therefore been
used to correct the orientation of star catalog reference frames (the star
positions were systematically adjusted) on the basis of
simultaneous observations of stars and planets.  In a sense,
the solar system is used as a gyrocompass.  However,
dynamical reference frames are not very practical for establishing
a coordinate system for day-to-day astronomical observations. 

Descriptions of reference frames and reference systems often refer
to three coordinate axes, which are simply the set of right-handed
cartesian axes that correspond to the usual celestial spherical
coordinate system.  The xy-plane is the equator, the z-axis points
toward the north celestial pole, and the x-axis points toward the
origin of right ascension.  Although in principle this allows us
to specify the position of any celestial object in rectangular
coordinates, the distance scale (based on stellar parallaxes) is not
established to high precision beyond the solar system.   What a
reference system actually defines is the way in which the two conventional
astronomical {\it angular\/} coordinates, right ascension and
declination, overlay real observable points in the sky.   (See
eqs. 5.1 \& 5.2 for the conversion between rectangular and
spherical celestial coordinates.)

\section{Recent Developments}\label{refsys.recent}

The establishment of celestial reference systems is coordinated
by the IAU.  The previous
astronomical reference system was based on the equator and
equinox of J2000.0 determined from observations of planetary
motions, together with the IAU (1976) System of Astronomical
Constants and related algorithms \citep{circ163}.  The reference frame that
embodied this system for practical purposes was the Fifth
Fundamental Catalogue (FK5).  The FK5 is a catalog of 1535
bright stars (to magnitude 7.5), supplemented by a fainter
extension of 3117 additional stars (to magnitude 9.5), compiled 
at the Astronomische Rechen-Institut in Heidelberg \citep{fk5,fk5ext}.
The FK5 was the successor to the FK3 and FK4 catalogs, all of which
were based on meridian observations taken in the visual band --- many
such observations were, in fact, taken by eye.  The formal
uncertainties in the star positions of the FK5 at the time of its
publication in 1988 were about 30--40
milliarcseconds over most of the sky, but the errors are
considerably worse when systematic trends are taken into account.

In recent years, the most precise wide-angle astrometry
has been conducted not in the optical regime but at radio
wavelengths, involving the techniques of very long baseline
interferometry (VLBI) and pulsar timing.  Uncertainties of radio
source positions listed in all-sky VLBI catalogs are now
typically less than one milliarcsecond, and often a factor of ten
better.  Furthermore, because these radio sources are very
distant extragalactic objects (mostly quasars) that are not
expected to show measurable intrinsic motion, a reference frame
defined by VLBI positions should be ``more inertial'' (less subject
to spurious rotation) than a reference frame defined by galactic
objects such as stars or pulsars.  The VLBI catalogs do have the
disadvantage that their origin of right ascension is somewhat
arbitrary; there is no real equinox in VLBI catalogs, since VLBI
has little sensitivity to the ecliptic plane.  However, this
problem has turned out to be more conceptual than practical,
since methods have been developed to relate the VLBI right
ascension origin to the equinox as conventionally defined.

Because of these considerations, since the mid 1980s,
astronomical measurements of the Earth's rotation --- from which
astronomical time is determined --- have depended heavily on
VLBI, with classical methods based on star transits being phased
out.  Hence the situation evolved to where the definition of the
fundamental astronomical reference frame (the FK5) became
irrelevant to some of the most precise and important astrometric
measurements.  VLBI revealed, in addition, that the models of the
Earth's precession and nutation that were part of the old system
were inadequate for modern astrometric precision.  In particular,
the ``constant of precession'' --- a measurement of the long-term
rate of change of the orientation of the Earth's axis in space
--- had been overestimated by about 0.3 arcseconds per century.
Moreover, the success of the European Space Agency's Hipparcos
astrometric satellite, launched in 1989, promised to provide a
new, very accurate set of star coordinates in the optical regime.

Thus, beginning in 1988, a number of IAU working groups began
considering the requirements for a new fundamental astronomical
reference system \citep{lieske90,hughes}.  The resulting
series of IAU resolutions, passed in 1991, 1994, 1997, and 2000
effectively form the specifications for the ICRS.  The axes of the ICRS are
defined by the adopted positions of a specific set of
extragalactic objects, which are assumed to have no measurable
proper motions.  The ICRS axes are consistent, to about 0.02
arcsecond, with the equator and equinox of J2000.0 defined by the
dynamics of the Earth.  However, the ICRS axes are meant to be
regarded as fixed directions in space that have an existence
independent of the dynamics of the Earth or the particular set of
objects used to define them at any given time.

\citet{feissel} have written a concise review of the
ICRS adoption and its implications.  \citet{seid02} published a broader
review of the ICRS and the new IAU Earth orientation models. 

The promotion, maintenance, extension, and use of the ICRS are the
responsibilities of IAU Division~1 (Fundamental Astronomy).

\section{ICRS Implementation}\label{refsys.icrs}

\subsection{The Defining Extragalactic Frame}\label{refsys.icrs.define}

The {\it International Celestial Reference Frame (ICRF)} is a
catalog of adopted positions of 608 extragalactic radio sources
observed with VLBI, all strong ($>$0.1~Jy) at S and X bands
(wavelengths 13 and 3.6~cm) \citep{ma97,url-icrf}.  Most have faint optical
counterparts (typically $m_V \gg 18$) and the majority are
quasars.  Of these objects, 212 are {\it defining sources\/} that
establish the orientation of the ICRS axes, with origin at the
solar system barycenter.  Typical position uncertainties for the
defining sources are of order 0.5 milliarcsecond; the orientation
of the axes is defined from the ensemble to an accuracy of about
0.02 milliarcsecond.  As described in section \ref{refsys.icrs.other}, these
axes correspond closely to what would conventionally be described
as ``the mean equator and equinox of J2000.0''.

The International Earth Rotation and Reference Systems Service
(IERS) monitors the radio sources involved in the ICRF.  This
monitoring is necessary because, at some level, most of the
sources are variable in both flux and structure and the centers
of emission can display spurious motions.  It is possible that,
eventually, the defining list of sources will have to be amended
to maintain the fixed orientation of the overall frame.

\subsection{The Frame at Optical Wavelengths}\label{refsys.icrs.optical}

The ICRS is realized at optical wavelengths by stars in the
Hipparcos Catalogue of 118,218 stars, some as faint as
visual magnitude 12 \citep{hip}.  Only
stars with uncomplicated and well-de\-ter\-mined proper motions (e.g., no
known binaries) are used for the ICRS realization.  This subset,
referred to as the Hipparcos Celestial Reference Frame (HCRF), comprises
85\% of the stars in the Hipparcos catalog.  Hipparcos star coordinates
and proper motions are given within the ICRS (J2000.0) coordinate system
but are listed for epoch J1991.25.  (That is, the catalog effectively
represents a snapshot of the positions and motions of the stars taken on
2~April 1991, a date that is near the mean epoch of the
Hipparcos observations.)    At the catalog epoch, Hipparcos uncertainties for
stars brighter than 9th magnitude have median values somewhat better
than 1~milli\-arc\-second in position and 1~milli\-arc\-second/year in
proper motion \citep{hip,mignard}.  Thus, pro\-jected to epoch
J2000.0, typical Hipparcos star position errors are in the range 5--10
milliarcseconds.

\subsection{Standard Algorithms}\label{refsys.icrs.algor} 

Chapters \ref{rel}, \ref{time}, \ref{prenut}, and \ref{erot} of this \pubname\ describe
IAU-sanctioned algorithms used in the construction, maintenance, and use of the ICRS.

The 2000 IAU resolutions provide the relativistic metric tensors for what it called the Barycentric Celestial Reference System (BCRS) and the Geocentric Celestial Reference System (GCRS), as well as
expressions for the transformation between the two systems; see Chapter~\ref{rel} and
res.~B1.3 of 2000.   As noted in Chapter~\ref{rel}, the resolutions specify only the relativistic
basis of the two reference systems, and there is no prescription given for establishing the axis directions.
The BCRS and GCRS could therefore be considered families of coordinate systems, each member differing from the others only in overall orientation.  The construction of the ICRS (in particular, the analysis of VLBI observations) was consistent with the definition of the BCRS in the resolutions.
Thus, the ICRS could
be considered one implementation of a BCRS; i.e., a member of the BCRS family.  Recently, the
IAU Working Group on Nomenclature for Fundamental Astronomy has recommended
that the orientation of the BCRS axes be understood to be that
of the ICRS/ICRF.   

In 2000, the IAU also adopted new models for the computation of the Earth's instantaneous orientation,
which affect the analysis of VLBI observations that are used to define and maintain the ICRS, as
well as the calculation of various observable quantities from ICRS-compatible reference data.
The new models include what is referred to as
the {\it IAU 2000A precession-nutation model}, a new definition of
the celestial pole, and two new reference points in the equatorial plane
for measuring the rotational angle of
the Earth around its instantaneous axis.  Despite the IAU action
in 2000, some aspects of the models were not finalized until late
2002 (mid 2005 for agreement on the final precession expressions).
These algorithms are described in Chapters~\ref{prenut} and~\ref{erot} of this \pubname\ and
in the \citet{iers03}.  

The new Earth orientation models are, of course, relevant only to
fundamental observations made from the surface of the Earth.
Astrometric observations taken from space platforms, or those
that are differential in nature (based on reference objects all located
within a small field), do not use these models.  There are, of course, other
effects that must be taken into account
in analyzing astrometric observations --- e.g., proper motion, parallax, aberration, and
gravitational light-bending --- and algorithms for these may be found in Volumes~1 and 3 of
the Hipparcos Cata\-logue docu\-mentation \citep{hip}.  For analysis of very high
accuracy satellite observations, see the de\-velop\-ment by \citet{klioner}.

As described in the Introduction, there are two collections of 
general-purpose computer subroutines that implement the new 
IAU-sanctioned algorithms for practical applications:  the Standards of Fundamental Astronomy (SOFA),
at \citet{url-sofa}, and the
Naval Observa\-tory Vec\-tor Astrometry Subroutines (NOVAS), at 
\citet{url-novas}.   NOVAS also implements
many of the Hipparcos algorithms, or the equivalent. 

For ground-based applications requiring accuracies of no better
than 50 milliarcseconds between about 1990 and 2010, the
algorithms described in Chapter~3 of the \citet{seid92} can still be used
with ICRS data.  (For such purposes, ICRS data can be treated as being
on the dynamical equator and equinox of J2000.0.)   A major revision of the 
Explanatory Supplement to reflect the adoption of the ICRS and all the
new models is in progress.

\subsection{Relationship to Other Systems}\label{refsys.icrs.other}

The orientation of the ICRS axes is consistent with the equator and
equinox of J2000.0 represented by the FK5, within the errors of the latter.
See \citet{feissel} for a short discussion.  Systematically,
the FK5 equator is tilted by 22~mas  and its origin of right ascension is
offset by 23~mas with respect to the ICRS.   But the uncertainties of
the FK5 equator and right ascension system with respect to the dynamical equator
and equinox of J2000.0 are 50 and 80~mas, respectively.     Since, at J2000.0, the
errors of the FK5 are significantly worse than
those of Hipparcos, the ICRS (as realized by the HCRF) can be considered to be
a refinement of the FK5 system \citep{hip} at (or near) that epoch.  

The ICRS can also be considered to be a good approximation (at
least as good as the FK5) to the conventionally defined dynamical
equator and equinox of J2000.0 \citep{feissel}, if the latter system is
considered to be barycentric.  This follows from an IAU resolution passed
in 1991 that provided the original specifications for the new fundamental
astronomical reference system based on extragalactic objects --- what
became the ICRS.  In fact, the equator is
well determined fundamentally from the VLBI observations that are the
basis for the ICRS, \pagebreak  
and the ICRS pole is within 20~milliarcseconds of the dynamical
pole.\footnote{The reason that the ICRS pole is not perfectly aligned with
the dynamical pole is complex.  The ICRF was created from almost 20 years
of VLBI observations from which a grand solution was made for the directions
to the extragalactic radio sources and the changing position of the
celestial pole.  A specific decision made in that analysis (see Figure~5.2
and footnote on page \pageref{notefig5.2}) resulted in an offset of the ICRF (hence
ICRS) pole at J2000.0.  With respect to the ICRS X and Y axes, the dynamical
mean pole has coordinates on the unit sphere, in milliarcseconds, of
approximately (-16.6,-6.8).}
The zero point of VLBI-derived right ascensions is arbitrary, but traditionally has
been set by assigning to the right ascension of source 3C~273B a value
derived from lunar occultation timings --- the Moon's ephemeris thus
providing an indirect link to the dynamical equinox.  The ICRS origin of
right ascension was made to be consistent with that in a group of VLBI
catalogs previously used by the IERS, each of which had been individually
aligned to the lunar occultation right ascension of 3C~273B.   The
difference between the ICRS
origin of right ascension and the dynamical equinox has been independently
measured by two groups that used different definitions of the equinox, but
in both cases the difference found was less than 0.1~arcsecond.

Because of its consistency with previous reference systems,
implementation of the ICRS will be transparent to any
applications with accuracy requirements of no better than 0.1
arcseconds near epoch J2000.0.  That is, for applications of this
accuracy, the distinctions between the ICRS, FK5, and dynamical
equator and equinox of J2000.0 are not significant.

\subsection{Data in the ICRS}\label{refsys.icrs.data} 

Although the ICRF and HCRF are its basic radio and optical
realizations, the ICRS is gradually being extended to fainter
magnitudes and other wavelengths.  Thus, an increasing amount of
fundamental astronomical data is being brought within the new
system.  A number of projects for the densification of the ICRS
have been completed or are in progress.

As described above, the ICRF consists of the adopted positions of
about 600 extragalactic radio sources, a third of which are
defining sources.  In its original presentation, the ICRF
contained 608 extragalactic radio sources, including 212 defining
sources.  All observational data were part of a common catalog
reduction \citep{ma97} and thus the adopted coordinates of all the sources
are in the ICRS.  Of the 396 non-defining sources, 294 are {\it
candidate sources\/} that do not meet all of the accuracy and
observing history requirements of the defining sources but which
may at some later time be added to the defining list.  The
remaining 102 {\it other sources\/} show excess apparent position
variation and are of lower astrometric quality.  ICRF Extension~2 
(ICRF-Ext.2) was issued in 2004 \citep{fey}; the positions of the candidate and other sources
were refined and 109 new sources were added.   The positions of the defining
sources were left unchanged. 

The VLBA Calibrator Survey is a list of radio sources,
 with positions in the ICRS, to be used as calibrators for the Very
Long Baseline Array and the Very Large Array.  The original
list was prepared by \citet{beasley}; the list has been extended several times
and the current version is known as VCS3 \citep{petrov}.
The ICRS is also being established at higher radio frequencies (24 and
43 GHz); see, e.g., \citet{jacobs}.

In the optical regime, the Tycho-2 Catalogue \citep{tyc2} (which supersedes the original
Tycho Catalogue and the ACT Reference Catalog) combines a re-analysis of the
Hipparcos star mapper observations with data from 144 
ground-based star catalogs.  The ground-based catalogs include
the Astrographic Catalogue (AC), a large photographic project
carried out near the beginning of the 20th~century involving 20
observatories worldwide.  Tycho-2 contains 2,539,913 stars, to about magnitude 12, 
and combines the accuracy of the recent Hipparcos position measurements
with proper motions derived from a time baseline of almost a
century.  Proper motion uncertainties are
1-3~milliarcseconds/year.  At epoch J2000.0, the Tycho-2 positions
of stars brighter than 9th magnitude will typically be in error
by 20~milliarcseconds.  However, the positional accuracy degrades
quite rapidly for magnitudes fainter than 9, so that 12th
magnitude stars may be expected to have a median J2000.0 position
error exceeding 100~milliarcseconds.

Also in the optical band, the UCAC catalog is nearing completion and 
will provide ICRS-compatible positions and proper motions for stars as faint as 
visual magnitude 16.  See \citet{zach} for information on the second
release of UCAC data. 

The ICRS has been extended to the near infrared through the 2MASS  
survey \citep{cutri}.  This ground-based program provides  
positions for 471 million point sources, most of which are stars,  
observed in the J, H, and K$_{\rm s}$ infrared bands.

The Jet Propulsion Laboratory DE405/LE405 planetary and lunar
ephemerides (usually just referred to as DE405) \citep{DE405} have been
aligned to the ICRS.  These ephemerides provide the positions and
velocities of the nine major planets and the Moon with respect to
the solar system barycenter, in rectangular coordinates.  The
data are represented in Chebyshev series form and Fortran
subroutines are provided to read and evaluate the series for any
date and time.  DE405 spans the years 1600 to 2200; a long
version, DE406, spans the years --3000 to +3000 with lower
precision.  See Chapter~\ref{eph}.

The barycentric data tabulated in \AsA\/ are in the ICRS beginning with the 2003 edition.
Planetary and lunar ephemerides are derived from DE405/LE405.  \AsA\/ for 2006
is the first edition fully to support the new ICRS-related algorithms, including
the new IAU Earth rotation models.   Geocentric coordinates are therefore
given with respect to the GCRS.

\section{Formulas}\label{refsys.formulas}

A matrix {\bf B} is required to convert ICRS data to the dynamical mean
equator and equinox of J2000.0 (the ``J2000.0 system''), the latter
considered for this purpose to be a barycentric system.  The same matrix is used
in the geocentric transformations described in Chapters~\ref{prenut} and \ref{erot}
to adjust vectors in the GCRS (the ``geocentric ICRS'') so that they can be
operated on by the conventional precession and nutation matrices.
The matrix {\bf B} is called the {\it frame bias matrix}, and it corresponds to a fixed
set of very small rotations.  In the barycentric case it is used as follows:
\begin{equation}
        {\bf r}_{\rm mean(2000)} = {\bf B} \; {\bf r}_{_{\rm ICRS}}
\end{equation}
where ${\bf r}_{_{\rm ICRS}}$ is a vector with respect to the ICRS and
${\bf r}_{\rm mean(2000)}$ is a vector with respect to the dynamical mean equator and
equinox of J2000.0.  Both of the {\bf r}'s are column vectors and, if they represent
a direction on the sky, are of the general form
\begin{equation}
{\bf r} = \left( \begin{array}{c}
                        \cos\delta \, \cos\alpha  \\
                        \cos\delta \, \sin\alpha   \\
                        \sin\delta
               \end{array} \right)
\end{equation}
where $\alpha$ is the right ascension and $\delta$ is the declination, with respect to
the ICRS or the dynamical system of J2000.0, as appropriate.  

In the geocentric case, ${\bf r}_{_{\rm ICRS}}$ is replaced by ${\bf r}_{_{\rm GCRS}}$ and
${\bf r}_{\rm mean(2000)}$ is then a geocentric vector.  This transformation must
be carried out, for example, before precession is
applied to GCRS vectors, since the precession algorithm assumes a dynamical
coordinate system.  That is, the above transformation is a necessary step in obtaining
coordinates with respect to the mean equator and equinox of date, if one starts with
ICRS reference data.  See Chapter~\ref{prenut} for more information.

The matrix {\bf B}  is, to first order,
\begin{equation}
{\bf B} = \left( \begin{array}{ccc}
                        1                  &  d\alpha_0 & -\xi_0      \\
                        -d\alpha_0 & 1                  & -\eta_0   \\
                        \xi_0            & \eta_0         & 1
               \end{array} \right) 
\end{equation}
where $d\alpha_0 = -14.6$ mas, $\xi_0 = -16.6170$ mas, and 
$\eta_0 = -6.8192$ mas, all converted to radians (divide by
$206\,264\,806.247$).   The values of the three small angular offsets
are taken from the \citet{iers03}.  They can be considered adopted values;
previous investigations of the dynamical-ICRS relationship obtained
results that differ at the mas level or more,
depending on the technique and assumptions.  See the discussion
in \citet{hilton}.  The angles $\xi_0$ and $\eta_0$ are the ICRS pole offsets,
and $d\alpha_0$ is the offset in the ICRS right ascension origin 
with respect to the dynamical equinox of J2000.0, as measured in an
inertial (non-rotating) system.

The above matrix can also be used to transform vectors from the ICRS to the
FK5 system at J2000.0.   Simply substitute $d\alpha_0 = -22.9$ mas, $\xi_0 = 9.1$ mas, and 
$\eta_0 = -19.9$ mas.   However, there is also a time-dependent rotation of the
FK5 with respect to the ICRS (i.e., a slow spin), reflecting the non-inertiality of
the FK5 proper motions; see \citet{mignard00}.

Although the above matrix is adequate for most applications,
a more precise result can be obtained by using the second-order version:
\begin{equation}
{\bf B} = \left( \begin{array}{ccc}
               1 - \frac12 ( d\alpha_0^2 + \xi_0^2)                             &  d\alpha_0    &   -\xi_0     \\
               -d\alpha_0 - \eta_0\xi_0 & 1 - \frac12 ( d\alpha_0^2 + \eta_0^2)       & -\eta_0   \\
                \xi_0 - \eta_0 d\alpha_0 & \eta_0 + \xi_0 d\alpha_0  & 1 - \frac12 ( \eta_0^2 + \xi_0^2 )  
             \end{array} \right)
\end{equation}
The above matrix, from \citet{slabinski}, is an excellent approximation to the 
set of rotations ${\bf R}_1(-\eta_0){\bf R}_2(\xi_0){\bf R}_3(d\alpha_0)$,
where ${\bf R}_1$, ${\bf R}_2$, and ${\bf R}_3$ are standard rotations
about the x, y, and z axes, respectively (see ``Abbreviations and Symbols
Frequently Used'' for precise definitions).  

\chapter{Ephemerides of the Major Solar System Bodies}\label{eph}
\markboth{EPHEMERIDES}{EPHEMERIDES}
\fbox{\parbox{6.5in}{
Relevant IAU resolutions:\quad (none)}}
\\
\addcontentsline{toc}{section}{Summary}
\begin{quotation}
\noindent{\bf Summary}\quad
The de facto standard source of accurate data on the positions and motions of the major solar system bodies  is currently the ephemeris designated DE405/LE405 (or simply DE405) developed at the Jet Propulsion Laboratory.  This ephemeris provides instantaneous position and velocity vectors
of the nine major planets and the Earth's Moon, with respect to the solar system barycenter, for any date and time between 1600 and 2201.  Lunar rotation angles are also provided. The ephemeris has been the basis for the tabulations in \AsA\/ since the 2003 edition.  The DE405 coordinate system has been aligned to the ICRS.  

IAU-standard data on the sizes, shapes, rotational parameters, and latitude-longitude systems for the major planets and their satellites are given in the reports of the IAU/IAG Working Group on Cartographic Coordinates and Rotational Elements, issued every three years.

\end{quotation}

\section{The JPL Ephemerides}\label{eph.jpl}

A list of positions of one or more solar system bodies as a function of time is called an {\it ephemeris\/} (pl. {\it ephemerides}).  An ephemeris can take many forms, including a printed tabulation, a sequential computer file, or a piece of software that, when interrogated, computes the requested data from series approximations or other mathematical schemes.

Ephemerides of the major solar system bodies, with respect to the solar system barycenter, have been calculated for many years at the Jet Propulsion Laboratory (JPL) to support various spacecraft missions.  These ephemerides have been widely distributed and, because of their quality, have become the de facto standard source of such data for applications requiring the highest accuracy.  Between the early 1980s and about 2000, the JPL ephemeris designated DE200/LE200 was most frequently used for such applications; it was the basis for the tabulations in \AsA\/ from the 1984 to 2002 editions.  A more recent JPL ephemeris, DE405/LE405, has now come into widespread use, and has been the basis for \AsA\/ since the 2003 edition.   These ephemerides are usually referred to as just DE200 and DE405, respectively.  Neither DE200 nor DE405 have been the subject of any IAU resolution, although they have been frequently reported on at various IAU-sponsored meetings, and DE405 is a recommended standard of the IERS \citep{iers03}.   A comparison of DE405 with DE200, with an estimate of their errors, has been given by \citet{standish04}.     

The JPL ephemerides are computed by an N-body numerical integration, carried out in a barycentric reference system which is consistent, except for the time scale used, with the Barycentric Celestial Reference System (BCRS) described in Chapter~\ref{rel}.   The equations of motion, the method of integration, and the techniques used to adjust the starting conditions of the integration so that the results are an optimal fit to observations are described in Chapter~5 of  the \citet{seid92}.   That chapter specifically describes DE200, but the basic procedures are largely the same for all of the JPL ephemerides.  

The position and velocity data provided by the JPL ephemerides represent  the centers of mass of each planet-satellite system (although data for the Earth and Moon can be extracted separately).  Therefore, the positions, when converted to geocentric apparent places --- angular coordinates as seen from Earth --- do not precisely indicate the center of the apparent planetary disk.   Displacements can amount to a few tens of milliarcseconds for Jupiter and Saturn, a few milliarcseconds for Uranus and Neptune, and about 0.1 arcsecond for Pluto.       

\section{DE405}\label{eph.de405}

The JPL\hspace{0.1em} DE405/LE405 ephemeris provides the coordinates and velocities of the major planets, the Sun, and the Earth's Moon for the period 1600 to 2200 \citep{DE405}.   The position and velocity 3-vectors are in equatorial rectangular coordinates referred to the solar system barycenter.   The reference frame for the DE405 is the ICRF; the alignment onto this frame, and therefore onto the ICRS, has an estimated accuracy of a few milliarcseconds, at least for the inner-planet data.  Optical, radar, laser, and spacecraft observations were analyzed to determine starting conditions for the numerical integration and values of fundamental constants such as the Earth/Moon mass ratio and the length of the astronomical unit in meters.  In addition to the planetary and lunar coordinates and velocities, the ephemerides, as distributed, include the nutation angles of the Earth and the rotation (libration) angles of the Moon.  (Note, however, that the nutation angles are not derived from the IAU 2000A theory described in Chapter~\ref{prenut}.)


As described in Chapter~\ref{time}, DE405 was developed in a barycentric
reference system using T$_{\rm eph}$, a barycentric coordinate time \citep{standish98}.
T$_{\rm eph}$ is rigorously equivalent to TCB in a mathematical sense, differing only in rate:
the rate of T$_{\rm eph}$ matches the average rate of TT, while the rate of TCB is defined by
the SI system.  The IAU time scale TDB, often (but erroneously) considered to be the
same as T$_{\rm eph}$,  is a quantity that cannot be physically realized, due to its
flawed definition.  So, in fact, the use of the name TDB actually refers to quantities based on or
created with T$_{\rm eph}$.\footnote{Because of this, the IAU Working Group on Nomenclature for Fundamental Astronomy has recommended changing the definition of TDB to be consistent
with that of T$_{\rm eph}$.}  Astronomical constants obtained from ephemerides
based on T$_{\rm eph}$ (or TDB) are not in the SI system of units and must
therefore be scaled for use with TCB or other SI-based time scales.

The ephemerides are distributed by JPL as plain-text (ASCII) computer files of Chebyshev series coefficients and Fortran source code.  Third-party C versions of the code are also available and, for Unix users, the data files can be downloaded in binary form.  Once the system is installed on a given computer, a Fortran subroutine named PLEPH can be called to provide the position and velocity of any requested body, at any date and time;  PLEPH supervises the process of reading the Chebyshev file and evaluating the appropriate series.  Normally the position and velocity vectors returned are expressed in units of AU and AU/day, respectively.  The date/time requested must be expressed as a T$_{\rm eph}$ or TDB Julian date.  (An entry named DPLEPH is provided that allows the input Julian date to be split into two parts for greater precision.)  And, since $|\mbox{TT}-\mbox{T}_{\rm eph}|<0.002$~s, a TT Julian date
may be used for applications not requiring the highest accuracy.  The data and software files can be obtained on CD-ROM from Willmann-Bell, Inc. (\citet{url-jplephem}a), or downloaded from a JPL ftp server (\citet{url-jplephem}b).   A ``README'' file provides export information and software documentation (available separately at \citet{url-jplreadme}).

An extended version of DE405/LE405, called DE406/LE406, is available that spans the years --3000 to +3000, but with coordinates given to lower precision (they are represented by shorter Chebyshev series).  The nutation angles and the lunar rotation angles are also omitted from the DE406 files.  DE406 is provided only in Unix binary format.  These files are about 1/3 the size of those for DE405 for a given span of time.  The additional error in the coordinates (DE406 -- DE405) may amount to 25~m for the planets and 1~m for the Moon, which may be significant for some applications. 

The NOVAS software package mentioned in the Introduction provides an interface to an existing DE405 or DE406 installation through Fortran subroutine SOLSYS or C function {\it ephemeris}.

\section{Sizes, Shapes, and Rotational Data}\label{eph.rot}

The IAU/IAG\footnote{IAG = International Association of Geodesy} Working Group on Cartographic Coordinates and Rotational Elements \citep{url-wgccre} produces a report every three years (for each IAU General Assembly) giving the best estimates for the dimensions and rotational parameters of the planets, satellites, and asteroids, as far as is known.  The working group is also responsible for establishing latitude-longitude coordinate systems for these bodies.  The rotational elements given in the report for the 2000 General Assembly \citep{seid02b} serve to orient these coordinate systems within the ICRS as a function of time.  (Note that the time scale in this report is TDB, not TCB as stated.)  The working group's reports are the basis for the physical ephemerides of the planets given in \AsA.  

Although the rotational elements of the Earth and Moon are given in each report for completeness, the expressions given there provide only an approximation to the known motions and should not be used for precise work (e.g., for the Earth, precession is accounted for only to first order and nutation is neglected).  Lunar rotation (libration) angles can be obtained from DE405, and Chapters~\ref{prenut}
and~\ref{erot} of this \pubname\ describe algorithms for the precise instantaneous alignment of the terrestrial coordinate system within the GCRS (the ``geocentric ICRS'').
   
\section{DE405 Constants}\label{eph.const}

The constants below were used in or determined from the DE405 ephemeris.  Many DE405 constants are expressed in what have been called ``TDB units'', rather than SI units.  (Perhaps it is more proper to say that
the constants have TDB-compatible values.)   A scaling factor, $K = 1/(1 - L_B)$, where $L_B$ is given below, is required to convert such constants, with dimensions of length or time, to SI units \citep{irwin}.  Dimensionless quantities such as mass ratios do not require scaling.  
\begin{equation}
    L_B = 1.550\,519\,767\,72\E{-8}  \quad\Longrightarrow\quad K = 1.000\,000\,015\,505\,198
\end{equation}
The value of $L_B$ is taken from note~3 of res.~B1.5 of 2000.

These conversions also depend on assumptions about what constants are truly fundamental and what their values are \citep{klioner05}; therefore, the SI value of the astronomical unit in meters is not given below.  The planetary masses below include contributions from satellites and atmospheres.\\[0.2in]  

 \par	
 \begin{tabular}{lcll}
{\bf Constant} & {\bf Symbol} & {\bf DE405 Value} & {\bf Equivalent SI value} \vspace{0.8ex}\\
astronomical unit in seconds & $\tau_A$ & 499.004\,783\,806\,1 s &  499.004\,786\,385\,2 s\\
astronomical unit in meters & $A$ or $c\tau_A$ & 149\,597\,870\,691 m\\  
heliocentric gravitational & $GS$ or $GM\!\ssun$ & 1.327\,124\,400\,2\E{20} & 1.327\,124\,420\,8\E{20}\\
\qquad constant & & \qquad\quad m$^3$s$^{-2}$ & \qquad\quad m$^3$s$^{-2}$\\
Moon/Earth mass ratio & $\mu$ & 1/81.30056 & \\ 
Sun/planet mass ratios: & & & \\
\qquad Mercury &  & 6\,023\,600 & \\
\qquad Venus & & 408\,523.71 & \\
\qquad Earth + Moon & & 328\,900.561400 &  \\
\qquad Earth & & 332\,946.050895\dots  & \\
\qquad Mars & & 3\,098\,708 &  \\
\qquad Jupiter &  & 1\,047.3486 & \\
\qquad Saturn & & 3\,497.898 & \\
\qquad Uranus & &  22\,902.98  & \\
\qquad Neptune & & 19\,412.24  & \\
\qquad Pluto & & 135\,200\,000  & \\
\end{tabular}
\chapter{Precession and Nutation}\label{prenut}
\markboth{PRECESSION \& NUTATION}{PRECESSION \& NUTATION}
\fbox{\parbox{6.5in}{ Relevant IAU resolutions:\quad B1.6, B1.7 of
2000}}
\\
\addcontentsline{toc}{section}{Summary}
\begin{quotation}
\noindent{\bf Summary}\quad Precession and nutation are really
two aspects of  a
single phenomenon, the overall response of the spinning, oblate,
elastic Earth to external gravitational torques from the Moon,
Sun, and planets.  As a result of these torques, the orientation
of the Earth's rotation axis is constantly changing with respect
to a space-fixed (locally inertial) reference system.   The motion of
the celestial pole among the stars is conventionally described as
consisting of a smooth long-term motion called precession upon
which is superimposed a series of small periodic components called
nutation.

The algorithms for precession used generally from about 1980
through 2000 (in \AsA\/ from the 1984 through 2005 editions) were
based on the IAU (1976) value for the rate of general precession
in ecliptic longitude (5029.0966 arcseconds per Julian century at
J2000.0).  Nutation over most of the same time period was given by
the 1980 IAU Theory of Nutation.  However, not long after these
algorithms were widely adopted, it became clear that the IAU
(1976) rate of precession had been overestimated by approximately
3~milliarcseconds per year. Further observations also revealed
periodic errors of a few milliarcseconds in the 1980 IAU Theory of
Nutation.  For many applications these errors are negligible, but
they are significant at the level of the best ground-based
astrometry and geodesy.

As part of the 2000 IAU resolutions, the IAU 2000A
precession-nutation model was introduced, based on an updated
value for the rate of precession and a completely new nutation
theory.  As before, the model actually consists of two
parts, a precession algorithm describing the smooth secular motion
of the celestial pole and a nutation algorithm describing the
small periodic variations in the pole's position. The precession
algorithm consists of short polynomial series for the values of
certain angles.  The sines and cosines of these angles, in
combination, then define the elements of a precession matrix, {\bf
P}. The nutation algorithm consists of a rather long series
expansion in Fourier terms for the angular offsets, in ecliptic
longitude and latitude, of the actual celestial pole (as modeled) from
the precession-only pole (true pole -- mean pole). The sines and
cosines of these offsets, in combination, then define the elements
of a nutation matrix, {\bf N}.  The {\bf P} and {\bf N} matrices
are applied to the coordinates of celestial objects, expressed as
3-vectors, to transform them from the equator and equinox of one
epoch to the equator and equinox of another.
\end{quotation}

\section{Aspects of Earth Rotation}\label{prenut.aspects}

The Earth is a relatively well-behaved rotating body, and
illustrates the three basic elements of classical spin dynamics:
precession, nutation, and Eulerian wobble.  In fact, to first
order, the Earth can be considered to be a rigid ``fast top'', and
very good approximations to its rotational motion can be obtained
from elementary developments.  Although the effects of the Earth's
liquid core, elastic mantle, and oceans are not negligible for
modern observations, they can be considered to be small
perturbations on the rigid-body motion.  Since the Earth is nearly
spherical and experiences relatively weak torques, its axis of
rotation moves slowly with respect to both the stars and the body
of the Earth itself.

The orientation of any rotating rigid body can be described as a
time series of three Euler angles that relate a body-fixed
coordinate system to a space-fixed coordinate system.  If the
body-fixed coordinate system can be defined such that the rate of
change of one of the three Euler angles is much greater than that
of the other two --- as is the case for the Earth --- then the
rotational kinematics are usually described in terms of the slowly
changing orientation of an axis of rotation passing through the
body's center of mass.   We can equivalently speak of the
kinematics of the pole: one of the points where the axis of
rotation intersects the body's surface or, extended to infinity,
the ``celestial sphere''.  For this kinematic construction to work well, 
the angular motion of the axis or pole should be small and nearly linear
over one rotation, predictable from theory, and observable.

However, as was pointed out by \citet{eubanks}, when we use such an
axis or pole, we need five angles, not three, to fully describe
the instantaneous orientation of the body: two angles to describe
the orientation of the body with respect to the axis, one to
describe the angle of the body's rotation about the axis, and two
more to describe the orientation of the axis in the fixed external
system (``inertial space''). For the Earth, these five angles
correspond to the five standard parameters of Earth orientation
disseminated by organizations such as the IERS: the coordinates
of the pole, $x_p$ and $y_p$, measured in a terrestrial coordinate
system; the Universal Time difference, UT1--UTC; and the celestial
pole offsets, $d\psi$ and $d\epsilon$, measured in a celestial
coordinate system.  Phenomenologically, the parameters divide up
as follows:  $x_p$ and $y_p$ describe polar motion, the variations in the
position of the pole with respect to the Earth's crust; UT1--UTC measures the integrated
variation in length of day, the departure from a constant angular rate of
rotation; and $d\psi$ and $d\epsilon$ are the errors in the
computed position of the celestial pole, reflecting deficiencies
in the adopted algorithms for precession and nutation.

What we call polar motion corresponds, in rigid-body rotation, to the free
Eulerian wobble of the figure axis about the rotation axis.  On the real
Earth, the phenomenon is not that simple.
From an Earth-fixed (rotating) frame of reference, polar motion
is a 10-meter (0.3~arcsecond) quasi-circular excursion in the pole position,
with principal periods of 12 and 14~months.  The 14-month
component corresponds to the Eulerian wobble, as modified by the Earth's
elasticity, while the 12-month component undoubtedly is a
seasonal effect.  Smaller, quasi-random variations are not well
understood.  None of the components is regular enough to permit
reliable predictions, and polar motion must be obtained from
observations.

Variations in the Earth's rotation rate are due to several causes.  There
are fortnightly, monthly, semiannual, and annual tidal effects, and
other short-term and seasonal changes
are largely due to exchange of angular momentum with the atmosphere.
Longer-term variations (decade fluctuations) are less well understood.
For a discussion of time scales that are based on the variable rotation
of the Earth, see sections \ref{time.flavors} and \ref{time.erot}.

Precession and nutation refer to the changing orientation of the
Earth's axis, with respect to a space-fixed (kinematically non-rotating) system,
in response to external torques.  The torques are due to the gravitational
attraction of the Moon and Sun (and, to a much lesser extent, the
planets) on the equatorial bulge of the Earth.   Precession
and nutation are really different aspects of a single physical phenomenon,
and it has become more common in recent years to write
``precession-nutation''.  Precession is simply the secular term in
the response, while nutation is the set of periodic terms. On the
celestial sphere, the celestial pole traces out a circle, about 
23\degr\ in radius, centered on the ecliptic pole (the direction
orthogonal to the ecliptic plane), taking about 26,000 years to
complete one circuit ($\approx$20~arcseconds/year).  Precession
theory describes this smooth, long-term motion, and the
precessional pole is referred to as the {\it mean pole} (the
orthogonal plane is the {\it mean equator}). But the pole also
undergoes a hierarchy of small epicyclic motions, the largest of
which is a 14$\times$18 arcsecond ellipse traced out every
18.6~years (see Fig. 1).  Nutation theory describes these periodic
motions.  To get the path of the {\it true pole} on the celestial sphere
(i.e., the direction of the Earth's axis in space), it is necessary to
compute both precession and nutation; conventionally, they are
described by separate time-dependent rotation matrices, ${\bf
P}(t)$ and ${\bf N}(t)$, which are either multiplied together or
applied sequentially.\\[0.20in]

\begin{center}
\includegraphics[width=5.0in]{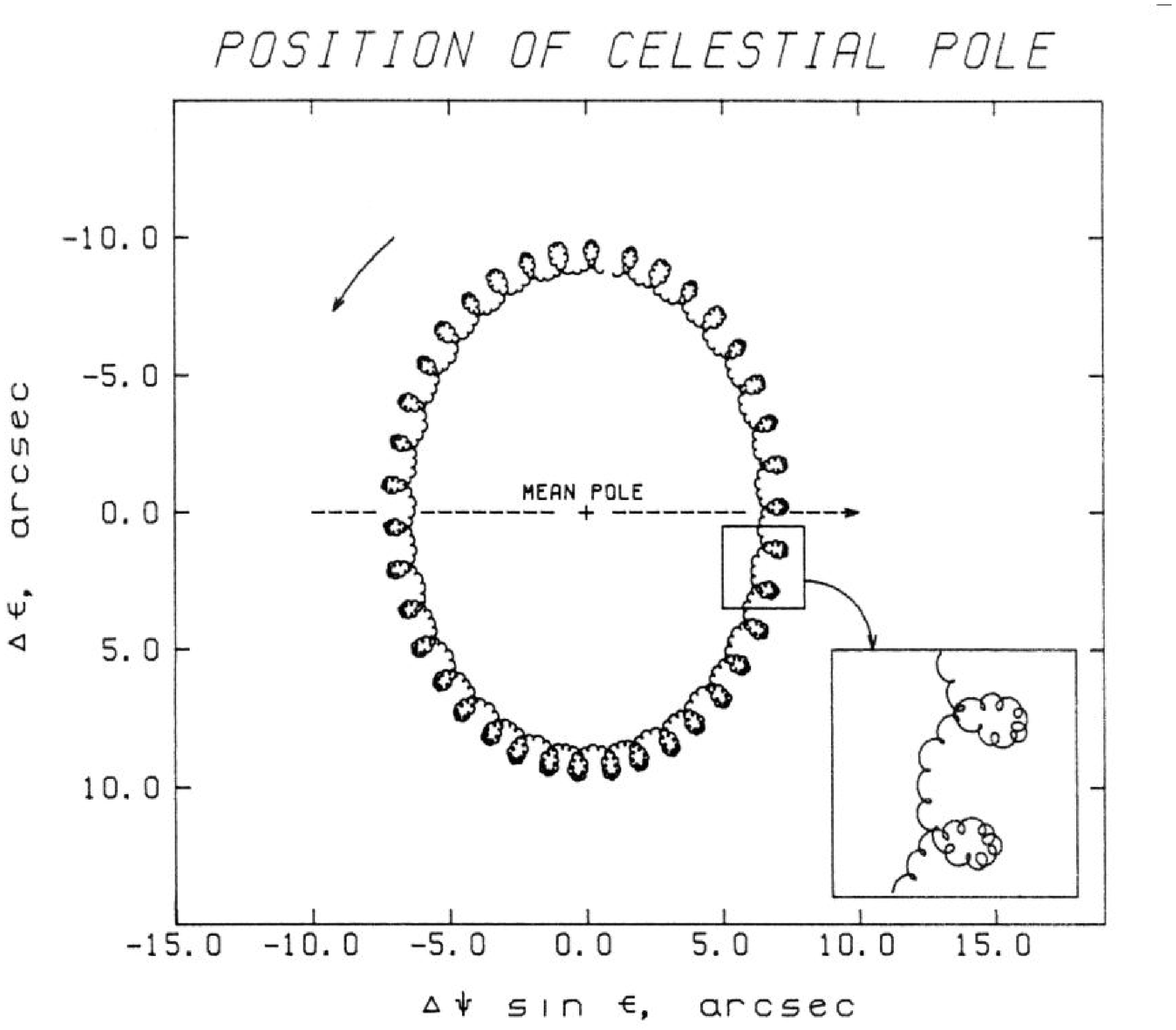}
\parbox{5.0in}{\small\vspace{0.2in}{\bf Figure 5.1}\quad The path of
the true celestial pole on the sky, over an 18-year period, compared to the
mean pole.  The mean pole moves along a smooth arc at a rate of 20
arcseconds per year due to precession only.  The complex epicyclic motion
of the true pole is nutation.  The inset shows the detail of one year's
motion.}\\[4ex]
\end{center}

\section{Which Pole?}\label{prenut.pole}

In theoretical developments of Earth rotation, the first issue
that must be confronted is the definition of the celestial pole.
If the Earth were a rigid oblate spheroid, there would be three
possible axes, and corresponding poles, to choose from:  the
angular momentum axis; the rotation axis, defined by the
instantaneous angular velocity vector; and the figure axis, which
is the body-fixed axis orthogonal to the geometric equator and
along the unique eigenvector of the Earth's inertia tensor.
The distinctions among the axes arise from the physics of rotation.
For example, as previously
noted, in a rotating rigid body, the free Eulerian wobble describes
the motion of the figure axis with respect to the rotation axis (or
vice versa).  The analog of this on the real Earth is polar motion.

In the previous section, precession-nutation was described
as the changing orientation of the Earth's axis in response to
external torques, expressed in a space-fixed (non-rotating) frame
of reference.  Which axis?  The principal components of the
response are rather large, amounting to many arcseconds over
the course of a year, and are nearly the same for all three axes.
(However, the three axes cannot coincide in the presence of
external torques.)  For a rigid Earth, the forced oscillations of the figure
and rotation axes differ by about 10~milliarcseconds, and those of
the angular momentum and rotation axes differ by only about
1~milliarcsecond.  Until the mid-20th century, observations were
not accurate enough to distinguish between the axes, so the choice
of the best axis for theory was academic.  But with improving
observational accuracy and new techniques coming online in the
1960s and 1970s, the question of which axis should be used for the
theoretical developments became important. After considerable
discussion, the consensus emerged that the forced motion of the
figure axis was the most relevant for observations, and therefore
also for theory.

At about the same time, new theoretical work was being undertaken
based on Earth models that were triaxial and contained a liquid
core and elastic mantle.  Such theories complicate the axis
question considerably, because the inertia tensor varies with
time as the Earth's shape responds to tidal forces, and the tidal
deformation results in large daily excursions of the
Earth's axis of figure.  These excursions do not, in general,
reflect the changing overall orientation of the Earth's crust in
space, which is relevant to astronomical observations.   That is,
for the elastic Earth, the figure axis as classically defined is not an
astronomically useful axis.  The solution is to construct a rotating
cartesian coordinate system tied to the elastic, rotating Earth in such a
way that (1) the net angular momentum of the tidal deformation,
relative to this system, is always zero;  and (2) for zero tidal
deformation, the axes correspond to the principal axes of the
Earth's mantle. These axes are the ``Tisserand mean axes of the
body'' \citep{munk}, and the Tisserand axis of the maximum moment
of inertia is referred to in res.~B1.7 of 2000 as ``the mean
surface geographic axis''.  Almost all modern theories of nutation
refer to the principal Tisserand axis; in the previously used
1980 IAU Theory of Nutation it was referred to as axis B, and the
corresponding pole called the ``Celestial Ephemeris Pole''.

However, even if we have chosen an axis that best reflects the
overall rotation of stations (observatories) on the Earth's
surface, a further complication arises as the observations and
theoretical developments become more sensitive to short-period
motions.  The problem is the small but non-negligible circular
components of nutation or polar motion with periods near one day.
One can imagine the geometric confusion that arises when the pole
undergoes a circular motion in one rotation period; in fact, it
becomes difficult to disentangle the various effects, and our
conventional labels become nearly meaningless. For example, any
prograde nearly-diurnal nutation is equivalent to a long-period
variation in polar motion, and any retrograde nearly-diurnal polar
motion appears as a long-period nutation component \citep{cap00a}.
In practice, this means a potential ``leakage'' or ``crossover''
from the Earth orientation parameters $x_p$ and $y_p$ to $d\psi$ and
$d\epsilon$ or vice versa.  The only practical solution is an
explicit (although somewhat arbitrary) cutoff in the periods of
what is considered precession-nutation, embodied in the definition
of the celestial pole.

Therefore, the new IAU definition of the celestial pole to be used
for the new precession-nutation model (res.~B1.7 of 2000) is
defined by the motions of Tisserand mean axis of the Earth with
periods greater than two days in the celestial reference system.
This pole is called the {\it Celestial Intermediate Pole} (CIP).
The position of the CIP is given by the adopted
precession-nutation model plus observational corrections.  The
word {\it intermediate} reminds us that the definition of the pole
is merely a convention, serving to impose a division between what
we call precession-nutation (the Earth orientation angles measured
in the celestial system) and polar motion (the Earth orientation
angles measured in the terrestrial system).   The CIP is the true pole,
orthogonal to the true equator of date.  Its motion is defined
within the Geocentric Celestial Reference System (GCRS) --- see
Chapter~\ref{rel}.  Therefore, the geometric transformations described
in this chapter (as well as those in Chapter~\ref{erot}) all apply within a
geocentric system.  The GCRS can be described loosely as the ``geocentric
ICRS'', since its axis directions are obtained from those of the ICRS.

\section{The New Models}\label{prenut.models}

The variables $d\psi$ and $d\epsilon$ are the small angular
offsets on the sky expressing the difference between the position of the
celestial pole that is observed and the position predicted by the
conventional precession and nutation theories. These angles are just
the differential forms of the angles $\Delta\psi$ and
$\Delta\epsilon$ in which nutation theories are conventionally
expressed ($d\psi$ and $d\epsilon$ are sometimes labeled
$\Delta\Delta\psi$ and $\Delta\Delta\epsilon$).  $\Delta\psi$ and
$\Delta\epsilon$ are in turn differential forms of the ecliptic
coordinates of the celestial pole (see Fig. 5.1).

Obviously the time series of $d\psi$ and $d\epsilon$ values, if
they show systematic trends, can be used to improve the theories
of precession and nutation.  In fact, 20 years of $d\psi$ and
$d\epsilon$ values from VLBI show significant patterns --- see
Fig.~5.2.  Most obvious is the overall downward slope in longitude
and an annual periodicity in both longitude and obliquity,
suggesting the need for substantial corrections to the precession
rate as well to the annual nutation term.  A long-period sinusoid
is also evident, and spectral analysis reveals the presence of a
number of periodic components.\footnote{\label{notefig5.2}The figure
indicates the origin of the ICRS ``frame bias'' discussed in Chapter~\ref{refsys}.
The pole offsets shown are taken from the solution for the ICRF
catalog.  The ICRS frame biases in longitude and obliquity are
essentially the values, at J2000.0 (Time=100), of the two curves
fitted to the data.  The data was arbitrarily zeroed near the
beginning of the data span, which led to non-zero values at J2000.0.}
Other techniques, particularly lunar laser ranging
(LLR), confirm the general trends.  As a result, there has been a
major multinational effort to improve the precession and nutation
formulation and obtain interesting geophysical information in the
process.  This project, coordinated by an IAU/IUGG\footnote{IUGG =
International Union of Geodesy and Geophysics} working group, has
involved dozens of investigators in several fields, and the
resulting algorithms, taken together, are referred to as the IAU
2000A precession-nutation model.

\begin{center}
\includegraphics[width=3.5in]{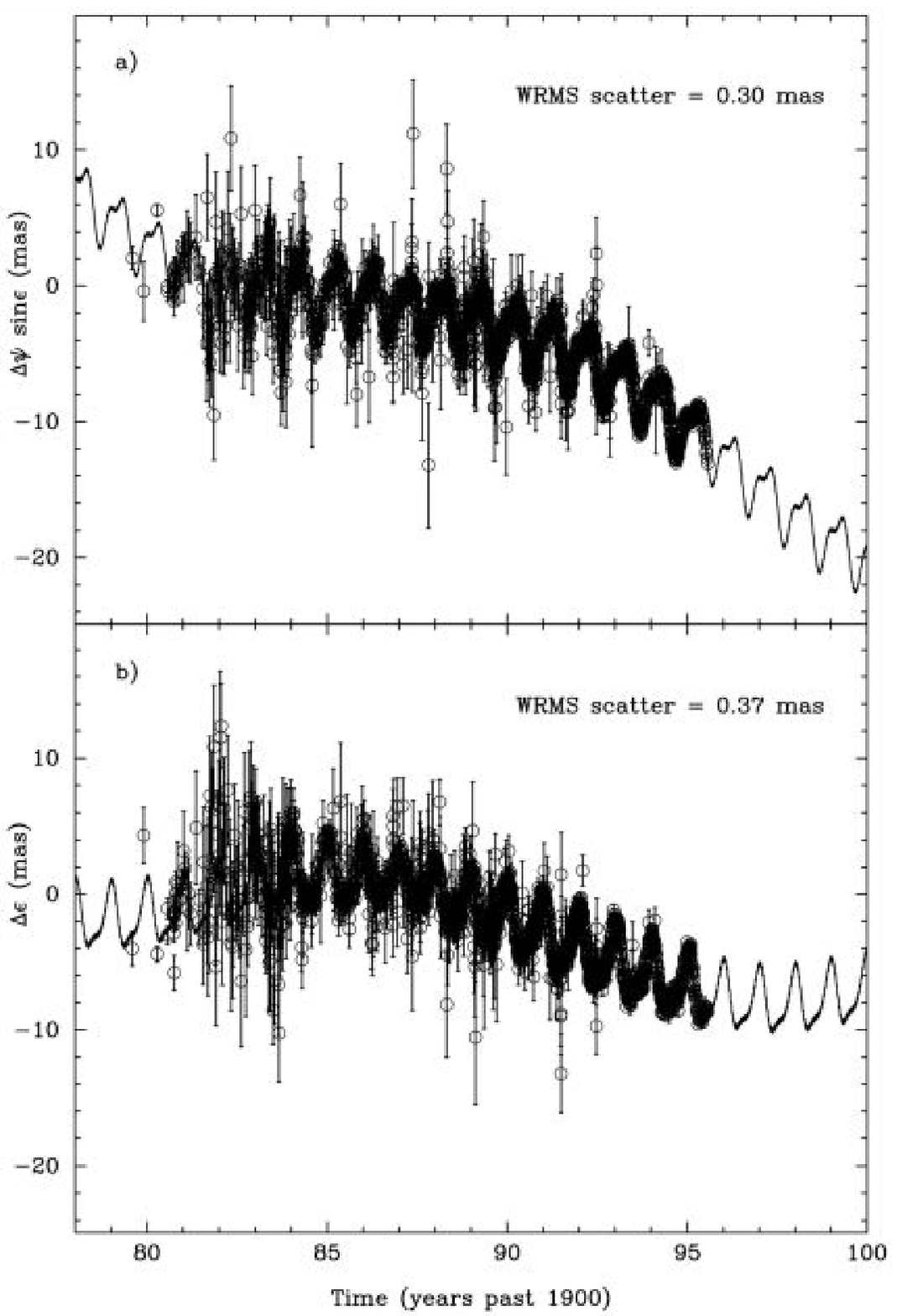}
\parbox{5.0in}{\small\vspace{0.2in}{\bf Figure 5.2}\quad Observed
values of celestial pole offsets from VLBI data.  Offsets in longitude
have been multiplied by the sine of the obliquity to allow the same
scale to be used for both components. Circled points with error bars
represent the offset of the observed pole with respect to the
computed pole, and the solid line in each plot is a curve fitted
to the data.  The computed pole is given by the \citet{lieske77}
precession expressions and the 1980 IAU Theory of Nutation.
These plots are from \citet{ma98}.}\\[4ex]
\end{center}

The VLBI observations of $d\psi$ and $d\epsilon$ indicate the
error in the computed position of the pole with respect to a
space-fixed system defined by the positions of extragalactic
objects.  However, the conventional expressions for precession and
nutation have used angles measured with respect to the ecliptic, a plane
to which VLBI is not sensitive.  The ecliptic plane has a slow
precessional movement of its own due to planetary perturbations on
the heliocentric orbital motion of the Earth-Moon barycenter.\footnote{The
{\it mean ecliptic} is always implied.  This is the smoothly moving plane
that does not undergo the periodic oscillations of the instantaneous
orbital plane of the Earth.}  In the theoretical developments it is necessary to
distinguish between {\it precession of the equator} and {\it precession
of the ecliptic}, which were formerly called, respectively,
{\it lunisolar precession} and {\it planetary precession}.  Both types
of precession are measured with respect to a space-fixed
system.  The algorithms for precession and nutation provide
the motion of the equator, as appropriate for most observations,
but generally use a moving ecliptic as a reference plane for
at least some of the angles involved (there are different
formulations of precession using different angle sets).  This
allows the precession and nutation transformations to properly
account for the motion of the equinox as well as that of the
equator.  The precession of the ecliptic is obtained from theory
(although indirectly tied to observations through the JPL DE405
ephemeris), as are the high-order (unobserved) components of the
precession of the equator.  However, because of the mix of theory
and observation that is involved in the final expressions, raw
corrections to rates of precession from VLBI observations will not
in general propagate exactly to the familiar precession
quantities.

The changes in the amplitudes of the nutation components are also
not directly taken from these observations; instead, a new
nutation theory is developed and fit to observations by allowing a
small number of geophysical constants to be free parameters. These
parameters are constants in a ``transfer function'' that modifies
the amplitudes of the terms from a rigid-Earth nutation
development. Since there are fewer solved-for geophysical
constants than the number of terms with observed amplitudes, the
fit cannot be perfect.  For the IAU 2000A model, 7~geo\-physical
parameters were determined based on the observed amplitudes of
21~nu\-ta\-tion terms (prograde and retrograde amplitudes for
each) together with the apparent change in the rate of precession
in longitude.  Note that the number of observational constraints
and the number of free parameters in the model are both quite small
compared to the 1365 terms in the new, full nutation series.

\begin{table}
\begin{center}
\hfil{\bf Table 5.1 \quad Precession-Nutation: Old \& New}\hfil\\
\hfil Values in arcseconds at J2000.0 \hfil\\[1.5ex]
\begin{tabular}{lrrr}
\hline\hfil{\bf Quantity}\hfil & {\bf Old value} &
     {\bf New value} & {\bf New--Old} \\
General precession in longitude (/cen) & $5029.0966$ &
$5028.796195$
                     & $-0.3004$ \\
Mean obliquity & \D$84381.448$\D & \D$84381.406$\DDD & $-0.042$\D \\
Mean obliquity rate (/cen) & $-46.8150$ & $-46.836769$  & $-0.0218$ \\
In-phase nutation amplitudes: & & & \\
\w{XXX}18.6-year longitude   & $-17.1966$ & $-17.2064161$ & $-0.0098$ \\
\w{XXX}18.6-year obliquity   & $  9.2025$ & $  9.2052331$ & $ 0.0027$ \\
\w{XXX}9.3-year longitude    & $  0.2062$ & $  0.2074554$ & $ 0.0013$ \\
\w{XXX}9.3-year obliquity    & $ -0.0895$ & $ -0.0897492$ & $-0.0002$ \\
\w{XXX}annual longitude      & $  0.1426$ & $  0.1475877$ & $ 0.0050$ \\
\w{XXX}annual obliquity      & $  0.0054$ & $  0.0073871$ & $ 0.0020$ \\
\w{XXX}semiannual longitude  & $ -1.3187$ & $ -1.3170906$ & $ 0.0016$ \\
\w{XXX}semiannual obliquity  & $  0.5736$ & $  0.5730336$ & $-0.0006$ \\
\w{XXX}122-day longitude     & $ -0.0517$ & $ -0.0516821$ & $ 0.0000$ \\
\w{XXX}122-day obliquity     & $  0.0224$ & $  0.0224386$ & $ 0.0000$ \\
\w{XXX}monthly longitude     & $  0.0712$ & $  0.0711159$ & $-0.0001$ \\
\w{XXX}monthly obliquity     & $ -0.0007$ & $ -0.0006750$ & $ 0.0000$ \\
\w{XXX}semimonthly longitude & $ -0.2274$ & $ -0.2276413$ & $-0.0002$ \\
\w{XXX}semimonthly obliquity & $  0.0977$ & $  0.0978459$ & $ 0.0001$ \\
\end{tabular}
\end{center}
\end{table}

Table 5.1 compares the old and new values, at epoch J2000.0, of
some of the primary quantities involved in the precession and
nutation algorithms. In the table, all quantities are in
arcseconds, and the rates (marked /cen) are per Julian century of
TDB (or TT).  The longitude components should be multiplied by the
sine of the obliquity ($\approx$0.3978) to obtain the
corresponding motion of the pole on the celestial sphere.  The new
mean obliquity at J2000.0 is $23\degr\, 26'\,21.\!''406$.  The
theories from which the values are taken are:

\begin{itemize}
\item Old precession: \citet{lieske77}, based on the IAU~(1976)
  values for general precession and the obliquity at J2000.0, shown
  in the table
\item Old nutation: 1980 IAU Theory of Nutation: \citet{wahr}, based on \citet{kino};  
    see report of the IAU working  group by \citet{seid82} 
\item New precession: P03 solution in \citet{cap03};  see report of the
   IAU working group by \citet{hilton05}
\item New nutation: \citet{mathews} (often referred to as MHB),
  based on \citet{souchay}; series listed at \citet{url-nuts}
\end{itemize}

The new precession development will probably be
formally adopted by the IAU in 2006.  The MHB nutation was adopted
in res.~B1.6 of 2000, even though the theory had not been
finalized at the time of the IAU General Assembly of that year.
Used together, these two developments yield the computed path of
the Celestial Intermediate Pole (CIP) as well as that of the true
equinox.  The formulas given below are based on these two developments.


\section{Formulas}\label{prenut.formulas}

In the development below, precession and nutation are represented
as 3$\times$3 rotation matrices that operate on column 3-vectors.
The latter are position vectors in a specific celestial coordinate
system --- which must be stated or understood --- with components
that are cartesian (rectangular) coordinates.  They have the general form
\begin{equation}
{\bf r} =  \left( \begin{array}{c} r_x \\ r_y \\ r_z \end{array} \right)
          =  \left( \begin{array}{c}
                     d \, \cos\delta \, \cos\alpha  \\
                     d \, \cos\delta \, \sin\alpha  \\
                     d \, \sin\delta
               \end{array} \right)
\end{equation}
where $\alpha$ is the right ascension, $\delta$ is the
declination, and $d$ is the distance from the specified origin.
For stars and other objects ``at infinity'' (beyond the solar
system), $d$ is often simply set to 1.  The celestial coordinate
system being used will be indicated by a subscript, e.g., ${\bf
r}_{_{\rm GCRS}}$.  If we have the vector ${\bf r}$ in some
coordinate system, then the right ascension and declination in
that coordinate system can be obtained from
\begin{eqnarray}
     \alpha & = & \arctan \, \left( r_y / r_x \right)          \nonumber\\
     \delta & = & \arctan \, \left( r_z / \sqrt{r_x^2 + r_y^2} \right)
\end{eqnarray}
where $r_x$, $r_y$, and $r_z$ are the three components of ${\bf
r}$.  A two-argument arctangent function (e.g., atan2) will return
the correct quadrant for $\alpha$ if $r_y$ and $r_x$ are provided
separately.

In the context of traditional equatorial celestial coordinate systems, the
adjective {\it mean} is applied to quantities (pole, equator,
equinox, coordinates) affected only by precession, while {\it
true} describes quantities affected by both precession and
nutation.  This is a computational distinction only, since
precession and nutation are simply different aspects of the same
physical phenomenon.  Thus, it is the {\it true} quantities that
are directly relevant to observations; {\it mean} quantities now usually
represent an intermediate step in the computations, or the final
step where only very low accuracy is needed (10~arcseconds or
worse) and nutation can be ignored.

Thus, a precession transformation is applied to celestial
coordinates to convert them from the mean equator and equinox of
J2000.0 to the mean equator and equinox of another date, $t$.
Nutation is applied to the resulting coordinates to transform them
to the true equator and equinox of $t$.  These transformations should be
understood to be inherently geocentric rotations and they originate in dynamical
theories.  Generally we will be starting with celestial coordinates in the GCRS,
which are obtained from basic ICRS data by applying the usual
algorithms for proper place.\footnote{Computing
{\it proper place}\/ involves adjusting the catalog place of a star or other extra-solar system object for proper motion and parallax (where known), gravitational light deflection within the solar system, and aberration due to the Earth's motions.  For a solar system object there are comparable adjustments to its position vector taken from a barycentric gravitational ephemeris.  See
section~\ref{rel.obs}.  In conventional usage, an {\it apparent place}\/ can
be considered to be a proper place that has been transformed to the true equator and equinox of date. The details of the proper place computations are beyond the scope of this \pubname\ but are described in detail in many textbooks on positional astronomy, and in \citet{hohenkerk}, \citet{kaplan89}, and \citet{klioner}.\label{placenote}}  As discussed in
Chapter~\ref{refsys}, the ICRS is not based on a dynamically defined equator and
equinox and so neither is the GCRS.  Therefore, before we apply precession
and nutation --- and if we require a final accuracy of
better than 0.02 arcsecond --- we must first apply the frame bias
correction (section~\ref{refsys.formulas}) to transform the GCRS coordinates to the dynamical
mean equator and equinox of J2000.0.  Schematically,

\begin{center}
\begin{tabular}{c}
                                                    \\
     \fbox{\D GCRS\D}                                         \\
     $\|$ \\ {\it frame bias} \\ $\Downarrow$       \\
     \fbox{\D mean equator \& equinox of J2000.0\D} \\
     $\|$ \\ {\it precession} \\ $\Downarrow$       \\
     \fbox{\D mean equator \& equinox of $t$\D}     \\
     $\|$ \\ {\it nutation}   \\ $\Downarrow$       \\
     \fbox{\D true equator \& equinox of $t$\D}     \\
                                                    \\
\end{tabular}
\end{center}

Mathematically, this sequence can be represented as follows:
\begin{equation}
    {\bf r}_{{\rm true}(t)} = {\bf N}(t)\;{\bf P}(t)\;{\bf B}\;
                              {\bf r}_{_{\rm GCRS}}
\end{equation}
where ${\bf r}_{_{\rm GCRS}}$ is a direction vector with respect to
the GCRS and ${\bf r}_{{\rm true}(t)}$ is the equivalent vector
with respect to the true equator and equinox of $t$.  ${\bf N}(t)$
and ${\bf P}(t)$ are the nutation and precession rotation
matrices, respectively.  The remainder of this chapter shows how
to compute the elements of these matrices.  ${\bf B}$ is the
(constant) frame-bias matrix given in section~\ref{refsys.formulas}.

The transformation from the mean equator and equinox of J2000.0 to
the mean equator and equinox of $t$ is simply
\begin{equation}
    {\bf r}_{{\rm mean}(t)} = {\bf P}(t)\;{\bf r}_{\rm mean(J2000.0)}
\end{equation}
and the reverse transformation is
\begin{equation}
    {\bf r}_{\rm mean(J2000.0)} = {\bf P}^{\rm T}(t)\;{\bf r}_{{\rm mean}(t)}
\end{equation}
where ${\bf P}^{\rm T}(t)$ is the transpose of ${\bf P}(t)$.

What is described above are all conventional, equinox-based transformations.
The equinox is the traditional origin of right ascension.
An alternative transformation has been developed
based on another point on the celestial equator called the Celestial
Intermediate Origin (CIO); see res.~B1.8 of 2000.  The alternative
scheme has been introduced because the equinox is based on
a barycentric concept (the ecliptic, the Earth's mean orbit)
that is not relevant to a geocentric system;  furthermore, the equinox suffers from
ambiguity of definition below the 0.1~arcsecond level.   Additionally, the
new scheme cleanly separates different aspects of Earth orientation in the
overall transformation between the terrestrial and celestial coordinate
systems.  The conventional transformations are
described in this chapter because of widespread current usage and the fact that
even the newest theories of precession and nutation provide the
angles needed for these transformations.
The new transformation, which combines frame bias, precession, and nutation into
a single matrix, is given in section \ref{prenut.formulas.alt} and is described more fully in
Chapter~\ref{erot}, where the CIO is introduced and explained.

The true celestial pole of date $t$ --- the Celestial Intermediate Pole (CIP) --- has,
by definition, unit vector coordinates (0,0,1) with respect to the true equator
and equinox of date.  Therefore we can obtain the computed coordinates of
the CIP with respect to the GCRS by simply reversing the transformation of
eq.~5.3:
\begin{eqnarray}
\mbox{Computed position of CIP:} \qquad {\bf r}_{_{\rm GCRS}} & = &
               \left( \begin{array}{c} X \\ Y \\ Z \end{array} \right)
             = {\bf B}^{\rm T} \; {\bf P}^{\rm T}(t) \; {\bf N}^{\rm T}(t) \;
                 \left( \begin{array}{c} 0 \\ 0 \\ 1 \end{array} \right)
                                                                  \nonumber\\
& = & \left( \begin{array}{c} \mbox{(NPB)}_{31} \\
                              \mbox{(NPB)}_{32} \\
                              \mbox{(NPB)}_{33} \end{array} \right) \\
 &  & \nonumber\\
\mbox{where} \qquad {\bf NPB} & = & {\bf N}(t)\;{\bf P}(t)\;{\bf B}
\nonumber
\end{eqnarray}
and where the superscript T's indicate that the transpose of the
matrix is used.  Daily values of the elements of the combined matrix {\bf NPB} are
listed in \AsA.

The \citet{iers03} list series expansions that directly provide
$X$ and $Y$, the two most rapidly changing components of the pole position
unit vector.  Daily values of $X$ and $Y$ are also listed in \AsA.  The values
of $X$ and $Y$ are given in arcseconds and are converted to dimensionless
unit vector components simply by dividing them by the number of arcseconds in one
radian, 206264.806247\dots. \quad Also, then, $Z = \sqrt{1 - X^2 - Y^2}$.  The
values of $X$ and $Y$ are used in the new transformation scheme discussed
in section~\ref{prenut.formulas.alt} and in several places in Chapter~\ref{erot}.

\subsection{Formulas for Precession}\label{prenut.formulas.pre}

To construct the precession matrix for the transformation of coordinates
from one date to another, we must evaluate short polynomials
for the angles involved.  The expressions for these angles in the IAU 2000A
model, given below, have only a single time argument, since precession from
or to J2000.0 (actually, the TDB equivalent of J2000.0) is assumed.  As
used in this \pubname\ (and \AsA), the matrix ${\bf P}(t)$ always
denotes precession from J2000.0 (TDB) to another date, $t$.  To precess in the
opposite direction, the angles are the same but the
transpose of the precession matrix, ${\bf P}^{\rm T}(t)$, is used.  To precess
coordinates from one arbitrary date, $t_1$, to another, $t_2$, it is necessary
to precess them from $t_1$ to J2000.0 (using ${\bf P}^{\rm T}(t_1)$), then
from J2000.0 to $t_2$ (using ${\bf P}(t_2)$).  Where high accuracy is not
required, and $t_1$ and $t_2$ are not more than a few years apart, a simpler
procedure for precession from $t_1$ to $t_2$ is available and is given at
the end of this subsection.

All expressions given in this subsection are from Section~7 (P03 solution) of
\citet{cap03} and all coefficients are expressed in arcseconds. This
is the theory of precession recommended by the IAU Working Group on Precession
and the Ecliptic \citep{hilton05}.

For a given TDB date and time $t$, let $T$ be the number of Julian
centuries of TDB since 2000 Jan~1, 12$^{\rm h}$ TDB.  If the dates
and times are expressed as Julian dates, then $T = (t - 2451545.0)
/ 36525$.  TT dates and times can be used equally well --- the resulting
error in precession is only a few \E{-9} arcseconds.

Then the mean obliquity of the ecliptic at J2000.0 (or the
equivalent TDB date) is $\epsilon_0 = 84381.406$ arcseconds and
let
\begin{eqnarray}
  \psi_A &=& 5038.481507\,T - 1.0790069\,T^2 - 0.00114045\,T^3
               + 0.000132851\,T^4 - 0.0000000951\,T^5   \nonumber\\
  \omega_A &=& \epsilon_0 - 0.025754\,T + 0.0512623\,T^2 - 0.00772503\,T^3
               - 0.000000467\,T^4 +  0.0000003337\,T^5 \nonumber\\
  \chi_A &=& 10.556403\,T - 2.3814292\,T^2 - 0.00121197\,T^3
             + 0.000170663\,T^4 -  0.0000000560\,T^5  \qquad
\end{eqnarray}
Equivalently, in notation appropriate for computer programs,
\begin{eqnarray}
      \psi_A   = ( ( ( ( &-& \DDD 0.0000000951 \w{)}  \; T \nonumber\\
                         &+& \DDD 0.000132851\D )     \; T \nonumber\\
                         &-& \DDD 0.00114045\DD )     \; T \nonumber\\
                         &-& \DDD 1.0790069\DDD )     \; T \nonumber\\
                         &+&  5038.481507\DD\DD )     \; T \nonumber\\
      \omega_A = ( ( ( ( &+&  \DDD 0.0000003337 \w{)} \; T \nonumber\\
                         &-&  \DDD 0.000000467\D )    \; T \nonumber\\
                         &-&  \DDD 0.00772503\DD )    \; T          \\
                         &+&  \DDD 0.0512623\DDD )    \; T \nonumber\\
                         &-&  \DDD 0.025754\DD\DD )   \; T
                                       + \epsilon_0        \nonumber\\
      \chi_A   = ( ( ( ( &-&  \DDD 0.0000000560 \w{)} \; T \nonumber\\
                         &+&  \DDD 0.000170663\D )    \; T \nonumber\\
                         &-&  \DDD 0.00121197\DD )    \; T \nonumber\\
                         &-&  \DDD 2.3814292\DDD )    \; T \nonumber\\
                         &+&  \DD 10.556403\DD\DD )   \; T \nonumber
\end{eqnarray}
The  precession matrix is then simply
${\bf P}(t) = {\bf R}_3(\chi_A)\,{\bf R}_1(-\omega_A)\,{\bf R}_3(-\psi_A)\,
{\bf R}_1(\epsilon_0)$, where ${\bf R}_1$ and ${\bf R}_3$ are standard
rotations about the x and z axes, respectively (see
``Abbreviations and Symbols Frequently Used'' for precise definitions).  This
4-angle precession formulation is comprised of
\begin{enumerate}
\item A rotation from the mean equator and equinox of J2000.0 to
the mean ecliptic and equinox of J2000.0.  This is simply a
rotation around the x-axis (the direction toward the mean equinox
of J2000.0) by the angle $\epsilon_0$, the mean obliquity of
J2000.0.  After the rotation, the fundamental plane is the
ecliptic of J2000.0. \item A rotation around the new z-axis (the
direction toward the ecliptic pole of J2000.0) by the angle
$-\psi_A$, the amount of precession of the equator from J2000.0 to
$t$. \item A rotation around the new x-axis (the direction along
the intersection of the mean equator of $t$ with the ecliptic of
J2000.0) by the angle $-\omega_A$, the obliquity of the mean
equator of $t$ with respect to the ecliptic of J2000.0.  After the
rotation, the fundamental plane is the mean equator of $t$. \item
A rotation around the new z-axis (the direction toward the mean
celestial pole of $t$) by the angle $\chi_A$, accounting for the
precession of the ecliptic along the mean equator of $t$.  After
the rotation, the new x-axis is in the direction of the mean
equinox of date.
\end{enumerate}

If we let
\begin{flushright}
\parbox{2.0in}{
\begin{eqnarray*}
      S_1 &=& \sin\,(\epsilon_0)   \\
      S_2 &=& \sin\,(-\psi_A)      \\
      S_3 &=& \sin\,(-\omega_A)    \\
      S_4 &=& \sin\,(\chi_A)
\end{eqnarray*} }
\parbox{4.0in}{
\begin{eqnarray}
      C_1 &=& \cos\,(\epsilon_0)   \nonumber\\
      C_2 &=& \cos\,(-\psi_A)               \\
      C_3 &=& \cos\,(-\omega_A)    \nonumber\\
      C_4 &=& \cos\,(\chi_A)       \nonumber
\end{eqnarray} }
\end{flushright}
then the precession matrix can also be written:
\begin{equation}
    {\bf P}(t) = \left( \begin{array}{ccc}
           C_4  C_2 - S_2   S_4  C_3                           &
           C_4  S_2   C_1 + S_4  C_3  C_2  C_1 - S_1  S_4  S_3 &
           C_4  S_2   S_1 + S_4  C_3  C_2  S_1 + C_1  S_4  S_3 \\
          -S_4  C_2 - S_2   C_4  C_3                           &
          -S_4  S_2   C_1 + C_4  C_3  C_2  C_1 - S_1  C_4  S_3 &
          -S_4  S_2   S_1 + C_4  C_3  C_2  S_1 + C_1  C_4  S_3 \\
           S_2  S_3                                            &
          -S_3  C_2   C_1 - S_1  C_3                           &
          -S_3  C_2   S_1 + C_3  C_1
              \end{array} \right)
\end{equation}

Existing applications that use the 3-angle precession formulation
of Newcomb and Lieske can be easily modified for the IAU 2000A
precession, by replacing the current polynomials for the angles
$\zeta_A$, $z_A$, and $\theta_A$ with the following:
\begin{eqnarray}
   \zeta_A &=&  2.650545 + 2306.083227\,T + 0.2988499\,T^2
       + 0.01801828\,T^3 - 0.000005971\,T^4                   \nonumber\\
     & & \w{9999999.} - 0.0000003173\,T^5                     \nonumber\\
      z_A  &=&  -2.650545 + 2306.077181\,T + 1.0927348\,T^2
           + 0.01826837\,T^3 - 0.000028596\,T^4               \nonumber\\
     & & \w{999999999} - 0.0000002904\,T^5                             \\
      \theta_A &=& 2004.191903\,T - 0.4294934\,T^2 - 0.04182264\,T^3
           - 0.000007089\,T^4                                  \nonumber\\
     & & \w{999999999999.} - 0.0000001274\,T^5                 \nonumber
\end{eqnarray}
The 3-angle precession matrix is ${\bf P}(t) = {\bf
R}_3(-z_A)\,{\bf R}_2(\theta_A)\,{\bf R}_3(-\zeta_A)$, but any
existing correct construction of ${\bf P}$ using these three
angles can still be used.

The expression for the mean obliquity of the ecliptic (the angle between
the mean equator and ecliptic, or, equivalently, between the ecliptic
pole and mean celestial pole of date)  is:
\begin{equation}
  \epsilon = \epsilon_0 - 46.836769\,T - 0.0001831\,T^2 + 0.00200340\,T^3
     -0.000000576\,T^4 - 0.0000000434\,T^5
\end{equation}
where, as stated above, $\epsilon_0 = 84381.406$ arcseconds.  This
expression arises from the precession formulation but is actually used
only for nutation.  (Almost all of the obliquity rate --- the term
linear in $T$ --- is due to the precession of the ecliptic.)

Where high accuracy is not required, the precession between two
dates, $t_1$ and $t_2$, not too far apart (i.e., where $|t_2 - t_1|
\ll$ 1~century), can be
approximated using the rates of change of right ascension and
declination with respect to the mean equator and equinox of date.
These rates are respectively
\begin{eqnarray}
        m &\approx& 4612.16 + 2.78\,T   \nonumber\\
        n &\approx& 2004.19 - 0.86\,T
\end{eqnarray}
where the values are in arcseconds per century and $T$ is the number
of centuries between J2000.0 and the midpoint of $t_1$ and $t_2$.  If
the dates are expressed as Julian dates, $T = ((t_1 + t_2)/2 -
2451545.0)/36525$.
Then, denoting the celestial coordinates at $t_1$ by $(\alpha_1,\delta_1)$
and those at $t_2$ by $(\alpha_2,\delta_2)$,
\begin{eqnarray}
      \alpha_2 &\approx& \alpha_1 + \tau \, (m
                       + n \,\sin\alpha_1 \tan\delta_1)      \nonumber\\
      \delta_2 &\approx& \delta_1 + \tau \, (n \,\cos\alpha_1)
\end{eqnarray}
where $\tau = t_2 - t_1$, expressed in centuries.  These formulas
deteriorate in accuracy at high (or low) declinations and should not be used at all
for coordinates close to the celestial poles (how close depends on the accuracy
requirement and the value of $\tau$) .

\subsection{Formulas for Nutation}\label{prenut.formulas.nut}

Nutation is conventionally expressed as two small angles, $\Delta\psi$, the
nutation in longitude, and $\Delta\epsilon$, the nutation in obliquity.  These
angles are measured in the ecliptic system of date, which is developed as part
of the precession formulation.  The angle $\Delta\psi$ is the small change in the position of
the equinox along the ecliptic due to nutation, so the effect of nutation on the ecliptic
coordinates of a fixed point in the sky is simply to add $\Delta\psi$ to its
ecliptic longitude.  The angle $\Delta\epsilon$ is the small change in the obliquity
of the ecliptic due to nutation.  The {\it true obliquity}
of date is $\epsilon' = \epsilon + \Delta\epsilon$.  Nutation in obliquity
reflects the orientation of the equator in space and does not affect the
ecliptic coordinates of a fixed point on the sky.

The angles $\Delta\psi$ and $\Delta\epsilon$ can also be thought of as small
shifts in the position of the celestial pole (CIP) with respect to the ecliptic
and mean equinox of date.  In that coordinate system, and assuming positive
values for $\Delta\psi$ and $\Delta\epsilon$, the nutation in longitude
shifts the celestial pole westward on the sky by the angle $\Delta\psi\,\sin\epsilon$,
decreasing the pole's mean ecliptic longitude by $\Delta\psi$.  Nutation in obliquity moves the
celestial pole further from the ecliptic pole, i.e., southward in ecliptic coordinates, by
$\Delta\epsilon$.  (Negative values of $\Delta\psi$ and $\Delta\epsilon$ move
the pole eastward and northward in ecliptic coordinates.)

The effect of nutation on the equatorial coordinates ($\alpha$,$\delta$) of
a fixed point in the sky is more complex and is best dealt with through the
action of the nutation matrix, ${\bf N}(t)$, on the equatorial position vector,
${\bf r}_{\rm mean(t)}$.  Where high accuracy is not required, formulas that
directly give the changes to $\alpha$ and $\delta$ as a function of $\Delta\psi$
and $\Delta\epsilon$ are available and are given at the end of this subsection.

The values of $\Delta\psi$ and $\Delta\epsilon$ are obtained by evaluating
rather lengthy trigonometric series, of the general form
\begin{eqnarray}
  \Delta\psi     &=& \sum_{i=1}^N \,\left( (S_i + \dot S_i T) \sin\,\Phi_i
                                    + C'_i \cos\,\Phi_i \right) \nonumber\\
  \Delta\epsilon &=& \sum_{i=1}^N \,\left( (C_i + \dot C_i T) \cos\,\Phi_i
                                    + S'_i \sin\,\Phi_i \right)          \\
   & & \mbox{where, in each term,}\quad \Phi_i = \sum_{j=1}^K M_{i,j}\,\phi_j(T)
\end{eqnarray}
For the IAU 2000A model, N=1365 and K=14.  The 14 $\phi_j(T)$ are
the fundamental arguments, which are, except for one, orbital angles.  The main
time dependence of the nutation series enters through these arguments.  The
expressions given below are all taken from \citet{simon} and all coefficients are
in arcseconds.

The first eight fundamental arguments are the mean heliocentric ecliptic
longitudes of the planets Mercury through Neptune:
\begin{eqnarray}
   \phi_1 &=&   \D 908103.259872  +    538101628.688982 \,T  \nonumber\\
   \phi_2 &=&   \D 655127.283060  +    210664136.433548 \,T  \nonumber\\
   \phi_3 &=&   \D 361679.244588  +    129597742.283429 \,T  \nonumber\\
   \phi_4 &=&     1279558.798488  +  \D 68905077.493988 \,T           \\
   \phi_5 &=&   \D 123665.467464  +  \D 10925660.377991 \,T  \nonumber\\
   \phi_6 &=&   \D 180278.799480  +  \DD 4399609.855732 \,T  \nonumber\\
   \phi_7 &=&     1130598.018396  +  \DD 1542481.193933 \,T  \nonumber\\
   \phi_8 &=&     1095655.195728  +  \DDD 786550.320744 \,T  \nonumber
\end{eqnarray}
In all of these expressions, $T$ is the number of Julian centuries of TDB since
2000 Jan~1, 12$^{\rm h}$ TDB (or, with negligible error, the number of Julian centuries of TT since
J2000.0).  In some implementations it may be necessary to reduce the resulting
angles, which are expressed in arcseconds, to radians in the range 0--$2\pi$.
The ninth argument is an approximation to the general precession in longitude:
\begin{equation}
      \phi_9 =  5028.8200 \,T + 1.112022 \,T^2
\end{equation}
The last five arguments are the same fundamental luni-solar arguments used in
previous nutation theories, but with updated expresssions.  They are,
respectively, $l$, the mean anomaly of the Moon; $l'$, the mean anomaly of the
Sun; $F$, the mean argument of latitude of the Moon; $D$, the mean elongation of
the Moon from the Sun, and $\Omega$, the mean longitude of the Moon's mean
ascending node:
\begin{eqnarray}
    &\phi_{10}& = l = 485868.249036
     + 1717915923.2178     \,T
     +         31.8792     \,T^2
     +          0.051635   \,T^3
     -          0.00024470 \,T^4  \nonumber\\
    &\phi_{11}& = l'= 1287104.79305
     +    129596581.0481   \,T
     -            0.5532   \,T^2
     +          0.000136   \,T^3
     -        0.00001149   \,T^4  \nonumber\\
    &\phi_{12}& = F = 335779.526232
     +   1739527262.8478   \,T
     -           12.7512   \,T^2
     -          0.001037   \,T^3
     +        0.00000417   \,T^4  \nonumber\\
    &\phi_{13}& = D = 1072260.70369
     +   1602961601.2090   \,T
     -            6.3706   \,T^2
     +          0.006593   \,T^3
     -        0.00003169   \,T^4  \nonumber\\
    &\phi_{14}& = \Omega = 450160.398036
     -      6962890.5431   \,T
     +            7.4722   \,T^2
     +          0.007702   \,T^3
     -        0.00005939   \,T^4  \qquad\qquad
\end{eqnarray}

The first step in evaluating the series for nutation for a given date is to
compute the values of all 14 fundamental arguments for the date of interest.
This is done only once.  Then the nutation terms are evaluated one by one.  For
each term $i$, first compute $\Phi_i$ according to eq.~5.16, using the 14 integer
multipliers, $M_{i,j}$, listed for the term; i.e., sum over $M_{i,j}\times\phi_j$
(where $j$=1--14).   Then the cosine and sine components for the term can be
evaluated, as per eq.~5.15, using the listed values of the
coefficients $S_i$, $\dot S_i$, $C'_i$, $C_i$, $\dot C_i$, and $S'_i$ for the
term.  Generally it is good practice to sum the terms from smallest to largest
to preserve precision in the sums.

The entire IAU 2000A nutation series is listed at the end of this \pubname.
About the first half of the series consists of lunisolar terms,  which depend
only on $l$, $l'$, $F$, $D$, and
$\Omega$ (= $\phi_{10}$ to $\phi_{14}$).  In all of these terms, the first nine multipliers
are all zero.   The generally smaller planetary terms
comprise the remainder of the series.  As an example of how the individual terms
are computed according to eqs.~5.15 and 5.16, term~6 would be evaluated
\begin{eqnarray*}
  \Delta\psi_6     &=& (     -0.0516821 + 0.0001226\,T) \sin\Phi_6
                                        - 0.0000524  \, \cos\Phi_6            \\
  \Delta\epsilon_6 &=& (\w{-} 0.0224386 - 0.0000667\,T) \cos\Phi_6
                                        - 0.0000174  \, \sin\Phi_6            \\
    & & \mbox{where}\quad \Phi_6 = \phi_{11} + 2\phi_{12} - 2\phi_{13} + 2\phi_{14}
\end{eqnarray*}
since $M_{6,1}$ through $M_{6,10}$ are zero, and only $\phi_{11}$ through $\phi_{14}$
are therefore relevant for this term.  It is assumed that all the $\phi_j$
have been pre-computed (for all terms) using the appropriate value of $T$ for the
date and time of interest.   A printed version of a 1365-term nutation series is
obviously not the most convenient form for computation; it is
given here only for the record, since the full series
has not previously appeared in print.  As noted earlier, the series is
available as a pair of plain-text computer files at \citet{url-nuts}, and the
SOFA and NOVAS software packages (\citet{url-sofa}, \citet{url-novas}) include
subroutines for evaluating it.  There are also shorter series available where the highest
precision is not required.  The IERS web site provides, in addition to the full
IAU~2000A series, an IAU~2000B series, which has
only 77 terms and duplicates the IAU~2000A results to within a milliarcsecond
for input times between 1995 and 2050.  NOVAS also provides a subroutine that evaluates
a truncated series, with 488 terms, that duplicates the full series to 0.1 milliarcsecond
accuracy between 1700 and 2300.

Once the nutation series has been evaluated and the values of $\Delta\psi$ and $\Delta\epsilon$
are available, the nutation matrix can be constructed.  The nutation matrix is
simply ${\bf N}(t) = {\bf R}_1(-\epsilon')\,{\bf R}_3(-\Delta\psi)\,{\bf R}_1(\epsilon)$,
where, again, ${\bf R}_1$ and ${\bf R}_3$ are standard rotations
about the x and z axes, respectively (see ``Abbreviations and
Symbols Frequently Used'' for precise definitions), and $\epsilon' = \epsilon
+ \Delta\epsilon$ is the true obliquity (compute $\epsilon$ using eq.~5.12).  This
formulation is comprised of
\begin{enumerate}
\item A rotation from the mean equator and equinox of $t$ to the
mean ecliptic and equinox of $t$.  This is simply a rotation
around the x-axis (the direction toward the mean equinox of $t$)
by the angle $\epsilon$, the mean obliquity of $t$.  After the
rotation, the fundamental plane is the ecliptic of $t$. \item A
rotation around the new z-axis (the direction toward the ecliptic
pole of $t$) by the angle $-\Delta\psi$, the amount of nutation in
longitude at $t$. After the rotation, the new x-axis is in the
direction of the true equinox of $t$. \item A rotation around the
new x-axis (the direction toward the true equinox of $t$) by the
angle $-\epsilon'$, the true obliquity of $t$.  After the
rotation, the fundamental plane is the true equator of $t$,
orthogonal to the computed position of the CIP at $t$.
\end{enumerate}

If we let
\begin{flushright}
\parbox{2.0in}{
\begin{eqnarray*}
      S_1 &=& \sin\,(\epsilon)                  \\
      S_2 &=& \sin\,(-\Delta\psi)               \\
      S_3 &=& \sin\,(-\epsilon-\Delta\epsilon)  \\
\end{eqnarray*} }
\parbox{4.0in}{
\begin{eqnarray}
      C_1 &=& \cos\,(\epsilon)                  \nonumber\\
      C_2 &=& \cos\,(-\Delta\psi)                        \\
      C_3 &=& \cos\,(-\epsilon-\Delta\epsilon)  \nonumber
\end{eqnarray} }
\end{flushright}
then the nutation matrix can also be written:
\begin{equation}
    {\bf N}(t) = \left( \begin{array}{ccc}
           C_2                       &
           S_2  C_1                  &
           S_2  S_1                  \\
         - S_2  C_3                  &
           C_3  C_2  C_1 - S_1  S_3  &
           C_3  C_2  S_1 + C_1  S_3  \\
           S_2  S_3                  &
          -S_3  C_2  C_1 - S_1  C_3  &
          -S_3  C_2  S_1 + C_3  C_1
              \end{array} \right)
\end{equation}

Where high accuracy is not required, coordinates corrected for nutation in
right ascension and declination can be obtained from
\begin{eqnarray}
  \alpha_{\rm t} &\approx& \alpha_{\rm m}
                          + \Delta\psi\, (\cos\epsilon'
                             + \sin\epsilon'\sin\alpha_{\rm m}\tan\delta_{\rm m})
                          - \Delta\epsilon\,\cos\alpha_{\rm m}\tan\delta_{\rm m}
                                                                        \nonumber\\
  \delta_{\rm t} &\approx& \delta_{\rm m}
                          + \Delta\psi\,\sin\epsilon'\cos\alpha_{\rm m}
                          + \Delta\epsilon\,\sin\alpha_{\rm m}
\end{eqnarray}
where $(\alpha_{\rm m},\delta_{\rm m})$ are coordinates with respect to the
mean equator and equinox of date (precession only), $(\alpha_{\rm t},\delta_{\rm t})$
are the corresponding coordinates with respect to the true equator and equinox of date
(precession + nutation), and $\epsilon'$ is the true obliquity.  Note the
$\tan\delta_{\rm m}$ factor in right ascension that makes these formulas unsuitable
for use close to the celestial poles.

The traditional formula for the equation of the equinoxes (the difference between
apparent and mean sidereal time) is
$\Delta\psi\cos\epsilon'$, but in recent years this has been superceded by the
more accurate version given in eq.~2.14.

\subsection{Alternative Combined Transformation}\label{prenut.formulas.alt}

The following matrix, ${\bf C}(t)$, combines precession, nutation, and frame bias and
is used to transform vectors from the GCRS to the Celestial Intermediate Reference
System (CIRS).   The CIRS is defined by the equator of the CIP and an origin
of right ascension called the Celestial Intermediate Origin
(CIO).  The CIO is discussed extensively in Chapter~\ref{erot}.   There, the CIRS is
symbolized \Ec;  it is analogous to the true equator and equinox of date, but with
a different right ascension origin.

The matrix ${\bf C}(t)$ is used in the sense
\begin{equation}
            {\bf r}_{_{\rm CIRS}} = {\bf C}(t) \;{\bf r}_{_{\rm GCRS}}
\end{equation}
and the components of ${\rm C}(t)$, as given in the \citet{iers03} and \AsA, are
\begin{equation}
 {\bf C}(t)  = \; {\bf R}_3(-s) \;\; \left( \begin{array}{ccc}
                        1-bX^2 & -bXY  & -X \\
                        -bXY & 1 - bY^2 & -Y  \\
                           X & Y & 1-b(X^2 + Y^2)
               \end{array} \right)
\end{equation}
where $X$ and $Y$ are the dimensionless coordinates of the CIP in the GCRS
(unit vector components), $b = 1/(1+Z)$,
 $Z = \sqrt{1-X^2-Y^2}$, and $s$ is the {\it CIO locator}, a small angle
described in Chapter~\ref{erot}.  All of these quantities are functions
of time.   ${\bf R}_3$ is a standard rotation around the z~axis; see
``Abbreviations and Symbols Frequently Used'' for a precise definition.\pagebreak

\subsection{Observational Corrections to Precession-Nutation}\label{prenut.formulas.obs}

The IERS still publishes daily values of the observed celestial pole offsets,
despite the vast improvement to the pole position predictions
given by the IAU 2000A precession-nutation model.  The offsets now have magnitudes
generally less than 1~mas.  The fact that they are non-zero is due in part to an
effect of unpredictable amplitude and phase called
the free core nutation (or nearly diurnal free wobble), caused by the
rotation of the fluid core of the Earth inside the ellipsoidal cavity
that it occupies.  The free core nutation appears as a very small
nutation component with a period of about 430 days.  Any other effects not
accounted for in the adopted precession-nutation model will also appear in the
celestial pole offsets.  In any event, the celestial pole offsets are now so
small that many users may now decide to ignore them.   However, it is worth
noting again that, by definition, the Celestial Intermediate Pole (CIP)
includes these observed offsets.

The IERS now publishes celestial pole offsets with respect to
the IAU 2000A precession-nutation model only as $dX$ and $dY$ --- corrections to
the pole's computed unit vector components $X$ and $Y$ in the GCRS (see eq.~5.6
and following notes).  The IERS pole offsets are published in units of
milliarcseconds but they can be converted to dimensionless quantities by dividing them 
by the number of milliarcseconds in one radian, 206264806.247... .  Then, the
observationally corrected values of $X$ and $Y$ are
\begin{equation}
          X_{\rm cor}  = X + dX \qquad\mbox{and}\qquad
          Y_{\rm cor}  = Y + dY
\end{equation}
The corrected values, expressed as dimensionless quantities (unit vector
components), are used, e.g., in the matrix ${\bf C}$ given
in sections \ref{prenut.formulas.alt} and \ref{erot.formulas.tercel}.  That is,
in eqs.~5.24 and 6.18, assume $X = X_{\rm cor}$ and $Y = Y_{\rm cor}$. 

The ecliptic-based pole celestial offsets, $d\psi$ and $d\epsilon$, which are used
to correct the nutation theory's output angles $\Delta\psi$ and $\Delta\epsilon$,
are no longer supplied (actually, they are supplied but only for the old pre-2000
precession-nutation model).  Software that has not been coded to use $X$ and $Y$
directly --- which includes all software developed prior to 2003 --- will
need a front-end to convert the IERS $dX$ and $dY$ values to
$d\psi$ and $d\epsilon$.  A derivation of a conversion algorithm
and several options for its implementation (depending on the accuracy desired)
are given by \citet{kaplan03}.  Succinctly, given dimensionless $dX$ and $dY$
values for a given date $t$, let
\begin{equation}
          \left( \begin{array}{c} dX' \\ dY' \\ dZ' \end{array} \right)
    = {\bf P}(t) \left( \begin{array}{c} dX \\ dY \\ dZ \end{array} \right)
\end{equation}
where  ${\bf P}(t)$
is the precession matrix from J2000.0 to date $t$, and we can set $dZ$= 0 in this
approximation, which holds for only a few centuries around J2000.0.  Then we 
compute the ecliptic-based correction angles in radians using
\begin{equation}
               d\psi        =  dX' / \sin\epsilon \qquad\mbox{and}\qquad
               d\epsilon = dY'
\end{equation}
where $\epsilon$ is the mean obliquity of the ecliptic of date $t$, computed
according to eq~5.12.  The observationally corrected values of $\Delta\psi$ and
$\Delta\epsilon$ are obtained simply by adding $d\psi$ and $d\epsilon$,
respectively:
\begin{equation}
               \Delta\psi_{\rm cor}          =  \Delta\psi         + d\psi     \qquad\mbox{and}\qquad
               \Delta\epsilon_{\rm cor}  =  \Delta\epsilon + d\epsilon
\end{equation}
where care must be taken to ensure that all angles are expressed in the same units.
The corrected values are used in forming
 the nutation matrix ${\bf N}(t)$ and in other nutation-related expressions.   That is,
in eqs.~5.20 and 5.22, assume $\Delta\psi =\Delta\psi_{\rm cor}$ and $\Delta\epsilon =
\Delta\epsilon_{\rm cor}$.   At the same time, the corrected value of $\Delta\psi$ should be used
in forming the equation of the equinoxes using eq.~2.14.

\chapter{Modeling the Earth's Rotation}\label{erot}
\markboth{MODELING THE EARTH'S ROTATION}{MODELING THE EARTH'S ROTATION}
\fbox{\parbox{6.5in}{ Relevant IAU resolutions:\quad B1.6, B1.7, B1.8 of
2000}}
\\
\addcontentsline{toc}{section}{Summary}
\begin{quotation}
\noindent{\bf Summary}\quad Res.~1.8 of 2000 establishes two new reference
points in the plane of the moving (instantaneous) equator for
the measurement of Earth rotation:  the point on the geocentric celestial
sphere is called the Celestial Intermediate
Origin (CIO) and the point on the surface of the Earth
is called the Terrestrial Intermediate Origin (TIO).
The CIO and TIO are specific examples of a concept
called a {\it non-rotating origin} that was first
described by \citet{guinot79,guinot81}.

The Earth Rotation Angle, $\theta$, is the geocentric angle between
the directions of the CIO and TIO, and provides a new way to represent the rotation of the Earth in the transformation from terrestrial to celestial
systems or vice versa.  Traditionally, Greenwich sidereal time,
which is the hour angle of the equinox with respect to the
Greenwich meridian, has served this purpose.  The CIO and TIO are
defined in such a way that $\theta$ is a linear function of
Universal Time (UT1) and independent of the Earth's precession
and nutation; it is a direct measure of the rotational angle of
the Earth around the Celestial Intermediate Pole (see Chapter~\ref{prenut}).
Since none of these statements holds for sidereal
time, the scheme based on the CIO, TIO, and $\theta$ represents
a simplification of the way the rotation of the Earth is treated.
In particular, the transformation between Earth-fixed and space-fixed reference
systems can now be specified by three rotation matrices that are independent
of each other: one for polar motion, one for ``pure'' rotation (i.e., $\theta$), and
one for precession-nutation.

The recent IAU resolutions do not eliminate sidereal time or the use of the
equinox as a fundamental reference point.  Instead, the resolutions establish an
alternative way of dealing with Earth rotation.  The comparison between the two
schemes can be illuminating.  For example, the CIO helps to clarify the
relationship between sidereal time and the Earth's rotation, since $\theta$ is now the
``fast term'' in the formula for sidereal time as a function of UT1.  The remaining terms
comprise the {\it equation of the origins} and represent the accumulated amount of
precession and nutation along the equator as a function of time.  The equation of
the origins is the length of the arc between the equinox and the CIO.
\end{quotation}

\section{A Messy Business}\label{erot.mess}

In the computation of the positions of celestial objects with respect to an Earth-fixed system --- or, equivalently, in the transformation between terrestrial and celestial coordinate systems --- sidereal time has conventionally represented the Earth's rotation about its axis.  For example, the hour angle of an object is simply the local apparent sidereal time minus the object's apparent right ascension with respect to the true equator and equinox of date (see section \ref{time.formulas.erot}).   Once its hour angle and
declination are available, the object's zenith distance and azimuth, or its coordinates with respect to some ground-based instrumental system, can be easily obtained.  The same result can be accomplished by a direct transformation between the celestial and terrestrial coordinate systems, conventionally represented  by a series of rotation matrices, one each for precession, nutation, sidereal time, and polar motion.

Yet there is something untidy about these procedures.  The computation of apparent
sidereal time mixes quantities related to Earth rotation, precession, and
 nutation (see eqs. 2.10--2.14).  Because sidereal time is defined as the hour angle of
 the equinox, the precession of the equinox in right ascension
must be a part of the expression for sidereal time (the terms in parentheses in eq.~2.12), and the mean sidereal day is thereby shorter than the rotation period of the Earth by
about 0.\ssec008.  Nutation also appears, in the equation of the equinoxes (eqs. 2.13 \& 2.14).   The result is that in the computation of hour angle, precession and nutation enter twice:  once in the sidereal time formula and again in the computation of the star's apparent
right ascension; the two contributions cancel for stars on the equator.  Similarly, in the transformation between the celestial and terrestrial coordinate systems, precession and nutation each enter into two of the rotation matrices, and none of the matrices represents Earth rotation alone.

A consequence of this way of doing things is that whenever improvements are made to
the theory of precession, the numerical coefficients in the expression for sidereal time must
also change.  This was not an issue for most of the twentieth century,
since no adjustments were made to the standard precession algorithm,
and the expression for mean sidereal time derived from Newcomb's
developments was used without much thought given to the matter.
It was the change to this expression, necessitated by the adjustment of the
precession constant in the IAU (1976) System of Astronomical
Constants, that first motivated the search for a fundamental
change of procedure.  At about the same time, new high-precision
observing techniques, such as VLBI and lunar laser ranging, were
being applied to the study of all components of the Earth's
rotation, and a review of the basic algorithms seemed
appropriate.  In particular, there was interest in constructing a
new geometrical picture and set of expressions for the
orientation of the Earth as a function of time that would cleanly
separate the effects of rotation, precession and nutation, and
polar motion.  Furthermore, since VLBI is not sensitive to the equinox, a procedure
that used a different reference point seemed desirable.

To bring the Earth's rotation period explicitly into the terrestrial--celestial
transformation, we must define an angle of rotation about the Earth's axis.  As described in
Chapter~\ref{prenut}, what we specifically mean by ``the Earth's axis'' is the line
through the geocenter in the direction of the Celestial Intermediate Pole (CIP).  The
angle of rotation about this axis must be measured with respect to some
agreed-upon direction in space.   Since the CIP moves
small amounts during one rotation of the Earth ($\sim$0.1 arcsecond with respect to the
stars and $\sim$0.005 arcsecond with respect to
the Earth's crust), the   \pagebreak     
reference direction cannot be simply a fixed vector or plane
in inertial space.   What we need is an appropriate azimuthal\footnote{The word
``azimuthal'' is used in its general sense, referring to an angle measured about the z-axis
of a  coordinate system.}
origin --- a point in the moving equatorial plane, which is orthogonal to the CIP.

\section{Non-Rotating Origins}\label{erot.nro}

The reference point that we define must be such that the rate of change of the Earth's
rotation angle, measured with respect to this point, is the angular velocity
of the Earth about the CIP.   As the CIP moves, the point
must move to remain in the equatorial plane;  but the point's motion
must be such that the measured rotation angle is not contaminated by some component
of the motion of the CIP itself.

The concept of a ``non-rotating origin'' (NRO) on the equator can be applied to any rotating
body.  The NRO idea was first described by Bernard Guinot \citep{guinot79,guinot81} and further developed by Nicole Capitaine and collaborators \citep{cap86,cap90,cap91,cap00,cap00a}.
The condition on the motion of such a point is simple:  as the
equator moves, the point's instantaneous motion
must always be orthogonal to the equator.  That is, the point's motion at some time $t$
must be directly toward or away from the position of the pole of rotation
at $t$.  Any other motion of the point would have a component {\it around\/} the
axis/pole and would thus introduce a spurious rate into
the measurement of the rotation angle of the body as a function
of time.  The point is not unique; any arbitrary point on the
moving equator could be made to move in the prescribed manner.
For the Earth, the difference between the motion of a non-rotating origin and
that of the equinox on the geocentric celestial sphere is illustrated in Fig.~6.1.

As illustrated in the figure, the motion of the non-rotating origin, $\sigma$, is always orthogonal to the equator, whereas the equinox has a motion along the equator (the precession in right ascension).
How do we specify the location of a non-rotating origin?  There are three possibilities, outlined in the
Formulas section of this chapter.  In the most straightforward scheme, one simply uses the GCRS right ascension of $\sigma$ obtained from a numerical integration (the GCRS is the ``geocentric ICRS'').
Alternatively, the position of $\sigma$ can be defined by a quantity, $s$, that
is the difference between the lengths of two arcs on the celestial sphere.  Finally, one can specify the
location of $\sigma$ with respect to the equinox, $\Upsilon$:  the equatorial arc
$\overline{\Upsilon\sigma}$ is called the {\it equation of the origins}.  Whatever geometry
is used, the position of $\sigma$ ultimately depends on an integral over time, because the defining property of $\sigma$ is its {\it motion} --- not a static geometrical relationship with other
points or planes.  The integral involved is fairly simple and
depends only on the coordinates of the pole and their derivatives with respect to time.  The initial
point for the integration can be any point on the moving equator at any time $t_0$.\\[0.3in]

\begin{center}
\includegraphics[width=6.1in,trim=1.0in 1.1in 1.0in 0.7in,clip=true]{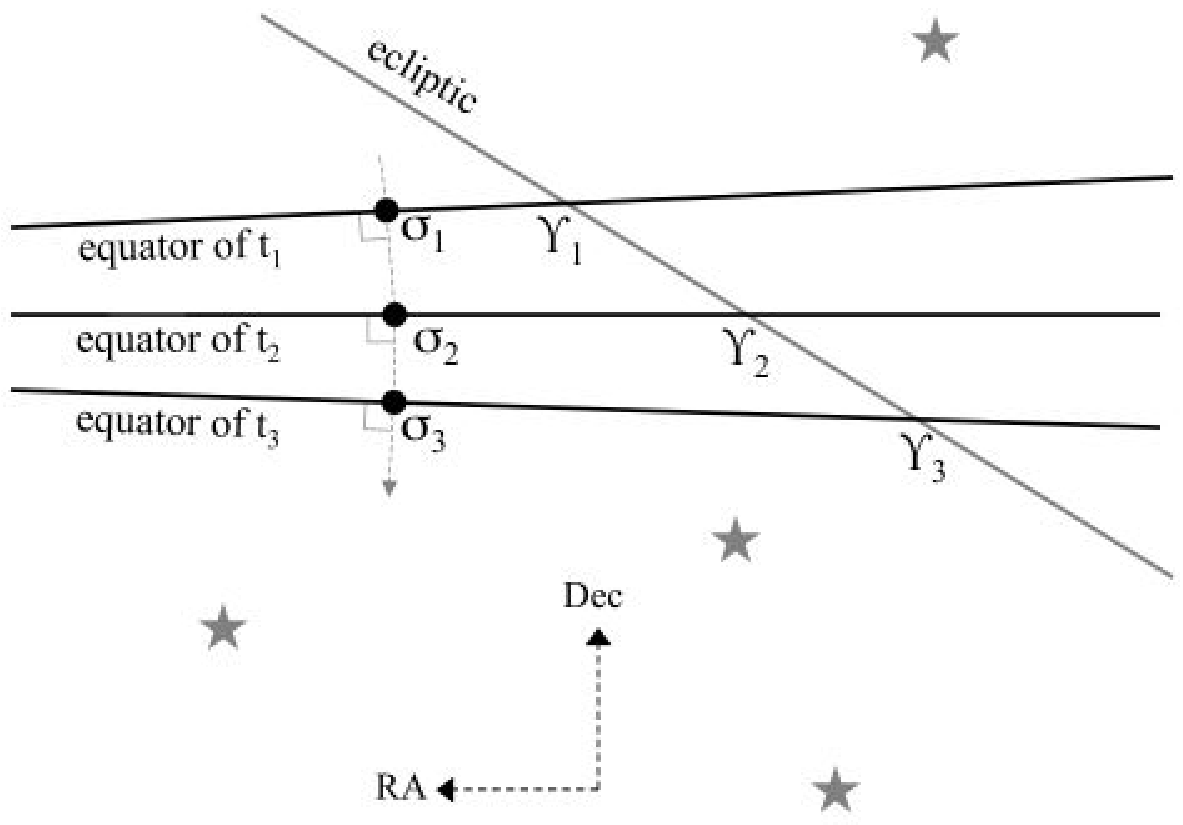}
\parbox{5.0in}{\small\vspace{0.2in}{\bf Figure 6.1}\quad Motion of a non-rotating origin, $\sigma$, compared with that of the true equinox, $\Upsilon$.  ``Snapshots'' of the positions of the points are shown at three successive times, $t_1$, $t_2$, and $t_3$.
The positions are shown with respect to a geocentric reference
system that has no systematic rotation with respect to a set of
extragalactic objects. The ecliptic is shown in the figure as
fixed, although it, too, has a small motion in inertial
space.}\\[2ex]
\end{center}

So far we have discussed a non-rotating origin only on the celestial sphere, required because of
the movement of the CIP in a space-fixed reference system.  But
there is a corresponding situation on the surface of the Earth.   The CIP has motions in both the
celestial and terrestrial reference systems.  Its motion in the celestial system is precession-nutation
and its motion in the terrestrial system is polar motion, or wobble.  From the point of view of a conventional geodetic coordinate system ``attached'' to the surface of the Earth (i.e., defined by the specified  coordinates of a group of stations), the CIP wanders around near the geodetic pole in a quasi-circular motion with an  amplitude of about 10~meters (0.3 arcsec) and two primary periods, 12~and 14~months.  Thus the equator of the CIP has a slight quasi-annual wobble around the geodetic equator.  Actually, it is better thought of in the opposite sense:  the geodetic equator has a slight wobble with respect to the equator of the CIP. That point of view makes it is a little clearer why a simple ``stake in the ground''  at the geodetic equator would not be suitable for measuring the Earth rotation angle around the CIP.  The situation is orders of magnitude less troublesome than that on the celestial sphere, but for completeness (and very precise applications) it is appropriate to define a terrestrial non-rotating origin, designated $\varpi$.  It stays on the CIP equator, and assuming that the current  amplitude of polar motion remains approximately constant, $\varpi$ will bob north and south by about 10~m in geodetic latitude every year or so and will have a secular eastward motion in longitude of about 1.5~mm/cen.  The exact motion of $\varpi$ depends, of course, on what polar motion, which is unpredictable, actually turns out to be.

The two non-rotating origins, $\sigma$ and $\varpi$, are called the Celestial Intermediate Origin (CIO) and the Terrestrial Intermediate Origin (TIO).  Both lie in the same plane --- the equator of the CIP.  The Earth Rotation Angle, $\theta$, is defined as the geocentric angle between these two points.  The angle $\theta$ is a linear function of Universal Time (UT1).  The formula, given in the note to res.~B1.8 of 2000, is simply  $\theta =  2\pi\,(0.7790572732640 + 1.00273781191135448 \; D_U)$, where $D_U$ is the number of UT1 days from JD 2451545.0 UT1.  The formula assumes a constant angular velocity of the Earth:  no attempt is made to model its secular decrease due to tidal friction, monthly tidal variations, changes due to the exchange of angular momentum between the atmosphere and the solid Earth, and other phenomena.  These effects will be reflected in the time series of UT1--UTC or $\Delta T$ values (see Chapter~\ref{time}) derived from precise observations.

The expression given above for $\theta$ is now the ``fast term'' in the formula for mean sidereal time; see eq.~2.12.  It accounts for the rotation of the Earth, while the other terms account for the motion of the equinox along the equator due to precession.

The plane defined by the geocenter, the CIP, and TIO is called the {\it TIO meridian}.  For most ordinary astronomical purposes the TIO meridian can be considered to be identical to what is often referred to as the Greenwich meridian.  The movement of this meridian with respect to a conventional geodetic system is important only for the most precise astrometric/geodetic applications.  It is worth noting that the TIO meridian, and the zero-longitude meridians of modern geodetic systems, are about 100~m from the old transit circle at Greenwich \citep{gebel}.  The term ``Greenwich meridian'' has ceased to have a technical meaning in the context of precise geodesy --- despite the nice line in the sidewalk at the old Greenwich observatory.  This has become obvious to tourists carrying GPS receivers!

\section{The Path of the CIO on the Sky}\label{erot.ciopath}

If we take the epoch J2000.0 as the starting epoch for evaluating the integral that provides the position of the CIO, the only mathematical requirement for the initial point is that it lie on the instantaneous (true) equator of that date --- its position along the equator is arbitrary.  By convention, however, the initial position of the CIO on the instantaneous equator of J2000.0 is set so that equinox-based and CIO-based computations of Earth rotation yield the same answers;  we want the hour angle of a given celestial object to be the same, as a function of UT1 (or UTC), no matter how the calculation is done.  For this to happen, the position of the CIO of J2000.0 must be at GCRS right ascension
 $0\degr\,0'\,00.\!''002012$.   This is about 12.8 arcseconds west of the true equinox of that date.

Since the CIO rides on the instantaneous equator, its primary motion over the next few millenia is southward at the rate of precession in declination, initially 2004 arcseconds per century.   Its rate of southward motion is modulated (but never reversed) by the nutation periodicities.  Its motion in GCRS right ascension is orders of magnitude less rapid; remember that the CIO has no motion along the instantaneous equator, and the instantaneous equator of J2000.0 is nearly co-planar  with the GCRS equator (xy-plane).   The motion of the CIO in GCRS right ascension over the next few millenia is dominated by a term  proportional to $t^3$;  the GCRS right ascension of the CIO at the beginning of year 2100 is only
$0.\!''068$; at the beginning of 2200 it is
$0.\!''573$; and at the beginning of 2300 it is
$1.\!''941$.   Nutation does produce a very slight wobble in the CIO's right ascension,  but the influence of the nutation terms is suppressed by several orders of magnitude relative to their effect on the position of the pole.  We can say, therefore, that to within a few arcseconds error,
the path of the CIO on the celestial sphere over the next few centuries is nearly a straight line
southward along the GCRS $\alpha$=0 hour circle.

\begin{center}
\includegraphics[width=7.5in,trim=0.7in 2.6in 0.0in 0.6in,clip=true]{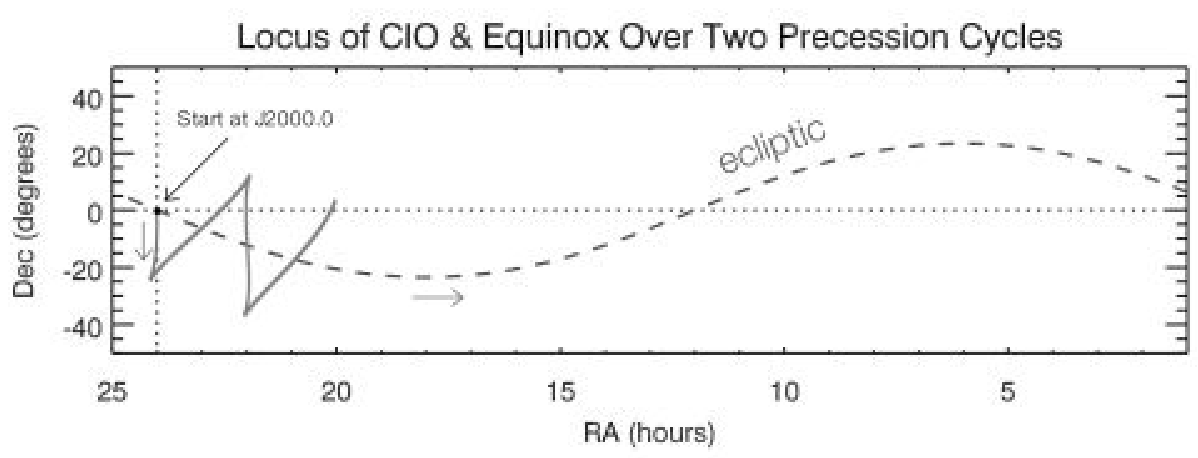}
\parbox{5.0in}{\small\vspace{0.2in}{\bf Figure 6.2}\quad Locus of the CIO (solid line) and equinox (dashed line) on the celestial sphere over 5\E4 years, with respect to space-fixed coordinates.  During this time the equinox wraps around the figure twice and ends up approximately
at the starting point.}\\[4ex]
\end{center}

The solid line on the left side of Figure 6.2 indicates the locus of the CIO in the GCRS
over 50,000 years --- about two precession cycles.  The ecliptic is shown as a dashed line.  The
initial nearly straight southward motion from the starting point at J2000.0 is clearly shown.
There are occasional ``cusps'' in the CIO's motion, where
its secular motion comes to a temporary halt before reversing.  The first of
these stationary points occurs in just over a quarter of a precession cycle, as the section of
the moving equator that is farthest south in ecliptic coordinates precesses to near the
GCRS $\alpha$=0 hour circle.  At that time, the CIO will exhibit only nutational oscillations
around a point that remains fixed on the celestial sphere to within 10~mas for almost
a decade.  Then its motion resumes, this time northward and westward.\footnote{There
is nothing profound about the stationary points or the dates on which they occur.  If
we had started the CIO at GCRS right ascension $6^{\rm h}$ or
$18^{\rm h}$ at J2000.0, it would have {\it started}\/ at a stationary point.}  The motion
of the equinox over the same 50,000-year time period begins
at nearly the same point as the CIO (on the plot scale used, the points overlap), but
smoothly follows the ecliptic westward (to the right on the plot), wrapping around twice and
ending up essentially at the starting point.

\section{Transforming Vectors Between Reference Systems}\label{erot.transform}

The reference points described above allow us to define three geocentric reference systems that share, as a common reference plane, the instantaneous, or true, equator of date.  The instantaneous equator is now defined as the plane through the geocenter orthogonal to the direction of the CIP at a given time, $t$.  The three reference systems are:\label{eqsys} \pagebreak        
\begin{enumerate}
\item True equator and equinox of $t$ --- azimuthal origin at the true equinox ($\Upsilon$) of $t$
\item Celestial Intermediate Reference System (CIRS) --- azimuthal origin at the
Celestial Intermediate Origin (CIO or $\sigma$) of $t$
\item Terrestrial Intermediate Reference System (TIRS) --- azimuthal origin at the Terrestrial Intermediate Origin (TIO or $\varpi$) of $t$
\end{enumerate}
In this \pubname\ we will often refer to these reference systems by the symbols \Eq, \Ec, and \Et, respectively; E denotes an equatorial system, and the subscript indicates the azimuthal origin. \Et\ rotates with the Earth whereas the orientations of \Eq\ and \Ec\ change slowly with respect to local inertial space.  The transformation between \Eq\ and \Et\ is just a rotation about the z-axis (which points toward the CIP) by GAST, the angular equivalent of Greenwich apparent sidereal time.  The transformation between \Ec\ and \Et\ is a similar rotation, but by $\theta$, the Earth Rotation Angle.  These two transformations reflect different ways --- old and new --- of representing the rotation of the Earth.

A short digression into terrestrial, i.e., geodetic, reference systems is in order here.  These systems all have their origin at the geocenter and rotate with the crust of the Earth.  Terrestrial latitude, longitude, and height are now most commonly given with respect to a {\it reference ellipsoid}, an oblate spheroid that approximates the Earth's overall shape (actually, that best fits the {\it geoid}, a gravitational equipotential surface).  The current standard reference elliposid for most purposes is that of the World Geodetic System 1984 (WGS~84), which forms the basis for the coordinates obtained from GPS.  The   WGS~84 ellipsoid has  an equatorial radius of 6,378,137 meters and a polar flattening of
1/298.257223563.   For the precise measurement of Earth rotation, however, the International Terrestrial Reference System (ITRS) is used, which was defined by the International Union of Geodesy and Geophysics (IUGG) in 1991.  The ITRS is realized for practical purposes by the adopted coordinates and velocities\footnote{The velocities are quite small and are due to plate tectonics and post-glacial rebound.} of a large group of observing stations.  These coordinates are expressed as geocentric rectangular 3-vectors and thus are not dependent on a reference ellipsoid.  The list of stations and their coordinates is referred to as the International Terrestrial Reference Frame (ITRF).  The fundamental terrestrial coordinate system is therefore defined in exactly the same way as the fundamental celestial coordinate system (see Chapter~\ref{refsys}):  a prescription is given for an idealized coordinate system (the ITRS or the ICRS), which is realized in practice by the adopted coordinates of a group of reference points (the ITRF stations or the ICRF quasars).   The coordinates may be refined as time goes on but the overall system is preserved.  It is important to know, however, that the ITRS/ITRF is consistent with WGS~84 to within a few centimeters;  thus for all astronomical purposes the GPS-obtained coordinates of instruments can be used with the algorithms presented here.

Our goal is to be able to transform an arbitrary vector (representing for example, an instrumental position, axis, boresight, or baseline) from the ITRS ($\approx$WGS~84$\approx$GPS) to the GCRS.   The three equatorial reference systems described above ---  \Eq, \Ec, and \Et\ --- are waypoints, or intermediate stops, in that process.  The complete transformations are:\label{transforms}

\parbox{3in}{
\begin{center}
\begin{tabular}{c}
                                                              \\
     {\bf Equinox-Based Transformation} \\
                                                              \\
     \fbox{\D ITRS or WGS 84\D}                               \\
     $\|$ \\ {\it polar motion} \\ $\Downarrow$       \\
     \fbox{\D \Et\ --- Terrestrial Intermediate Ref. System\D} \\
     $\|$ \\ {\it Greenwich apparent sidereal time} \\ $\Downarrow$       \\
     \fbox{\D \Eq\ --- true equator \& equinox\D}     \\
     $\|$ \\ {\it equinox-based rotation for} \\
               {\it nutation + precession + frame bias}   \\ $\Downarrow$       \\
     \fbox{\D GCRS\D}     \\
                                                    \\
\end{tabular}
\end{center}
}
\hfill
\parbox{3in}{
\begin{center}
\begin{tabular}{c}
                                                      \\
     {\bf CIO-Based Transformation} \\
                                                      \\
     \fbox{\D ITRS or WGS 84\D}                               \\
     $\|$ \\ {\it polar motion} \\ $\Downarrow$       \\
     \fbox{\D \Et\ --- Terrestrial Intermediate Ref. System\D} \\
     $\|$ \\ {\it Earth Rotation Angle} \\ $\Downarrow$       \\
     \fbox{\D \Ec\ --- Celestial Intermediate Ref. System\D}  \\
     $\|$ \\ {\it CIO-based rotation for} \\
               {\it nutation + precession + frame bias}   \\ $\Downarrow$       \\
     \fbox{\D GCRS \D}     \\
                                                    \\
\end{tabular}
\end{center}
}

\noindent which are equivalent. That is, given the same input vector, the same output vector will result
from the two procedures.  In the CIO-based transformation, the three sub-transformations (for polar motion, Earth Rotation Angle, and nutation/precession/frame bias) are independent.  That is not true for the equinox-based method, because apparent sidereal time incorporates precession and nutation.    Each of the two methods could be made into a single matrix, and the two matrices must be numerically identical.  That means that the use of the CIO in the second method does not increase the precision of the result but simply allows for a mathematical redescription of the overall transformation --- basically, a re-sorting of the effects to be taken into account.  This redescription of the transformation provides a clean separation of the three main aspects of Earth rotation, and recognizes that the
observations defining modern reference systems are not sensitive to the equinox.  It thus yields a more straightforward conceptual view and facilitates a simpler observational analysis for Earth-rotation measurements and Earth-based astrometry.

These transformations are all rotations that pivot around a common point, the geocenter.
Although developed for observations referred to the geocenter, the same set of rotations can be
applied to observations made from a specific point on the surface of the Earth.  (This follows from
the assumption that all points in or on the Earth are rigidly attached to all other points.  The
actual non-rigidity --- e.g., Earth tides --- is handled as a separate correction.)   In such a case,  the computations for parallax, light-bending, aberration, etc., must take into account
the non-geocentric position and velocity of the observer.  Then the final computed coordinates
are referred, not to the GCRS, but rather to a proper reference system (in the terminology
of relativity) of the observer.
\\[0.1in]
\indent We return to the more familiar problem mentioned at the beginning of the chapter:  the computation of local hour angle.  In the usual equinox-based scheme, the apparent place\footnote{See footnote~1 on page~\pageref{placenote}.} of the star or planet is expressed with respect to the true equator and equinox of date (\Eq).  The local hour angle is just h = GAST -- $\alpha_{_\Upsilon}$ + $\lambda$, where GAST is Greenwich apparent sidereal time, $\alpha_{_\Upsilon}$ is the apparent right ascension of the object, measured with respect to the true equinox, and $\lambda$ is the longitude of
the observer (corrected, where necessary, for polar motion).  Obviously these quantities must all be
given in the same units.  In the CIO-based scheme, the apparent place would be expressed in the
Celestial Intermediate Reference System (\Ec), and h = $\theta$ -- $\alpha_{\sigma}$ + $\lambda$, where $\theta$ is the Earth Rotation Angle and $\alpha_{\sigma}$ is the apparent right ascension of the
object, measured with respect to the CIO.  (The recommended terminology is {\it intermediate place}\/ for
the object position in the \Ec\ and {\it intermediate right ascension}\/ for the quantity $\alpha_{\sigma}$,
although some people hold that the term right ascension should refer only to an origin at the equinox.)
In the CIO-based formula, precession and nutation come into play only once, in expressing the object's
right ascension in the \Ec\ system.  See section \ref{erot.formulas.ha} for more details.

\section{Formulas}\label{erot.formulas}

The formulas below draw heavily on the developments presented previously.  In particular, the 3$\times$3 matrices ${\bf P}$, ${\bf N}$, and ${\bf B}$ represent the transformations for precession, nutation, and frame bias, respectively, and are taken directly from Chapters~\ref{prenut} and \ref{refsys}.   The matrices ${\bf P}$ and ${\bf N}$ are functions of time, $t$.  The time measured in Julian centuries of TDB (or TT) from J2000.0 is denoted $T$ and is given by $T$ = (JD(TDB) -- 2451545.0)/36525.  The elementary
rotation matrices ${\bf R}_1$, ${\bf R}_2$, and ${\bf R}_3$ are defined in ``Abbreviations and Symbols
Frequently Used''.   Formulas from Chapter~\ref{time} for sidereal time and the Earth Rotation Angle are used.   Explanations of, and formulas for the time scales UT1, TT, and TDB are also found in
Chapter~\ref{time}.

The ultimate objective is to express ``local'' Earth-fixed vectors,
representing geographic positions, baselines, instrumental axes and boresights, etc.,
in the GCRS, where they can be related to the proper coordinates\footnote{See footnote~1 on page~\pageref{placenote}.} of celestial objects.
As mentioned above, the ``GCRS'' is a reference system that can be thought of as the
``geocentric ICRS''.  Celestial coordinates in the GCRS are obtained from basic
ICRS data (which are barycentric) by applying the usual algorithms for proper place; see
section \ref{rel.obs}.   In this chapter we will be working entirely in a geocentric system and
the GCRS will be obtained from a series of rotations that start with an ordinary Earth-fixed
geodetic system.

\subsection{Location of Cardinal Points}\label{erot.formulas.cardinal}

We will start by establishing the positions of three cardinal points within the GCRS:  the Celestial Intermediate Pole (CIP), the true equinox of date ($\Upsilon$), and the Celestial Intermediate Origin (CIO).   The unit vectors toward these points will be designated \vcip, \veqx, and \vcio, respectively.  As the Earth precesses and nutates in local inertial space, these points are in continual motion.

The CIP and the equinox can easily be located in the GCRS at any time $t$ simply by recognizing that they are, respectively, the z- and x-axes of the true equator and equinox of date system (\Eq) at $t$.  The unit vectors therefore are:
\begin{eqnarray}
\mbox{CIP\w{xxxx}:} \qquad  \vcip(t) & = &
               {\bf B}^{\rm T} \; {\bf P}^{\rm T}(t) \; {\bf N}^{\rm T}(t) \;
                 \left( \begin{array}{c} 0 \\ 0 \\ 1 \end{array} \right)
                                                                  \nonumber\\
  & & \\
\mbox{Equinox:} \qquad \veqx(t) & = &
               {\bf B}^{\rm T} \; {\bf P}^{\rm T}(t) \; {\bf N}^{\rm T}(t) \;
                 \left( \begin{array}{c} 1 \\ 0 \\ 0 \end{array} \right)
                                                                  \nonumber
\end{eqnarray}
where  the matrix {\bf B} accounts for the GCRS frame bias (same as for the ICRS) and the matrices {\bf P}$(t)$ and {\bf N}$(t)$ provide the transformations for precession and nutation, respectively, at time $t$.  These matrices were developed in sections \ref{refsys.formulas} and \ref{prenut.formulas}; the superscript T's above indicate that the transpose of each of these matrices as previously developed is used (i.e., we are using the ``reverse'' transformations here, from the true equator and equinox of $t$ to the GCRS).  The first equation above is simply eq.~5.6 rewritten.   Note that \veqx\ is orthogonal to \vcip\ at each time $t$.

The components of the unit vector in the direction of the pole, \vcip, are denoted $X$, $Y$, and $Z$, and another approach to determining \vcip\ is to use the series expansions for $X$ and $Y$ given in the \citet{iers03}.  There is a table of daily values of $X$ and $Y$ in Section~B of \AsA\ (there labeled
$\mathcal{X}$ and $\mathcal{Y}$).  Once $X$ and $Y$ are converted to dimensionless values, $Z = \sqrt{1 - X^2 - Y^2}$. \quad(The IERS series for $X$ and $Y$ are part of a data analysis approach adopted by the IERS that avoids any explicit reference to the ecliptic or the equinox, although the underlying theories are those described in Chapter~\ref{prenut}.)

There are three possible procedures for obtaining the location of the Celestial Intermediate Origin on the celestial sphere at a given time: (1) following the arc on the instantaneous equator from the equinox to the CIO;  (2) directly computing the position vector of the CIO in the GCRS by numerical integration;  or (3) using the quantity $s$, representing the difference in two arcs on the celestial sphere, one of which ends at the CIO.  These procedures will be described in the three subsections below.  Figure 6.3 indicates the geometric relationships among the points mentioned.

\begin{center}
\includegraphics[width=6.5in,trim=0.0in 1.7in 0.0in 0.9in,clip=true]{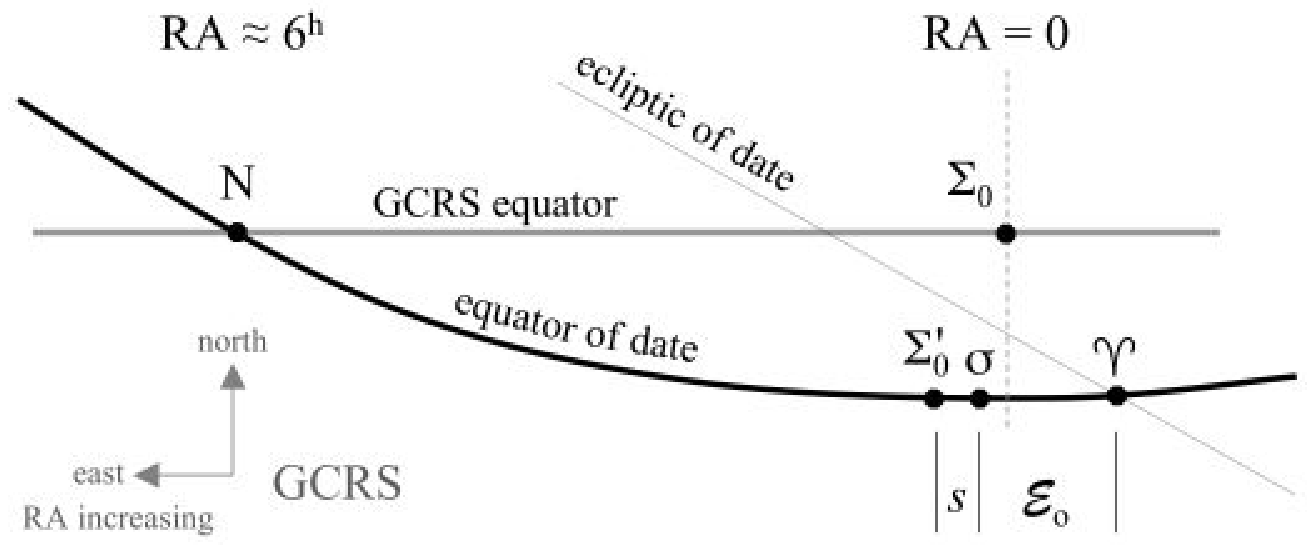}
\parbox{5.3in}{\small\vspace{0.2in}{\bf Figure 6.3}\quad Relationship between various points involved
 in locating the CIO.  The figure approximates the relative positions at 2020.0, although the spacings are not to scale.  The point labeled $\sigma$ is the CIO; $\Upsilon$ is the true equinox; $\Sigma_0$ is the GCRS right ascension origin; N is the ascending node of the instantaneous (true) equator of date on the GCRS equator; and $\Sigma'_0$ is the point on the instantaneous equator that is the same distance from N as $\Sigma_0$. As shown, the quantities $s$ and $\mathcal{E}_o$ are respectively positive and negative. The motion of the instantaneous equator (which is orthogonal to the CIP) is generally southward (down
 in the figure) near RA=0, which tends to move the equinox westward (right) and the CIO very slightly eastward (left) with respect to the GCRS.}\\[4ex]
\end{center}

\subsubsection{CIO Location Relative to the Equinox}\label{erot.formulas.cardinal.cio1}

The arc on the instantaneous (true) equator of date $t$ from the CIO to the equinox is called the {\it equation of the origins} and is the right ascension of the true equinox relative to the CIO (or, minus the true right ascension of the CIO).  The equation of the origins is also the difference $\theta - \mbox{GAST}$.  It therefore equals the accumulated precession-nutation of the equinox in right ascension, given by the sum of the terms in parentheses from eq.~2.12 and the equation of the equinoxes given in eq.~2.14 (all times --1).  The equation of the origins in arcseconds therefore is:
\begin{eqnarray}
       \mathcal{E}_o = & - & 0.014506 - 4612.156534\,T - 1.3915817\,T^2
                                                                             + 0.00000044\,T^3  \nonumber\\
             & + & 0.000029956\,T^4 + 0.0000000368\,T^5
                          \quad  -  \Delta\psi \, \cos\epsilon                              \nonumber\\
             & - &  0.00264096 \,\sin \,( \Omega )                                     \nonumber\\
             & - &  0.00006352 \,\sin \,( 2 \Omega )                                  \nonumber\\
             & - & 0.00001175 \,\sin \,( 2 F - 2 D + 3 \Omega )                  \nonumber\\
             & - & 0.00001121 \,\sin \,( 2 F - 2 D + \Omega )                                      \\
             & + &  0.00000455 \,\sin \,( 2 F - 2 D + 2 \Omega )                \nonumber\\
             & - & 0.00000202 \,\sin \,( 2 F            + 3 \Omega )               \nonumber\\
             & - & 0.00000198 \,\sin \,( 2 F            + \Omega )                  \nonumber\\
             & + & 0.00000172 \,\sin \,( 3 \Omega )                                   \nonumber\\
             & + &  0.00000087 \, T \,\sin \,( \Omega ) + \cdots                 \nonumber
\end{eqnarray}
where $T$ is the number of centuries of TDB (or TT) from J2000.0;  $\Delta\psi$ is the nutation in longitude, in arcseconds; $\epsilon$ is the mean obliquity of the ecliptic; and $F$, $D$, and $\Omega$ are fundamental luni-solar arguments.  All of the angles are functions of time; see Chapter~\ref{prenut}
for expressions (esp. eqs.~5.12, 5.15, \& 5.19).  There is a table of daily values of $\mathcal{E}_o$ in Section~B of \AsA.

To transform an object's celestial coordinates from the true equator and equinox of $t$ to the Celestial Intermediate System (i.e., from \Eq\ to \Ec), simply add $\mathcal{E}_o$ to the object's true right ascension.  To similarly transform the components of a position vector, apply the rotation ${\bf R}_3(-\mathcal{E}_o)$.  Since many existing software systems are set up to produce positions with respect to the equator and equinox of date, this is a relatively easy way to convert those positions to the Celestial Intermediate Reference System if desired.   Note that in such a case there is no computational difference in using either the equinox-based or CIO-based methods for computing hour angle: $\mathcal{E}_o$ is computed in both methods and is just applied to different quantities.  In the equinox-based method, $\mathcal{E}_o$ is subtracted from $\theta$ to form sidereal time;  in the CIO-based method, $\mathcal{E}_o$ is added to the object's true right ascension so that $\theta$ can be used in place of sidereal time.

The position of the CIO in the GCRS, \vcio, can be established by taking the position vector of the equinox in the GCRS, \veqx, and
rotating it counterclockwise by the angle $-\mathcal{E}_o$ (i.e., clockwise by
$\mathcal{E}_o$) about the axis \vcip.  Equivalently,
establish the orthonormal basis triad of the equator-and-equinox system within the GCRS:
\veqx, (\vcip $\times$ \veqx), and \vcip.  Then
\begin{equation}
          \vcio = \veqx\,\cos\mathcal{E}_o - (\vcip \times \veqx) \sin\mathcal{E}_o
\end{equation}

\subsubsection{CIO Location from Numerical Integration}\label{erot.formulas.cardinal.cio2}

As described above, a non-rotating origin can be described as a point on
the moving equator whose instantaneous motion is always orthogonal
to the equator.  A simple geometric construction
based on this definition yields the following differential
equation for the motion of a non-rotating origin:
\begin{equation}
     \dot{\bm{\sigma}}(t) =  -\Big(\,\bm{\sigma}(t) \cdot \dot{\bf n}(t) \,\Big) \; {\bf n}(t)
\end{equation}
That is, if we have a model for the motion of the pole, ${\bf n}(t)$,
the path of the non-rotating origin is described by $\bm{\sigma}(t)$,
once an initial point on the equator, $\bm{\sigma}(t_0)$, is
chosen.  Conceptually and practically, it is simple to integrate
this equation, using, for example, a standard 4th-order
Runge-Kutta integrator.  For the motions of the real Earth, fixed
step sizes of order 0.5~day work quite well, and the integration
is quite robust.  This is actually a one-dimensional problem
carried out in three dimensions, since we know the non-rotating origin
remains on the equator; we really need only to know where along
the equator it is.  Therefore, two constraints can be applied at
each step: $|\bm{\sigma}| = 1$ and $\bm{\sigma}\cdot{\bf n} = 0$.  See
\citet{kaplan05}.

The above equation is quite general, and to get the specific motion of the CIO,
each of the vectors in the above equation is expressed with respect to the
GCRS, i.e., $\bm{\sigma}(t)\rightarrow\vcio(t)$,
${\bf n}(t)\rightarrow\vcip(t)$, and $\dot{\bf n}(t)\rightarrow\dot{\bf n}_{_{\rm GCRS}}(t)$.
The pole's position, $\vcip(t)$, is given by the first expression in eq.~6.1.
The pole's motion, $\dot{\bf n}_{_{\rm GCRS}}(t)$, can be obtained by numerical
differentiation of the pole's position.
By numerically integrating the above equation, we obtain a time series
of unit vectors, $\vcio(t_i)$, where each $i$ is an integration step.
The fact that this is actually just a one-dimensional problem means that it
is sufficient to store as output the CIO right ascensions (with respect to the
GCRS), using eq.~5.2 to decompose the $\vcio(t_i)$ vectors.  In this way,
the integration results in a tabulation of CIO right ascensions at discrete times.
For example, see \citet{url-ciora},
where the times are expressed as TDB Julian dates and the right ascensions are in
arcseconds.    This file runs from years 1700 to 2300 at 1.2-day intervals.

When we need to obtain the position of the CIO for some specific time, the file of CIO
right ascensions can be interpolated to that time.  The CIO's unit vector, if required,
can be readily computed:  generally, given a fixed coordinate system within which a
pole and its equator move, a point of interest on the equator has a position vector
in the fixed system given by
\begin{equation}
   {\bf r} = \left( \begin{array}{c} Z \cos\alpha \\ Z \sin\alpha \\
                 -X \cos\alpha - Y \sin\alpha \end{array} \right)
\end{equation}
where $\alpha$ is the right ascension of the point, and $X$, $Y$, and $Z$ are the
components of the pole's instantaneous unit vector, and all of these quantities
are measured relative to the fixed coordinate system.  The
vector {\bf r} is not in general of unit length but it can be readily normalized.  This
formula allows us to reconstruct the unit vector toward the
CIO from just its GCRS right ascension value at the time of interest, since we already
know how to obtain the pole's position vector in the GCRS for that time.

Eq.~6.4 can be also made to yield the locus of
the Terrestrial Intermediate Origin (TIO), simply by referring all the vectors to the
ITRS --- a rotating geodetic system --- rather than the GCRS.   In this case,
therefore, $\bm{\sigma}(t)\rightarrow\bm{\varpi}_{_{\rm ITRS}}(t)$,
${\bf n}(t)\rightarrow{\bf n}_{_{\rm ITRS}}(t)$, and $\dot{\bf n}(t)\rightarrow\dot{\bf n}_{_{\rm ITRS}}(t)$.
The path of the CIP within the ITRS ($\,{\bf n}_{_{\rm ITRS}}(t)\,$) is what we call
polar motion (usually specified by the parameters $x_p$ and $y_p$), and is
fundamentally unpredictable.  The integration can therefore only be
accurately done for past times, using observed  pole
positions.  A computed future path of the TIO on the surface of the Earth
depends on the assumption that the two major periodicities observed
in polar motion will continue at approximately the current amplitude.

\subsubsection{CIO Location from the Arc-Difference $s$}\label{erot.formulas.cardinal.cio3}

On the celestial sphere, the Earth's instantaneous (moving) equator intersects the GCRS equator at two nodes.    Let N be the ascending node of the instantaneous equator on the GCRS equator.  We can define a scalar quantity $s(t)$ that represents the {\it difference} between the length of the arc from N  westward to the CIO (on the instantaneous equator) and the length of the arc from N westward to the GCRS origin of right ascension (on the GCRS equator).  The quantity $s$ is called the {\it CIO locator}.
If $\sigma$ represents the CIO and $\Sigma_0$ represents the right ascension origin of the GCRS (the direction of the GCRS x-axis), then
\begin{equation}
          s = \overline{\sigma\mbox{N}} - \overline{\Sigma_0\mbox{N}}
\end{equation}
See Fig.~6.3, where the points $\Sigma_0$ and $\Sigma_0'$ are equidistant from the node N. The quantity $s$ is seen to be the ``extra'' length of the arc on the instantaneous equator from N to $\sigma$,
the position of the CIO.  The value of $s$ is fundamentally obtained from an integral,
\begin{equation}
   s(t) = - \int_{t_0}^t \frac{X(t)\dot{Y}(t) - Y(t)\dot{X}(t)}{1+Z(t)} \, dt \; + s_0
\end{equation}
where $X(t)$, $Y(t)$, and $Z(t)$ are the three components of the unit vector,
$\vcip(t)$, toward the celestial pole (CIP).  See, e.g., \citet{cap00} or the \citet{iers03}.
The constant of integration, $s_0$, has been set to ensure that the equinox-based and CIO-based compu\-ta\-tions of Earth ro\-ta\-tion yield the same answers; $s_0$=$94\,\uas$ \citep{iers03}.
Effectively, the constant adjusts the position of the CIO on the equator and is thus part of the arc $\overline{\sigma\mbox{N}}$.   For practical purposes, the value of $s$ at any given time is provided by
a series expansion, given in Table 5.2c of the \citet{iers03}.  Software to evaluate this series is available at the IERS Conventions web site and is also part of the SOFA package.  There is a table of daily values of $s$ in Section~B of \AsA.

At any time, $t$, the the unit vector toward the node N is simply
${\bf N}_{_{\rm GCRS}} = (-Y,X,0)/\sqrt{X^2+Y^2}$ (where we are no longer explicitly indicating the time
dependence of $X$ and $Y$).
To locate the CIO, we rearrange eq.~6.6 to yield the arc length $\overline{\sigma\mbox{N}}$:
\begin{equation}
     \overline{\sigma\mbox{N}} = s + \overline{\Sigma_0\mbox{N}}  = s + \arctan(X/(-\! Y))
\end{equation}
The location of the CIO is then obtained by starting at the node N and moving along the instantaneous equator of $t$ through the arc $\overline{\sigma\mbox{N}}$.  That is, \vcio\ can be constructed by taking the position vector of the node N in the GCRS, ${\bf N}_{_{\rm GCRS}}$, and rotating it counterclockwise by the angle
$-\overline{\sigma\mbox{N}}$ (i.e., clockwise by $\overline{\sigma\mbox{N}}$) about
the axis \vcip.  Equivalently,
\begin{equation}
     \vcio = {\bf N}_{_{\rm GCRS}}\,\cos(\overline{\sigma\mbox{N}}) -  (\vcip \times {\bf N}_{_{\rm GCRS}})
                  \sin(\overline{\sigma\mbox{N}})
\end{equation}
\\
\begin{center} \rule[1ex]{60mm}{0.1mm} \end{center}
\par The three methods for determining the position of the CIO in the GCRS are numerically the
same to within several microarcseconds (\uas) over six centuries centered on J2000.0.  We
now have formulas in hand for obtaining the positions  of the three cardinal points on the
sky --- the CIP, the CIO, and the equinox --- that are involved in the ITRS-to-GCRS
(terrestrial-to-celestial) transformations.   In the following, it is assumed that \vcip, \vcio,
and \veqx\ are known vectors for some time $t$ of interest.

\subsection{Geodetic Position Vectors and Polar Motion}\label{erot.formulas.geod}

Vectors representing the geocentric positions of points on or near the
surface of the Earth are of the general form
\begin{equation}
{\bf r} = \left( \begin{array}{c}
                     (aC + h) \, \cos\phi_G \, \cos\lambda_G  \\
                     (aC + h) \, \cos\phi_G \, \sin\lambda_G  \\
                     (aS + h) \, \sin\phi_G \, \w{\sin\lambda_G}
               \end{array} \right)
\end{equation}
where $\lambda_G$ is the geodetic longitude, $\phi_G$ is the
geodetic latitude, and $h$ is the height.  These coordinates are measured
with respect to a reference ellipsoid, fit to
the equipotential surface that effectively defines mean sea level.   The ellipsoid
has an equatorial radius of $a$ and a flattening factor $f$.   The quantities $C$ and $S$
depend on the flattening:
\begin{equation}
  C = 1/\sqrt{\cos^2\phi_G + (1-f)^2 \sin^2\phi_G} \qquad\qquad\qquad  S = (1-f)^2\,C
\end{equation}
A complete description of geodetic concepts, reference ellipsoids,
and computations is beyond the scope of this \pubname, but a brief
summary can be found in Section~K of \AsA\ and a more thorough
account is given in Chapter~4 of the \citet{seid92}.  More information
can be found in any introductory textbook on geodesy. Suffice it
here to say that the reference ellipsoid for GPS is WGS~84, with
$a = 6378137$~m and $f = 1/298.257223563$.  For astronomical
purposes it can be assumed that WGS~84 is a good approximation to
the International Terrestrial Reference System (ITRS) described
previously in this chapter.  That is, GPS provides a realization
of the ITRS.

It is worth noting that modern space techniques often measure geocentric positions
in rectangular coordinates directly, without using a reference
ellipsoid.   Also, not all vectors of interest
represent geographic locations.  Vectors representing instrumental axes, baselines,
and boresights are often of more interest to astronomers and these can
usually be easily expressed in the same geodetic system as the instrument location.
All Earth-fixed vectors, regardless of what they represent, are subject to the same
transformations described below.  The ITRS is the assumed starting point for
these transformations, even though in most cases astronomers will be using
vectors in some system that approximates the ITRS.

For astronomical applications, we must correct ITRS vectors for
polar motion (also called wobble).  In current terminology, this
is a transformation from the ITRS to the Terrestrial Intermediate
Reference System (TIRS; in this \pubname\ it is designated \Et),
and is the first transformation shown in the
flowcharts on page~\pageref{transforms}.  Polar motion is the
small quasi-periodic excursion of the geodetic pole from the pole of rotation,
or, more precisely stated, the excursion of the ITRS z-axis from the CIP.
It is described by the parameters $x_p$ and $y_p$, which are the
coordinates of the CIP in the ITRS, and which generally amount to
a few tenths of an arcsecond.   Daily values of $x$ and $y$, the observed
pole coordinates, are published by the IERS (see, e.g., \citet{IERSBullA}).
These published values should, for the most precise applications
($<$1~mas), be augmented by very small predictable
components to polar motion, with periods $\le$1.2~days.  These
extra components are evaluated and added after interpolation of the published
$x$ and $y$ values --- see section 5.4.2 of the \citet{iers03}.
The sum (total polar motion) is  $x_p$ and $y_p$.   However, most users will be
able to assume $x_p = x$ and $y_p = y$.

The transformation we seek not only must reorient the pole from
the ITRS z-axis
to the CIP, it also must move the origin of longitude very slightly from the ITRS
x-axis to the Terrestrial Intermediate Origin (TIO).  The latter
shift is so tiny that its magnitude can be given by an approximate
formula, linear in time, based on the two main circular components
of polar motion as observed over the last few decades.  That shift
is $s' =  -47\,\uas\,T$, where $T$ is the time (either TT or TDB)
in centuries from J2000.0 \citep{lambert}.  Since 47~\uas\ amounts
to 1.5~mm on the surface of the Earth, this correction is entirely
negligible for most purposes.  Nevertheless, we include it here
for completeness.

The ITRS to \Et\ transformation then is accomplished using the
polar motion (wobble) matrix ${\bf W}(t)$:
\begin{equation}
       {\bf r}_{_{{\rm E}_{\varpi}}} = {\bf W}(t) \; {\bf r}_{_{\rm ITRS}}
\end{equation}
where
\begin{equation}
    {\bf W}(t) = {\bf R}_3(-s')\;{\bf R}_2(x_p)\;{\bf R}_1(y_p)
\end{equation}
If we let
\begin{flushright}
\parbox{2.0in}{
\begin{eqnarray*}
      S_x &=& \sin\,(x_p)             \\
      S_y &=& \sin\,(y_p)             \\
      S_s &=& \sin\,(-s')
\end{eqnarray*} }
\parbox{4.0in}{
\begin{eqnarray}
      C_x &=& \cos\,(x_p)    \nonumber\\
      C_y &=& \cos\,(y_p)                        \\
      C_s &=& \cos\,(-s')       \nonumber
\end{eqnarray} }
\end{flushright}
then the wobble matrix can also be written:
\begin{equation}
    {\bf W}(t) = \left( \begin{array}{ccc}
           C_x  C_s                          &
           S_x  S_y  C_s + C_y  S_s          &
          -S_x  C_y  C_s + S_y  S_s          \\
          -C_x  S_s                          &
          -S_x  S_y  S_s + C_y  C_s          &
           S_x  C_y  S_s + S_y  C_s          \\
           S_x                               &
          -C_x  S_y                          &
           C_x  C_y
                          \end{array} \right)
           \approx \left( \begin{array}{rrr}
            1\w{.}   &
           -s'          &
           -x_p       \\
            s'          &
            1\w{.}   &
            y_p        \\
            x_p       &
           -y_p       &
            1
                        \end{array} \right)
\end{equation}
where the form on the right is a first-order approximation.  Due to
the smallness of the angles involved, the first-order matrix is
quite adequate for most applications --- further, $s'$
can be set to zero.

\subsection{Complete Terrestrial to Celestial Transformation}\label{erot.formulas.tercel}

The transformations corresponding to the two flowcharts on page~\pageref{transforms}
are
\begin{eqnarray}
     \mbox{Equinox-based transformation:} \qquad
     {\bf r}_{_{\rm GCRS}} &=& {\bf B}^{\rm T}\;{\bf P}^{\rm T}\; {\bf N}^{\rm T}\;{\bf R}_3(-\mbox{GAST})\;{\bf W}\;\;{\bf r}_{_{\rm ITRS}} \nonumber\\
     \mbox{CIO-based transformation:} \qquad
     {\bf r}_{_{\rm GCRS}} &=& {\bf C}^{\rm T}\;{\bf R}_3(-\theta)\;{\bf W}\;\;{\bf r}_{_{\rm ITRS}}
\end{eqnarray}
where all of the matrices except ${\bf B}$ are time-dependent.  GAST is Greenwich apparent
sidereal time and $\theta$ is the Earth Rotation Angle; formulas are
given in section \ref{time.formulas.erot} (eqs.~2.10--2.14).  The matrices, working from right
to left, perform the following sub-transformations:
\begin{tabbing}
\w{XXXXXXXXXXXXXXXXXX}\= ${\bf W}$ \w{XXXXXXXXXX} \= ITRS to \Et        \\
                              \>  ${\bf R}_3(-\mbox{GAST})$   \> \Et\ to \Eq                                  \\
                              \>  ${\bf B}^{\rm T}\;{\bf P}^{\rm T}\; {\bf N}^{\rm T}$ \> \Eq\ to GCRS \\
                             \> ${\bf R}_3(-\theta)$ \> \Et\ to \Ec      \\
                             \> ${\bf C}^{\rm T}$ \> \Ec\ to GCRS
\end{tabbing}
The three ``E'' reference systems were described on page~\pageref{eqsys}.

The matrix ${\bf C}$ has not yet been developed in this chapter.  One form of the matrix ${\bf C}$ was introduced in section~\ref{prenut.formulas.alt} for the transformation from the
GCRS to \Ec.  We use the transpose here because we are interested in the opposite
transformation.  ${\bf C}$ or ${\bf C}^{\rm T}$ is easy to construct because we
already have the three basis vectors of \Ec\ expressed in the GCRS: the z-axis is toward
\vcip, the CIP; the x-axis is toward \vcio, the CIO; and the y-axis is toward $\vcip \times \vcio$.  Call the latter vector ${\bf y}_{_{\rm GCRS}}$.  Then:
\begin{equation}
      {\bf C}^{\rm T} = \Big(\;\vcio\;\;{\bf y}_{_{\rm GCRS}}\;\;\vcip\;\Big)
              = \left( \begin{array}{ccc}
                        \sigma_1 & y_1 & n_1 \\
                        \sigma_2 & y_2 & n_2 \\
                        \sigma_3 & y_3 & n_3
                           \end{array} \right)
           = \left( \begin{array}{ccc}
                        \sigma_1 & y_1 & X \\
                        \sigma_2 & y_2 & Y \\
                        \sigma_3 & y_3 & Z
                           \end{array} \right)
 \end{equation}
where $X$, $Y$, and $Z$ are the CIP coordinates, expressed as dimensionless quantities.  As in section~\ref{prenut.formulas.alt}, the matrix can also be constructed using only $X$ and $Y$, together with the CIO locator,~$s$:
\begin{equation}
      {\bf C}^{\rm T}  = \left( \begin{array}{ccc}
                        1-bX^2 & -bXY  &  X \\
                        -bXY & 1 - bY^2 & Y  \\
                          -X & -Y & 1-b(X^2 + Y^2)
               \end{array} \right) \;\; {\bf R}_3(s)
\end{equation}
where $b = 1/(1+Z)$ and $Z = \sqrt{1-X^2-Y^2}$.  The latter form is taken from the \citet{iers03},
Chapter~5, where ${\bf C}^{\rm T}$ is called $Q(t)$.  The two constructions of ${\bf C}^{\rm T}$
are numerically the same.

\subsection{Hour Angle}\label{erot.formulas.ha}

The local hour angle of a celestial object is given by
\begin{eqnarray}
      \mbox{Equinox-based formula:} \qquad
       \mbox{h} &=&  \mbox{GAST} - \alpha_{_\Upsilon} + \lambda   \nonumber\\
       \mbox{CIO-based formula:} \qquad
       \mbox{h} &=&  \theta               - \alpha_{\sigma}      + \lambda
\end{eqnarray}
where GAST is Greenwich apparent sidereal time, $\theta$ is the Earth Rotation Angle,
and $\lambda$ is the longitude of the observer.  The quantities involved can be expressed
in either angle or time units as long as they are consistent.  The
right ascension in the two cases is expressed with respect to different origins:
$\alpha_{_\Upsilon}$ is the apparent right ascension of the object, measured with respect to the true equinox, and $\alpha_{\sigma}$ is the apparent right ascension of the object, measured with respect to the CIO.  That is, the coordinates of the object are expressed in system \Eq\ in the equinox-based
formula and in system \Ec\ in the CIO-based formula.  Since both systems share the same
equator --- the instantaneous equator of date, orthogonal to the CIP --- the apparent declination of
the object is the same in the two cases.

The two formulas in 6.19 are equivalent, which can be seen by substituting, in the equinox-based formula,  GAST = $\theta - \mathcal{E}_o$ and $\alpha_{_\Upsilon}$ = $\alpha_{\sigma} - \mathcal{E}_o$, where $\mathcal{E}_o$ is the equation of the origins.

The longitude of the observer, $\lambda$, is expressed in the \Et\ system,
that is, it is corrected for polar motion.  Using the first-order form of the matrix {\bf W}, given
in eq.~6.15 (and assuming $s'$=0), it is straightforward to derive eq.~2.16 for $\lambda$.
Using notation consistent with that used in this chapter, this equation is
\begin{equation}
            \lambda \equiv
            \lambda_{_{{\rm E}_{\varpi}}}  = \lambda_{_{\rm ITRS}} + \Big( x_p \sin\lambda_{_{\rm ITRS}} +
                                                                             y_p \cos\lambda_{_{\rm ITRS}}\Big)
                                                        \, \tan \phi_{_{\rm ITRS}} \, / 3600
\end{equation}
where $\lambda_{_{\rm ITRS}}$ and $\phi_{_{\rm ITRS}}$ are the ITRS (geodetic) longitude
and latitude of the observer, with $\lambda_{_{\rm ITRS}}$ in degrees; and $x_p$ and $y_p$ are the
coordinates of the pole (CIP), in arcseconds.  This formula is approximate and
should not be used for places at polar latitudes.

The corresponding equation for the latitude, $\phi$, corrected for polar motion is
\begin{equation}
            \phi \equiv
            \phi_{_{{\rm E}_{\varpi}}}  = \phi_{_{\rm ITRS}} + \Big( x_p \cos\lambda_{_{\rm ITRS}} -
                                                                             y_p \sin\lambda_{_{\rm ITRS}}\Big)
                                                                                                    / 3600
\end{equation}
although this equation is not needed in hour angle computations; it is given here only for completeness.

The common notion of hour angle can be expressed more precisely using concepts introduced
in Chapters~\ref{prenut} and \ref{erot}.  The local hour angle of an object is the angle between two planes:  the plane
containing the geocenter, the CIP, and the observer; and the plane containing the geocenter,
the CIP, and the object.  Hour angle increases with time and is positive when the object is west
of the observer as viewed from the geocenter.  The two planes define meridians on the celestial
sphere that meet at the CIP.  From the point of view of the observer, the CIP is not, in general,
exactly at the geodetic north point, which is the direction toward the ITRS z-axis.  The azimuths of
the two directions differ by as much as $\sqrt{x_p^2+y_p^2}/\cos\phi_{_{\rm ITRS}}$, depending on
time of day.  This difference is small (usually $<$1~arcsecond) and often negligible for practical
applications.  The plane defining the astronomical Greenwich meridian (from which
Greenwich hour angles are measured) can be understood to contain the geocenter, the CIP,
and TIO;  there,  $\lambda \equiv \lambda_{_{{\rm E}_{\varpi}}}=0$.  This plane is now called the
TIO meridian.

The CIO-based formula for hour angle is quite simple to use (since $\theta$ is linear with time)
if we have the coordinates of the object
expressed in system \Ec.  Fortunately, this is straightforward if we have the object's coordinates
expressed in the GCRS, because we also have the basis vectors of \Ec\ expressed in the GCRS.  If the object's vector in the GCRS is ${\bf r}_{_{\rm GCRS}}$, then the object's vector in \Ec\ is simply
\begin{equation}
        {\bf r}_{_{{\rm E}_{\sigma}}} = {\bf C} \, {\bf r}_{_{\rm GCRS}} =
     \left( \begin{array}{c}
                {\bf r}_{_{\rm GCRS}} \cdot \vcio                              \\
                {\bf r}_{_{\rm GCRS}} \cdot {\bf y}_{_{\rm GCRS}}  \\
                {\bf r}_{_{\rm GCRS}} \cdot \vcip
                   \end{array}\right)
            \qquad \mbox{where} \quad {\bf y}_{_{\rm GCRS}} = \vcip \times \vcio
\end{equation}
Then
\begin{equation}
          \alpha_{\sigma} = \arctan \left(  \frac{{\bf r}_{_{\rm GCRS}} \cdot {\bf y}_{_{\rm GCRS}}}
                                                                            {{\bf r}_{_{\rm GCRS}} \cdot \vcio} \right)
\end{equation}
As a specific case, we know the position vector of the equinox, \veqx.  Applying eq.~6.23 to \veqx\ and
using it in the second formula of 6.19, with $\lambda$=0, we obtain the hour angle of the equinox
at the Greenwich (or TIO) meridian.  But this is the definition of Greenwich Apparent Sidereal Time.  Therefore,
\begin{equation}
       \mbox{GAST} =  \theta - \arctan \left(  \frac{\veqx \cdot {\bf y}_{_{\rm GCRS}}}
                                                                                 {\veqx \cdot \vcio} \right)
\end{equation}
Evidently, then,
\begin{equation}
         \mathcal{E}_o =  \arctan \left(  \frac{\veqx \cdot {\bf y}_{_{\rm GCRS}}}
                                                                      {\veqx \cdot \vcio} \right)
\end{equation}
which merely restates the definition of $\mathcal{E}_o$ --- the equatorial angle from the CIO to the
equinox, i.e., the right ascension of the equinox in system \Ec.

\backmatter
\chapter*{Text of IAU Resolutions of 1997\\
{\Large\bf Adopted at the XXIIIrd General Assembly, Kyoto}}
\markboth{IAU RESOLUTIONS 1997}{IAU RESOLUTIONS 1997}
\addcontentsline{toc}{chapter}{Text of IAU Resolutions of 1997}

\setlength{\parskip}{2ex}
\setlength{\parindent}{0em}

{\Large Resolution B2 \quad On the International Celestial Reference System (ICRS)}

The XXIIIrd International Astronomical Union General Assembly

Considering

(a) That Recommendation VII of Resolution A4 of the 21st General Assembly
specifies the coordinate system for the new celestial reference frame and, in
particular, its continuity with the FK5 system at J2000.0;

(b) That Resolution B5 of the 22nd General Assembly specifies a list of
extragalactic sources for consideration as candidates for the realization of the
new celestial reference frame;

(c) That the IAU Working Group on Reference Frames has in 1995 finalized the
positions of these candidate extragalactic sources in a coordinate frame aligned
to that of the FK5 to within the tolerance of the errors in the latter (see note
1);

(d) That the Hipparcos Catalogue was finalized in 1996 and that its coordinate
frame is aligned to that of the frame of the extragalactic sources in (c) with
one sigma uncertainties of $\pm$ 0.6~milliarcseconds (mas) at epoch J1991.25 and
$\pm$ 0.25~mas per year in rotation rate;

Noting

That all the conditions in the IAU Resolutions have now been met;

Resolves

(a) That, as from 1 January 1998, the IAU celestial reference system shall be
the International Celestial Reference System (ICRS) as specified in the 1991 IAU
Resolution on reference frames and as defined by the International Earth
Rotation Service (IERS) (see note 2);

(b) That the corresponding fundamental reference frame shall be the
International Celestial Reference Frame (ICRF) constructed by the IAU Working
Group on Reference Frames;

(c) That the Hipparcos Catalogue shall be the primary realization of the ICRS at
optical wavelengths;

(d) That IERS should take appropriate measures, in conjunction with the IAU
Working Group on reference frames, to maintain the ICRF and its ties to the
reference frames at other wavelengths.

Note 1: IERS 1995 Report, Observatoire de Paris, p. II-19 (1996).

Note 2: ``The extragalactic reference system of the International Earth Rotation
Service (ICRS)'', Arias, E.F. et al. A \& A 303, 604 (1995).

\bigskip\hrule\bigskip

{\Large Resolution B4 \quad On Non-Rigid Earth Nutation Theory}

The XXIIIrd International Astronomical Union General Assembly

Recognizing

that the International Astronomical Union and the International Union of Geodesy
and Geophysics Working Group (IAU-IUGG WG) on Non-rigid Earth Nutation Theory
has met its goal by identifying the remaining geophysical and astronomical
phenomena that must be modeled before an accurate theory of nutation for a
non-rigid Earth can be adopted, and

that, as instructed by IAU Recommendation C1 in 1994, the International Earth
Rotation Service (IERS) has published in the IERS Conventions (1996) an interim
precession-nutation model that matches the observations with an uncertainty of
$\pm$ 1~milliarcsecond (mas),

endorses

the conclusions of the IAU-IUGG WG on Non-rigid Earth Nutation Theory given in
the appendix,

requests

the IAU-IUGG WG on Non-rigid Earth Nutation Theory to present a detailed report
to the next IUGG General Assembly (August 1999), at which time the WG will be
discontinued,

and urges

the scientific community to address the following questions in the future:

- completion of a new rigid Earth nutation series with the additional terms
necessary for the theory to be complete to within $\pm$ 5~microarcseconds, and

- completion of a new non-rigid Earth transfer function for an Earth initially
in non-hydrostatic equilibrium, incorporating mantle inelasticity and a Free
Core Nutation period in agreement with the observations, and taking into account
better modeling of the fluid parts of the planet, including dissipation.

APPENDIX

The WG on Non-rigid Earth Nutation Theory has quantified the problems in the
nutation series adopted by the IAU in 1980 by noting:

(1) that there is a difference in the precession rate of about $-3.0$~milliarcseconds
per year (mas/year) between the value observed by Very Long
Baseline Interferometry (VLBI) and Lunar Laser Ranging (LLR) and the adopted
value,

(2) that the obliquity has been observed (by VLBI and LLR) to change at a rate of
about $-0.24$~mas/year, although there is no such change implied by the 1980
precession-nutation theory,

(3) that, in addition to these trends, there are observable peak-to-peak
differences of up to 20 milliarcseconds (mas) between the nutation observed by
VLBI and LLR and the nutation adopted by the IAU in 1980,

(4) that these differences correspond to spectral amplitudes of up to several
mas, and

(5) that the differences between observation and theory are well beyond the
present observational accuracy.

The WG has recognized the improvements made in the modeling of these quantities,
and recommends, in order to derive a more precise nutation model, at the mas
level in spectral amplitudes and at a few mas level in the peak to peak
analysis, the use of models:

(1) based on a new non-rigid Earth transfer function for an Earth initially in
non-hydrostatic equilibrium, incorporating mantle inelasticity, a
core-mantle-boundary flattening giving a Free Core Nutation (FCN) period in
agreement with the observed value, and a global Earth dynamical flattening in
agreement with the observed precession, and

(2) based on a new rigid Earth nutation series which takes into account the
following perturbing effects:
\begin{quotation}
\noindent1. in lunisolar ephemerides: indirect planetary effects, lunar inequality,
J2-tilt, planetary-tilt, secular variations of the amplitudes, effects of
precession and nutation,\\[0.7ex]
2. in the perturbing bodies to be considered: in addition to the Moon and the
Sun, the direct planetary effects of Venus, Jupiter, Mars, and Saturn, should be
included,\\[0.7ex]
3. in the order of the external potential to be considered: J3 and J4 effects
for the Moon, and\\[0.7ex]
4. in the theory itself: effects of the tri-axiality of the Earth, relativistic
effects and second order effects.
\end{quotation}

The WG recognizes that this new generation of models still has some
imperfections, the principal one being poor modeling of the dissipation in the
core and of certain effects of the ocean and the atmosphere, and urges the
scientific community to address these questions in the future.

The WG recognizes that, due to the remaining imperfections of the present
theoretical nutation models, the nutation series published in the IERS
Conventions (1996), following 1994 IAU recommendation C1, still provides the
users with the best nutation series. This IERS model being based on observations
of the celestial pole offset, the WG supports the recommendation that the
scientific community continue VLBI and LLR observations to provide accurate
estimations of nutation, precession and rate of change in obliquity.

\chapter*{Text of IAU Resolutions of 2000\\
{\Large\bf Adopted at the XXIVth General Assembly, Manchester}}
\markboth{IAU RESOLUTIONS 2000}{IAU RESOLUTIONS 2000}
\addcontentsline{toc}{chapter}{Text of IAU Resolutions of 2000}

\setlength{\parskip}{2ex}
\setlength{\parindent}{0em}

{\Large Resolution B1.1 \quad Maintenance and Establishment of
Reference Frames and Systems}

The XXIVth International Astronomical Union

Noting

{1.} that Resolution B2 of the XXIIIrd General Assembly (1997)
specifies that ``the fundamental reference frame shall be the
International Celestial Reference Frame (ICRF) constructed by the
IAU Working Group on Reference Frames,"

{2.} that Resolution B2 of the XXIIIrd General Assembly (1997)
specifies ``That the Hipparcos Catalogue shall be the primary
realization of the ICRS at optical wavelengths", and

{3.} the need for accurate definition of reference systems brought
about by unprecedented precision, and

Recognizing

{1.} the importance of continuing operational observations made
with Very Long Baseline Interferometry (VLBI) to maintain the
ICRF,

{2.} the importance of VLBI observations to the operational
determination of the parameters needed to specify the
time-variable transformation between the International Celestial
and Terrestrial Reference Frames,

{3.} the progressive shift between the Hipparcos frame and the
ICRF, and

{4.} the need to maintain the optical realization as close as
possible to the ICRF

Recommends

{1.} that IAU Division I maintain the Working Group on Celestial
Reference Systems formed from Division I members to consult with
the International Earth Rotation Service (IERS) regarding the
maintenance of the ICRS,

{2.} that the IAU recognize the International VLBI service (IVS)
for Geodesy and Astrometry as an IAU Service Organization,

{3.} that an official representative of the IVS be invited to
participate in the IAU Working Group on Celestial Reference
Systems,

{4.} that the IAU continue to provide an official representative
to the IVS Directing Board,

{5.} that the astrometric and geodetic VLBI observing programs
consider the requirements for maintenance of the ICRF and linking
to the Hipparcos optical frame in the selection of sources to be
observed (with emphasis on the Southern Hemisphere), design of
observing networks, and the distribution of data, and

{6.} that the scientific community continue with high priority
ground- and space-based observations (a) for the maintenance of
the optical Hipparcos frame and frames at other wavelengths and
(b) for the links of the frames to the ICRF.

\bigskip\hrule\bigskip
{\Large Resolution B1.2 \quad Hipparcos Celestial Reference Frame}

The XXIVth International Astronomical Union

Noting

{1.} that Resolution B2 of the XXIIIrd General Assembly (1997)
specifies, ``That the Hipparcos Catalogue shall be the primary
realization of the International Celestial Reference System (ICRS)
at optical wavelengths,"

{2.} the need for this realization to be of the highest precision,

{3.} that the proper motions of many of the Hipparcos stars known,
or suspected, to be multiple are adversely affected by uncorrected
orbital motion,

{4.} the extensive use of the Hipparcos Catalogue as reference for
the ICRS in extension to fainter stars,

{5.} the need to avoid
confusion between the International Celestial Reference Frame
(ICRF) and the Hipparcos frame, and

{6.} the progressive shift
between the Hipparcos frame and the ICRF,

Recommends

{1.} that Resolution B2 of the XXIIIrd IAU General Assembly (1997)
be amended by excluding from the optical realization of the ICRS
all stars flagged C, G, O, V and X in the Hipparcos Catalogue, and

{2.} that this modified Hipparcos frame be labeled the Hipparcos
Celestial Reference Frame (HCRF).

\bigskip\hrule\bigskip
\pagebreak
{\Large Resolution B1.3 \quad Definition of Barycentric Celestial
Reference System and Geocentric Celestial Reference System}

The XXIVth International Astronomical Union

Considering

{1.} that the Resolution A4 of the XXIst General Assembly (1991)
has defined a system of space-time coordinates for (a) the solar
system (now called the Barycentric Celestial Reference System,
(BCRS)) and (b) the Earth (now called the Geocentric Celestial
Reference System (GCRS)), within the framework of General
Relativity,

{2.} the desire to write the metric tensors both in the BCRS and
in the GCRS in a compact and self-consistent form, and

{3.} the fact that considerable work in General Relativity has
been done using the harmonic gauge that was found to be a useful
and simplifying gauge for many kinds of applications,

Recommends

{1.} the choice of harmonic coordinates both for the barycentric
and for the geocentric reference systems.

{2.} writing the time-time component and the space-space component
of the barycentric metric g$_{\mu\nu}$ with barycentric
coordinates (t, {\bf x}) (t = Barycentric Coordinate Time (TCB))
with a single scalar potential w(t,{\bf x}) that generalizes the
Newtonian potential, and the space-time component with a vector
potential w$^i$(t, {\bf x}); as a boundary condition it is assumed
that these two potentials vanish far from the solar system,

explicitly,

\indent\indent $g_{00} = -1 + {2w \over c^2} - {2w^2 \over c^4}$,

\indent\indent $g_{0i} = -{4 \over c^3} w^i$,

\indent\indent $g_{ij} = \delta_{ij}\biggl(1 + {2 \over c^2}w \biggr)$,

with

\indent\indent$w(t,{\bf x}) = G \int d^3 x' {\sigma(t,{\bf x'}) \over |{\bf x}
- {\bf x'}|} + {1 \over 2c^2} G {\partial^2 \over \partial t^2} \int d^3 x'
\sigma(t,{\bf x'}) |{\bf x} - {\bf x'}|$

\indent\indent$w^i(t,{\bf x}) = G \int d^3 x' {\sigma^i(t,{\bf x'}) \over
|{\bf x} - {\bf x'}|}$.

Here, $\sigma$ and $\sigma^i$ are the gravitational mass and current densities,
respectively.

{3.} writing the geocentric metric tensor G$_{\alpha\beta}$ with
geocentric coordinates (T, {\bf X}) (T= Geocentric Coordinate Time
(TCG)) in the same form as the barycentric one but with potentials
W(T, {\bf X}) and W$^a$(T, {\bf X}); these geocentric potentials
should be split into two parts
--- potentials W and W$^a$ arising from the gravitational action of the
Earth and external parts W$_{ext}$ and W$^a_{ext}$ due to tidal and inertial
effects; the external parts of the metric potentials are assumed to vanish at
the geocenter and admit an expansion into positive powers of {\bf X},

explicitly,

\indent\indent$G_{00} = -1 + {2W \over c^2} - {2W^2 \over c^4}$,

\indent\indent$G_{0a} = -{4 \over c^3} W^a$,

\indent\indent$G_{ab} = \delta_{ab}\biggl(1 + {2 \over c^2}W \biggr)$.

The potentials W and W$^a$ should be split according to

\indent\indent$W(T,{\bf X}) = W_E(T,{\bf X}) + W_{ext}(T,{\bf X})$,

\indent\indent$W^a(T,{\bf X}) = W_E^a(T,{\bf X}) + W_{ext}^a(T,{\bf X})$.

The Earth's potentials W$_E$ and W$_E^a$ are defined in the same
way as w and w$^i$ but with quantities calculated in the GCRS with
integrals taken over the whole Earth.

{4.} using, if accuracy requires, the full post-Newtonian
coordinate transformation between the BCRS and the GCRS as induced
by the form of the corresponding metric tensors,

explicitly, for the kinematically non-rotating GCRS (T=TCG, t=TCB,
$r_E^i \equiv x^i - x^i_E(t)$ and a summation from 1 to 3 over
equal indices is implied),

\indent\indent$T = t - {1 \over c^2} \bigl[ A(t) + v_E^i r_E^i\bigr] + {1
\over c^4} \bigl[B(t) + B^i(t)r_E^i + B^{ij}(t)r_E^i r_E^j + C(t,{\bf x})
\bigr ] + O(c^{-5})$,

\indent\indent$X^a = \delta_{ai} \biggl[ r_E^i + {1 \over c^2} \biggl({1 \over
2} v_E^i v_E^j r_E^j + w_{ext}({\bf x_E})r_E^i + r_E^i a_E^j r_E^j - {1 \over
2} a_E^i r_E^2 \biggr) \biggr] + O(c^{-4})$,

where

\indent\indent${d \over dt}A(t) = {1 \over 2} v_E^2 + w_{ext}({\bf x_E})$,

\indent\indent${d \over dt}B(t) = -{1 \over 8} v_E^4 - {3 \over 2} v_E^2
w_{ext}({\bf x_E}) + 4v_E^iw_{ext}^i({\bf x_E}) + {1 \over 2}w_{ext}^2
({\bf x_E})$,

\indent\indent$B^i(t) = -{1 \over 2}v_E^2 v_E^i + 4 w_{ext}^i({\bf x_E}) -
3v_E^iw_{ext} ({\bf x_E})$,

\indent\indent$B^{ij}(t) = -v_E^i\delta_{aj}Q^a + 2{\partial \over \partial
x^j}w_{ext}^i ({\bf x_E}) - v_E^i{\partial \over \partial x^j} w_{ext}
({\bf x_E}) +  {1 \over 2}\delta^{ij}\dot{w}_{ext}({\bf x_E})$,

\indent\indent$C(t,{\bf x}) = -{1 \over 10}r_E^2(\dot{a}_E^i r_E^i)$.

{}Here $x_E^i$, $v_E^i$, and $a_E^i$ are the barycentric position,
velocity and acceleration vectors of the Earth, the dot stands for
the total derivative with respect to t, and

\indent\indent$Q^a = \delta_{ai} \biggl[{\partial \over \partial x_i} w_{ext}
({\bf x_E}) - a_E^i\biggr ]$.

The external potentials, $w_{ext}$ and $w_{ext}^i$, are given by

\indent\indent$w_{ext}=\sum_{A\not= E} w_A, \quad w_{ext}^i = \sum_{A\not= E}
w_A^i$,

where E stands for the Earth and $w_A$ and $w_A^i$ are determined
by the expressions for w and w$^i$ with integrals taken over body
A only.

Notes

It is to be understood that these expressions for w and w$^i$ give
g$_{00}$ correct up to O(c$^{-5}$), g$_{0i}$ up to O(c$^{-5}$),
and g$_{ij}$ up to O(c$^{-4}$).  The densities $\sigma$ and
$\sigma^i$ are determined by the components of the energy momentum
tensor of the matter composing the solar system bodies as given in
the references. Accuracies for G$_{\alpha\beta}$ in terms of
c$^{-n}$ correspond to those of g$_{\mu\nu}$.

The external potentials W$_{ext}$ and W$_{ext}^a$ can be written
in the form

$W_{ext} = W_{tidal} + W_{iner}$,

$W_{ext}^a = W_{tidal}^a + W_{iner}^a$.

W$_{tidal}$ generalizes the Newtonian expression for the tidal
potential.  Post-Newtonian expressions for W$_{tidal}$ and
W$_{tidal}^a$ can be found in the references. The potentials
W$_{iner}$, W$_{iner}^a$ are inertial contributions that are
linear in X$^a$. The former is determined mainly by the coupling
of the Earth's nonsphericity to the external potential. In the
kinematically non-rotating Geocentric Celestial Reference System,
W$_{iner}^a$ describes the Coriolis force induced mainly by
geodetic precession.

Finally, the local gravitational potentials W$_E$ and W$_E^a$ of
the Earth are related to the barycentric gravitational potentials
w$_E$ and w$_E^i$ by

$W_E(T,{\bf X}) = w_e(t,{\bf x})\biggl(1 + {2 \over c^2}v_E^2 \biggr) -
{4 \over c^2} v_E^i w_E^i(t,{\bf x}) + O(c^{-4})$,

$W_E^a(T,{\bf X}) = \delta_{ai}(w_E^i(t,{\bf x}) - v_E^i w_E(t,{\bf x}))
+ O(c^{-2})$.

References

Brumberg, V. A., Kopeikin, S. M., 1989, {\it Nuovo Cimento}, {\bf
B103}, 63.

Brumberg, V. A., 1991, {\it Essential Relativistic Celestial
Mechanics}, Hilger, Bristol.

Damour, T., Soffel, M., Xu, C., {\it Phys. Rev. D}, {\bf 43}, 3273
(1991); {\bf 45}, 1017 (1992); {\bf 47}, 3124 (1993); {\bf 49},
618 (1994).

Klioner, S. A., Voinov, A. V., 1993, {\it Phys Rev. D}, {\bf 48},
1451.

Kopeikin, S. M., 1988, {\it Celest. Mech.}, {\bf 44}, 87.

\bigskip\hrule\bigskip
{\Large Resolution B1.4 \quad Post-Newtonian Potential
Coefficients}

The XXIVth International Astronomical Union

Considering

{1.} that for many applications in the fields of celestial
mechanics and astrometry a suitable parametrization of the metric
potentials (or multipole moments) outside the massive solar-system
bodies in the form of expansions in terms of potential
coefficients are extremely useful, and

{2.} that physically meaningful post-Newtonian potential
coefficients can be derived from the literature,

Recommends

{1.} expansion of the post-Newtonian potential of the Earth in the
Geocentric Celestial Reference System (GCRS) outside the Earth in
the form

\indent\indent$W_E(T,{\bf X}) = {GM_E \over R} \biggl[1 + \sum_{l=2}^\infty
\sum_{m=0}^{+l} \bigl({R_E \over R} \bigr )^l P_{lm}(\cos\theta)
\bigl(C_{lm}^E(T)\cos m\phi + S_{lm}^E(T)\sin m\phi\bigr ) \biggr]$,

where $C_{lm}^E$ and $S_{lm}^E$ are, to sufficient accuracy,
equivalent to the post-Newtonian multipole moments introduced in
(Damour {\it et al.}, {\it Phys. Rev. D}, {\bf 43}, 3273, 1991),
$\theta$ and $\phi$ are the polar angles corresponding to the
spatial coordinates X$^a$ of the GCRS and $R=|X|$, and

{2.} expression of the vector potential outside the Earth, leading to
the well-known Lense-Thirring effect, in terms of the Earth's
total angular momentum vector ${\bf S_E}$ in the form

\indent\indent$W_E^a(T,{\bf X}) = -{G \over 2}{({\bf X}\times {\bf S_E})^a
\over R^3}$.

\bigskip\hrule\bigskip
{\Large Resolution B1.5 \quad Extended relativistic framework for
time transformations and realization of coordinate times in the
solar system}

The XXIVth International Astronomical Union

Considering

{1.} that the Resolution A4 of the XXIst General Assembly (1991)
has defined systems of space-time coordinates for the solar system
(Barycentric Reference System) and for the Earth (Geocentric
Reference System), within the framework of General Relativity,

{2.} that Resolution B1.3 entitled ``Definition of Barycentric
Celestial Reference System and Geocentric Celestial Reference
System" has renamed these systems the Barycentric Celestial
Reference System (BCRS) and the Geocentric Celestial Reference
System (GCRS), respectively, and has specified a general framework
for expressing their metric tensor and defining coordinate
transformations at the first post-Newtonian level,

{3.} that, based on the anticipated performance of atomic clocks,
future time and frequency measurements will require practical
application of this framework in the BCRS, and

{4.} that theoretical work requiring such expansions has already
been performed,

Recommends

that for applications that concern time transformations and
realization of coordinate times within the solar system,
Resolution B1.3 be applied as follows:

{1.} the metric tensor be expressed as

\indent\indent$g_{00} = -\biggl ( 1 - {2 \over c^2} (w_0(t,{\bf x}) +
w_L(t,{\bf x})) + {2 \over c^4}(w_0^2(t,{\bf x}) + \Delta(t,{\bf x}))\biggr )$,

\indent\indent$g_{0i} = -{4 \over c^3} w^i(t,{\bf x})$,

\indent\indent$g_{ij} = \biggl ( 1 + {2w_0(t,{\bf x}) \over c^2} \biggr )
\delta_{ij}$,

where (t $\equiv$ Barycentric Coordinate Time (TCB),{\bf x}) are
the barycentric coordinates, $w_0=G\sum_A M_A/r_A$ with the
summation carried out over all solar system bodies A, ${\bf r}_A =
{\bf x} - {\bf x}_A, {\bf x}_A$ are the coordinates of the center
of mass of body A, r$_A$ = $|{\bf r_A}|$, and where $w_L$ contains
the expansion in terms of multipole moments [see their definition
in the Resolution B1.4 entitled ``Post-Newtonian Potential
Coefficients"] required for each body.  The vector potential
$w^i(t,{\bf x})=\sum_A w_A^i(t,{\bf x})$ and the function
$\Delta(t,{\bf x}) = \sum_A \Delta_A (t,{\bf x})$ are given in
note 2. 

{2.} the relation between TCB and Geocentric Coordinate
Time (TCG) can be expressed to sufficient accuracy by

\indent\indent$TCB-TCG = c^{-2}\Biggl[\int_{t_0}^t\biggl({v_E^2
\over 2} + w_{0 ext}({\bf x}_E)\biggr ) dt + v_E^i r_E^i\Biggr ]$
\indent$-c^{-4}\Biggl[\int_{t_0}^t\biggl(-{1 \over 8} v_E^4 - {3
\over 2} v_E^2 w_{0 ext}({\bf x}_E) + 4v_E^iw_{ext}^i({\bf x}_E) +
{1 \over 2}w_{0 ext}^2({\bf x}_E)\biggr)dt -\biggl(3w_{0 ext}({\bf
x}_E) + {v_E^2 \over 2}\biggr)v_E^i r_E^i \Biggr ]$,

where $v_E$ is the barycentric velocity of the Earth and where the
index ext refers to summation over all bodies except the Earth.

Notes

{1.} This formulation will provide an uncertainty not larger than
$5 \times 10^{-18}$ in rate and, for quasi-periodic terms, not
larger than $5 \times 10^{-18}$ in rate amplitude and 0.2 ps in
phase amplitude, for locations farther than a few solar radii from
the Sun. The same uncertainty also applies to the transformation
between TCB and TCG for locations within 50000 km of the Earth.
Uncertainties in the values of astronomical quantities may induce
larger errors in the formulas.

{2.} Within the above mentioned uncertainties, it is sufficient to
express the vector potential $w_A^i(t,{\bf x})$ of body A as

$w_A^i(t,{\bf x}) = G \Biggl[{-({\bf r}_A \times {\bf S}_A)^i \over 2r_A^3} +
{M_A v_A^i \over r_A} \Biggr ]$,

where ${\bf S}_A$ is the total angular momentum of body A and
$v_A^i$ is the barycentric coordinate velocity of body A. As for
the function $\Delta_A(t,{\bf x})$ it is sufficient to express it
as

$\Delta_A(t,{\bf x}) = {GM_A \over r_A} \Biggl [-2v_a^2 + \sum_{B\not=A}{GM_B
\over r_{BA}} + {1 \over 2}\biggl({(r_A^kv_A^k)^2 \over r_A^2} + r_A^k a_A^k
\biggr ) \Biggr ] + {2Gv_A^k({\bf r}_A \times {\bf S}_A)^k \over r_A^3}$,

where $r_{BA}=|{\bf x}_B-{\bf x}_A|$ and $a_A^k$ is the
barycentric coordinate acceleration of body A. In these formulas,
the terms in ${\bf S}_A$ are needed only for Jupiter ($S \approx
6.9 \times 10^{38} m^2s^{-1}kg$) and Saturn ($S \approx 1.4 \times
10^{38} m^2s^{-1}kg$), in the immediate vicinity of these planets.

{3.} Because the present recommendation provides an extension of
the IAU 1991 recommendations valid at the full first
post-Newtonian level, the constants L$_C$ and L$_B$ that were
introduced in the IAU 1991 recommendations should be defined as
$<TCG/TCB>$ = 1 - L$_C$ and $<TT/TCB>$ = 1 - L$_B$, where TT
refers to Terrestrial Time and $<>$ refers to a sufficiently long
average taken at the geocenter. The most recent estimate of L$_C$
is (Irwin, A. and Fukushima, T., {\it Astron. Astroph.}, {\bf
348}, 642--652, 1999)

$L_C = 1.48082686741 \times 10^{-8} \pm 2 \times 10^{-17}$.

From Resolution B1.9 on ``Redefinition of Terrestrial Time TT",
one infers $L_B = 1.55051976772  \times 10^{-8} \pm 2 \times
10^{-17}$ by using the relation $1-L_B=(1-L_C)(1-L_G)$. $L_G$ is
defined in Resolution B1.9.

Because no unambiguous definition may be provided for $L_B$ and
$L_C$, these constants should not be used in formulating time
transformations when it would require knowing their value with an
uncertainty of order $1 \times 10^{-16}$ or less.

{4.} If TCB$-$TCG is computed using planetary ephemerides which
are expressed in terms of a time argument (noted T$_{eph}$) which
is close to Barycentric Dynamical Time (TDB), rather than in terms
of TCB, the first integral in Recommendation 2 above may be
computed as

$\int_{t_0}^t \biggl({v_E^2 \over 2} + w_{0 ext}({\bf x}_E) \biggr) dt =
\Biggl[ \int_{T_{eph_0}}^{T_{eph}} \biggl({v_E^2 \over 2} + w_{0 ext}
({\bf x}_E) \biggr ) dt \Biggr ] / (1-L_B)$.

\bigskip\hrule\bigskip
{\Large Resolution B1.6 \quad IAU 2000 Precession-Nutation Model}

The XXIVth International Astronomical Union

Recognizing

{1.} that the International Astronomical Union and the
International Union of Geodesy and Geophysics Working Group
(IAU-IUGG WG) on `Non-rigid Earth Nutation Theory' has met its
goals by

\begin{quotation}
{a.} establishing new high precision rigid Earth nutation series,
such as (1) SMART97 of Bretagnon {\it et al.}, 1998, {\it Astron.
Astroph.}, {\bf 329}, 329--338; (2) REN2000 of Souchay {\it et
al.}, 1999, {\it Astron. Astroph. Supl. Ser.}, {\bf 135},
111--131; (3) RDAN97 of Roosbeek and Dehant 1999, {\it Celest.
Mech.}, {\bf 70}, 215--253;

{b.} completing the comparison of new non-rigid Earth transfer
functions for an Earth initially in non-hydrostatic equilibrium,
incorporating mantle anelasticity and a Free Core Nutation period
in agreement with observations,

{c.} noting that numerical integration models are not yet ready to
incorporate dissipation in the core,

{d.} noting the effects of other geophysical and astronomical
phenomena that must be modelled, such as ocean and atmospheric
tides, that need further development;
\end{quotation}

{2.} that, as instructed by IAU Recommendation C1 in 1994, the
International Earth Rotation Service (IERS) will publish in the
IERS Conventions (2000) a precession-nutation model that matches
the observations with a weighted rms of 0.2 milliarcsecond (mas);

{3.} that semi-analytical geophysical theories of forced nutation
are available which incorporate some or all of the following ---
anelasticity and electromagnetic couplings at the core-mantle and
inner core-outer core boundaries, annual atmospheric tide,
geodesic nutation, and ocean tide effects;

{4.} that ocean tide corrections are necessary at all nutation
frequencies; and

{5.} that empirical models based on a resonance formula without
further corrections do also exist;

Accepts

the conclusions of the IAU-IUGG WG on Non-rigid Earth Nutation
Theory published by Dehant {\it et al.}, 1999, {\it Celest. Mech.}
{\bf 72(4)}, 245--310 and the recent comparisons between the
various possibilities, and

Recommends

that, beginning on 1 January 2003, the IAU 1976 Precession Model
and IAU 1980 Theory of Nutation, be replaced by the
precession-nutation model IAU 2000A (MHB2000, based on the
transfer functions of Mathews, Herring and Buffett, 2000 ---
submitted to the {\it Journal of Geophysical Research}) for those
who need a model at the 0.2~mas level, or its shorter version IAU
2000B for those who need a model only at the 1 mas level, together
with their associated precession and obliquity rates, and their
associated celestial pole offsets at J2000.0, to be published in the IERS
Conventions 2000, and

Encourages

{1.} the continuation of theoretical developments of non-rigid
Earth nutation series,

{2.} the continuation of VLBI observations to increase the
accuracy of the nutation series and the nutation model, and to
monitor the unpredictable free core nutation, and 

{3.} the development of new expressions for precession
consistent with the IAU 2000A model.

\bigskip\hrule\bigskip
{\Large Resolution B1.7 \quad Definition of Celestial Intermediate
Pole}

The XXIVth International Astronomical Union

Noting

the need for accurate definition of reference systems brought
about by unprecedented observational precision, and

Recognizing

{1.} the need to specify an axis with respect to which the Earth's
angle of rotation is defined,

{2.} that the Celestial Ephemeris Pole (CEP) does not take account
of diurnal and higher frequency variations in the Earth's
orientation,

Recommends

{1.} that the Celestial Intermediate Pole (CIP) be the pole, the
motion of which is specified in the Geocentric Celestial Reference
System (GCRS, see Resolution B1.3) by motion of the Tisserand mean
axis of the Earth with periods greater than two days,

{2.} that the direction of the CIP at J2000.0 be offset from the
direction of the pole of the GCRS in a manner consistent with the
IAU 2000A (see Resolution B1.6) precession-nutation model,

{3.} that the motion of the CIP in the GCRS be realized by the IAU
2000A model for precession and forced nutation for periods greater
than two days plus additional time-dependent corrections provided
by the International Earth Rotation Service (IERS) through
appropriate astro-geodetic observations,

{4.} that the motion of the CIP in the International Terrestrial
Reference System (ITRS) be provided by the IERS through
appropriate astro-geodetic observations and models including
high-frequency variations,

{5.} that for highest precision, corrections to the models for the
motion of the CIP in the ITRS may be estimated using procedures
specified by the IERS, and

{6.} that implementation of the CIP be on 1 January 2003.

Notes

{1.} The forced nutations with periods less than two days are
included in the model for the motion of the CIP in the ITRS.

{2.} The Tisserand mean axis of the Earth corresponds to the mean
surface geographic axis, quoted B axis, in Seidelmann, 1982, {\it
Celest. Mech.}, {\bf 27}, 79--106.

{3.} As a consequence of this
resolution, the Celestial Ephemeris Pole is no longer necessary.

\bigskip\hrule\bigskip
{\Large Resolution B1.8 \quad Definition and use of Celestial and
Terrestrial Ephemeris Origin}

The XXIVth International Astronomical Union

Recognizing

{1.} the need for reference system definitions suitable for modern
realizations of the conventional reference systems and consistent
with observational precision,

{2.} the need for a rigorous definition of sidereal rotation of
the Earth,

{3.} the desirability of describing the rotation of the Earth
independently from its orbital motion, and

Noting

that the use of the ``non-rotating origin" (Guinot, 1979) on the
moving equator fulfills the above conditions and allows for a
definition of UT1 which is insensitive to changes in models for
precession and nutation at the microarcsecond level,

Recommends

{1.} the use of the ``non-rotating origin" in the Geocentric
Celestial Reference System (GCRS) and that this point be
designated as the Celestial Ephemeris Origin (CEO) on the equator
of the Celestial Intermediate Pole (CIP),

{2.} the use of the ``non-rotating origin" in the International
Terrestrial Reference System (ITRS) and that this point be
designated as the Terrestrial Ephemeris Origin (TEO) on the
equator of the CIP,

{3.} that UT1 be linearly proportional to the Earth Rotation Angle
defined as the angle measured along the equator of the CIP between
the unit vectors directed toward the CEO and the TEO,

{4.} that the transformation between the ITRS and GCRS be
specified by the position of the CIP in the GCRS, the position of
the CIP in the ITRS, and the Earth Rotation Angle,

{5.} that the International Earth Rotation Service (IERS) take
steps to implement this by 1 January 2003, and

{6.} that the IERS will continue to provide users with data and
algorithms for the conventional transformations.

Note

{1.} The position of the CEO can be computed from the IAU 2000A
model for precession and nutation of the CIP and from the current
values of the offset of the CIP from the pole of the ICRF at
J2000.0 using the development provided by Capitaine {\it et al.}
(2000).

{2.} The position of the TEO is only slightly dependent on polar
motion and can be extrapolated as done by Capitaine {\it et al.}
(2000) using the IERS data.

{3.} The linear relationship between the Earth's rotation angle
$\theta$ and UT1 should ensure the continuity in phase and rate of
UT1 with the value obtained by the conventional relationship
between Greenwich Mean Sidereal Time (GMST) and UT1.  This is
accomplished by the following relationship:

$\theta(UT1)=2\pi(0.7790572732640+1.00273781191135448\times(Julian\ UT1\ date
-2451545.0))$

References

Guinot, B., 1979, in D.D. McCarthy and J.D. Pilkington (eds.),
{\it Time and the Earth's Rotation}, D. Reidel Publ., 7--18.

Capitaine, N., Guinot, B., McCarthy, D.D., 2000, ``Definition of
the Celestial Ephemeris Origin and of UT1 in the International
Celestial Reference Frame", {\it Astron. Astrophys.}, {\bf 355},
398--405.

\bigskip\hrule\bigskip
\pagebreak
{\Large Resolution B1.9 \quad Re-definition of Terrestrial Time
TT}

The XXIVth International Astronomical Union

Considering

{1.} that IAU Resolution A4 (1991) has defined Terrestrial Time
(TT) in its Recommendation 4, and

{2.} that the intricacy and temporal changes inherent to the
definition and realization of the geoid are a source of
uncertainty in the definition and realization of TT, which may
become, in the near future, the dominant source of uncertainty in
realizing TT from atomic clocks,

Recommends

that TT be a time scale differing from TCG by a constant rate:
dTT/dTCG = 1--L$_G$, where L$_G$ = 6.969290134$\times10^{-10}$ is a
defining constant,

Note

L$_G$ was defined by the IAU Resolution A4 (1991) in its
Recommendation 4 as equal to U$_G$/c$^2$ where U$_G$ is the
geopotential at the geoid. L$_G$ is now used as a defining
constant.

\bigskip\hrule\bigskip
{\Large Resolution B2 \quad Coordinated Universal Time}

The XXIVth International Astronomical Union

Recognizing

{1.} that the definition of Coordinated Universal Time (UTC)
relies on the astronomical observation of the UT1 time scale in
order to introduce leap seconds,

{2.} that the unpredictable leap seconds affects modern
communication and navigation systems, 

{3.} that astronomical
observations provide an accurate estimate of the secular
deceleration of the Earth's rate of rotation

Recommends

{1.} that the IAU establish a working group reporting to Division
I at the General Assembly in 2003 to consider the redefinition of
UTC,

{2.} that this study discuss whether there is a requirement for
leap seconds, the possibility of inserting leap seconds at
pre-determined intervals, and the tolerance limits for UT1$-$UTC,
and

{3.} that this study be undertaken in cooperation with the
appropriate groups of the International Union of Radio Science
(URSI), the International Telecommunications Union (ITU-R), the
International Bureau for Weights and Measures (BIPM), the
International Earth Rotation Service (IERS) and relevant
navigational agencies.

\chapter*{IAU 2000A Nutation Series}                                            
\markboth{NUTATION SERIES}{NUTATION SERIES}                                     
\addcontentsline{toc}{chapter}{IAU 2000A Nutation Series}                       
                                                                                
\setlength{\oddsidemargin}{-0.2in}                                              
\setlength{\evensidemargin}{-0.2in} 
\setlength{\parskip}{0in}
\setlength{\parindent}{1.5em}
                                                                                
The nutation series adopted by the IAU, developed by \cite{mathews} (MHB),      
is listed below in its entirety.  It is also available from the IERS            
as a pair of plain-text computer files at \citet{url-nuts},                     
although the arrangement of the columns differs from what is presented here.    
The IERS also provides a Fortran subroutine for evaluating the nutation 
series, written by P.~Wallace, at \citet{url-nutsub}.                               
The NOVAS software package includes this subroutine, and the SOFA               
package contains the same code in a subroutine of a different name.  There are  
also subroutines available that evaluate only a subset of the series terms      
for applications that do not require the highest accuracy.                      
                                                                                
There are 1365 terms in the series.  The term numbers are arbitrary and 
are not involved in the computation.  As listed below, the first 678 are                                              
lunisolar terms and the remaining 687 are planetary terms.  In the              
lunisolar terms, the only fundamental argument multipliers that are             
non-zero are $M_{i,10}$ through $M_{i,14}$, corresponding to the                
arguments $l$, $l'$, $F$, $D$, and $\Omega$, respectively.  In the              
planetary terms, there are no rates of change of the coefficients,              
i.e., $\dot S_i$ and $\dot C_i$ are zero.                                       
                                                                                
The formulas for evaluating the series are given in
section~\ref{prenut.formulas.nut};  
see eqs.~5.15--5.16 and the following text.\\ \\                                                           
\rule{6.8in}{0.1mm}                                                             
\vspace{0.1in}                                                                  
                                                                                
\tiny                                                                           
\setlength{\tabcolsep}{0in}

\begin{tabular}{rrrrrrrrrrrrrrrrrrrrr}
Term & \multicolumn{14}{c}{Fundamental Argument Multipliers $M_{i,j}$}& \multicolumn{3}{c}{$\Delta\psi$ Coefficients}& \multicolumn{3}{c}{$\Delta\epsilon$ Coefficients} \\
$i$ & \qquad$j$= 1 & 2 & 3 & 4 & 5 & 6 & 7 & 8 & 9 & 10 & 11 & 12 & 13 & 14 
& $S_i$\w{999} & $\dot S_i$\w{999} & $C'_i$\w{999} & $C_i$\w{999} & $\dot C_i$\w{999} & $S'_i$\w{999} \\
\w{9999}  & \w{99999} & \w{-999} & \w{-999} & \w{-999} & \w{-999} & \w{-999} & \w{-999} & \w{-999} & \w{-999} & \w{-999} & \w{-999} & \w{-999} & \w{-999} & \w{-999} 
& \w{+9999}$_{''}$\w{999999} & \w{+99}$_{''}$\w{999999} & \w{+99}$_{''}$\w{999999}
& \w{+9999}$_{''}$\w{999999} & \w{+99}$_{''}$\w{999999} & \w{+99}$_{''}$\w{999999} \\

   1 &   0 &   0 &   0 &   0 &   0 &   0 &   0 &   0 &   0 &   0 &   0 &   0 &   0 &   1 &  -17.2064161 &   -0.0174666 &    0.0033386 &    9.2052331 &    0.0009086 &    0.0015377 \\
   2 &   0 &   0 &   0 &   0 &   0 &   0 &   0 &   0 &   0 &   0 &   0 &   2 &  -2 &   2 &   -1.3170906 &   -0.0001675 &   -0.0013696 &    0.5730336 &   -0.0003015 &   -0.0004587 \\
   3 &   0 &   0 &   0 &   0 &   0 &   0 &   0 &   0 &   0 &   0 &   0 &   2 &   0 &   2 &   -0.2276413 &   -0.0000234 &    0.0002796 &    0.0978459 &   -0.0000485 &    0.0001374 \\
   4 &   0 &   0 &   0 &   0 &   0 &   0 &   0 &   0 &   0 &   0 &   0 &   0 &   0 &   2 &    0.2074554 &    0.0000207 &   -0.0000698 &   -0.0897492 &    0.0000470 &   -0.0000291 \\
   5 &   0 &   0 &   0 &   0 &   0 &   0 &   0 &   0 &   0 &   0 &   1 &   0 &   0 &   0 &    0.1475877 &   -0.0003633 &    0.0011817 &    0.0073871 &   -0.0000184 &   -0.0001924 \\
\end{tabular}
\newpage

\begin{tabular}{rrrrrrrrrrrrrrrrrrrrr}
Term & \multicolumn{14}{c}{Fundamental Argument Multipliers $M_{i,j}$}& \multicolumn{3}{c}{$\Delta\psi$ Coefficients}& \multicolumn{3}{c}{$\Delta\epsilon$ Coefficients} \\
$i$ & \qquad$j$= 1 & 2 & 3 & 4 & 5 & 6 & 7 & 8 & 9 & 10 & 11 & 12 & 13 & 14 
& $S_i$\w{999} & $\dot S_i$\w{999} & $C'_i$\w{999} & $C_i$\w{999} & $\dot C_i$\w{999} & $S'_i$\w{999} \\
\w{9999}  & \w{99999} & \w{-999} & \w{-999} & \w{-999} & \w{-999} & \w{-999} & \w{-999} & \w{-999} & \w{-999} & \w{-999} & \w{-999} & \w{-999} & \w{-999} & \w{-999} 
& \w{+9999}$_{''}$\w{999999} & \w{+99}$_{''}$\w{999999} & \w{+99}$_{''}$\w{999999}
& \w{+9999}$_{''}$\w{999999} & \w{+99}$_{''}$\w{999999} & \w{+99}$_{''}$\w{999999} \\

  31 &   0 &   0 &   0 &   0 &   0 &   0 &   0 &   0 &   0 &   0 &   0 &  -2 &   2 &   0 &    0.0021783 &    0.0000000 &    0.0000013 &   -0.0000167 &    0.0000000 &    0.0000013 \\
  32 &   0 &   0 &   0 &   0 &   0 &   0 &   0 &   0 &   0 &   1 &   0 &   0 &  -2 &   1 &   -0.0012873 &   -0.0000010 &   -0.0000037 &    0.0006953 &    0.0000000 &   -0.0000014 \\
  33 &   0 &   0 &   0 &   0 &   0 &   0 &   0 &   0 &   0 &   0 &  -1 &   0 &   0 &   1 &   -0.0012654 &    0.0000011 &    0.0000063 &    0.0006415 &    0.0000000 &    0.0000026 \\
  34 &   0 &   0 &   0 &   0 &   0 &   0 &   0 &   0 &   0 &  -1 &   0 &   2 &   2 &   1 &   -0.0010204 &    0.0000000 &    0.0000025 &    0.0005222 &    0.0000000 &    0.0000015 \\
  35 &   0 &   0 &   0 &   0 &   0 &   0 &   0 &   0 &   0 &   0 &   2 &   0 &   0 &   0 &    0.0016707 &   -0.0000085 &   -0.0000010 &    0.0000168 &   -0.0000001 &    0.0000010 \\
\end{tabular}
\newpage

\begin{tabular}{rrrrrrrrrrrrrrrrrrrrr}
Term & \multicolumn{14}{c}{Fundamental Argument Multipliers $M_{i,j}$}& \multicolumn{3}{c}{$\Delta\psi$ Coefficients}& \multicolumn{3}{c}{$\Delta\epsilon$ Coefficients} \\
$i$ & \qquad$j$= 1 & 2 & 3 & 4 & 5 & 6 & 7 & 8 & 9 & 10 & 11 & 12 & 13 & 14 
& $S_i$\w{999} & $\dot S_i$\w{999} & $C'_i$\w{999} & $C_i$\w{999} & $\dot C_i$\w{999} & $S'_i$\w{999} \\
\w{9999}  & \w{99999} & \w{-999} & \w{-999} & \w{-999} & \w{-999} & \w{-999} & \w{-999} & \w{-999} & \w{-999} & \w{-999} & \w{-999} & \w{-999} & \w{-999} & \w{-999} 
& \w{+9999}$_{''}$\w{999999} & \w{+99}$_{''}$\w{999999} & \w{+99}$_{''}$\w{999999}
& \w{+9999}$_{''}$\w{999999} & \w{+99}$_{''}$\w{999999} & \w{+99}$_{''}$\w{999999} \\

 121 &   0 &   0 &   0 &   0 &   0 &   0 &   0 &   0 &   0 &   0 &   3 &   2 &  -2 &   2 &   -0.0000438 &    0.0000000 &    0.0000000 &    0.0000188 &    0.0000000 &    0.0000000 \\
 122 &   0 &   0 &   0 &   0 &   0 &   0 &   0 &   0 &   0 &   0 &   0 &   0 &   1 &   1 &   -0.0000390 &    0.0000000 &    0.0000000 &    0.0000205 &    0.0000000 &    0.0000000 \\
 123 &   0 &   0 &   0 &   0 &   0 &   0 &   0 &   0 &   0 &  -1 &   0 &   2 &   2 &   0 &    0.0000639 &   -0.0000011 &   -0.0000002 &   -0.0000019 &    0.0000000 &    0.0000000 \\
 124 &   0 &   0 &   0 &   0 &   0 &   0 &   0 &   0 &   0 &   2 &   1 &   2 &   0 &   2 &    0.0000412 &    0.0000000 &   -0.0000002 &   -0.0000176 &    0.0000000 &   -0.0000001 \\
 125 &   0 &   0 &   0 &   0 &   0 &   0 &   0 &   0 &   0 &   1 &   1 &   0 &   0 &   1 &   -0.0000361 &    0.0000000 &    0.0000000 &    0.0000189 &    0.0000000 &    0.0000000 \\
\end{tabular}
\newpage

\begin{tabular}{rrrrrrrrrrrrrrrrrrrrr}
Term & \multicolumn{14}{c}{Fundamental Argument Multipliers $M_{i,j}$}& \multicolumn{3}{c}{$\Delta\psi$ Coefficients}& \multicolumn{3}{c}{$\Delta\epsilon$ Coefficients} \\
$i$ & \qquad$j$= 1 & 2 & 3 & 4 & 5 & 6 & 7 & 8 & 9 & 10 & 11 & 12 & 13 & 14 
& $S_i$\w{999} & $\dot S_i$\w{999} & $C'_i$\w{999} & $C_i$\w{999} & $\dot C_i$\w{999} & $S'_i$\w{999} \\
\w{9999}  & \w{99999} & \w{-999} & \w{-999} & \w{-999} & \w{-999} & \w{-999} & \w{-999} & \w{-999} & \w{-999} & \w{-999} & \w{-999} & \w{-999} & \w{-999} & \w{-999} 
& \w{+9999}$_{''}$\w{999999} & \w{+99}$_{''}$\w{999999} & \w{+99}$_{''}$\w{999999}
& \w{+9999}$_{''}$\w{999999} & \w{+99}$_{''}$\w{999999} & \w{+99}$_{''}$\w{999999} \\

 211 &   0 &   0 &   0 &   0 &   0 &   0 &   0 &   0 &   0 &  -1 &   0 &   4 &  -2 &   2 &   -0.0000093 &    0.0000000 &    0.0000000 &    0.0000040 &    0.0000000 &    0.0000000 \\
 212 &   0 &   0 &   0 &   0 &   0 &   0 &   0 &   0 &   0 &   0 &  -2 &   2 &   0 &   2 &   -0.0000083 &    0.0000000 &    0.0000010 &    0.0000040 &    0.0000000 &   -0.0000002 \\
 213 &   0 &   0 &   0 &   0 &   0 &   0 &   0 &   0 &   0 &  -1 &   0 &   2 &   1 &   2 &    0.0000083 &    0.0000000 &    0.0000000 &   -0.0000036 &    0.0000000 &    0.0000000 \\
 214 &   0 &   0 &   0 &   0 &   0 &   0 &   0 &   0 &   0 &   2 &   0 &   0 &   0 &   2 &   -0.0000091 &    0.0000000 &    0.0000000 &    0.0000039 &    0.0000000 &    0.0000000 \\
 215 &   0 &   0 &   0 &   0 &   0 &   0 &   0 &   0 &   0 &   0 &   0 &   2 &   0 &   3 &    0.0000128 &    0.0000000 &    0.0000000 &   -0.0000001 &    0.0000000 &    0.0000000 \\
\end{tabular}
\newpage

\begin{tabular}{rrrrrrrrrrrrrrrrrrrrr}
Term & \multicolumn{14}{c}{Fundamental Argument Multipliers $M_{i,j}$}& \multicolumn{3}{c}{$\Delta\psi$ Coefficients}& \multicolumn{3}{c}{$\Delta\epsilon$ Coefficients} \\
$i$ & \qquad$j$= 1 & 2 & 3 & 4 & 5 & 6 & 7 & 8 & 9 & 10 & 11 & 12 & 13 & 14 
& $S_i$\w{999} & $\dot S_i$\w{999} & $C'_i$\w{999} & $C_i$\w{999} & $\dot C_i$\w{999} & $S'_i$\w{999} \\
\w{9999}  & \w{99999} & \w{-999} & \w{-999} & \w{-999} & \w{-999} & \w{-999} & \w{-999} & \w{-999} & \w{-999} & \w{-999} & \w{-999} & \w{-999} & \w{-999} & \w{-999} 
& \w{+9999}$_{''}$\w{999999} & \w{+99}$_{''}$\w{999999} & \w{+99}$_{''}$\w{999999}
& \w{+9999}$_{''}$\w{999999} & \w{+99}$_{''}$\w{999999} & \w{+99}$_{''}$\w{999999} \\

 301 &   0 &   0 &   0 &   0 &   0 &   0 &   0 &   0 &   0 &   3 &   1 &   2 &  -2 &   2 &    0.0000032 &    0.0000000 &    0.0000000 &   -0.0000013 &    0.0000000 &    0.0000000 \\
 302 &   0 &   0 &   0 &   0 &   0 &   0 &   0 &   0 &   0 &   0 &  -1 &   0 &   4 &   0 &    0.0000065 &    0.0000000 &    0.0000000 &   -0.0000002 &    0.0000000 &    0.0000000 \\
 303 &   0 &   0 &   0 &   0 &   0 &   0 &   0 &   0 &   0 &   2 &  -1 &   0 &   2 &   0 &    0.0000047 &    0.0000000 &    0.0000000 &   -0.0000001 &    0.0000000 &    0.0000000 \\
 304 &   0 &   0 &   0 &   0 &   0 &   0 &   0 &   0 &   0 &   0 &   0 &   4 &   0 &   1 &    0.0000032 &    0.0000000 &    0.0000000 &   -0.0000016 &    0.0000000 &    0.0000000 \\
 305 &   0 &   0 &   0 &   0 &   0 &   0 &   0 &   0 &   0 &   2 &   0 &   4 &  -2 &   2 &    0.0000037 &    0.0000000 &    0.0000000 &   -0.0000016 &    0.0000000 &    0.0000000 \\
\end{tabular}
\newpage

\begin{tabular}{rrrrrrrrrrrrrrrrrrrrr}
Term & \multicolumn{14}{c}{Fundamental Argument Multipliers $M_{i,j}$}& \multicolumn{3}{c}{$\Delta\psi$ Coefficients}& \multicolumn{3}{c}{$\Delta\epsilon$ Coefficients} \\
$i$ & \qquad$j$= 1 & 2 & 3 & 4 & 5 & 6 & 7 & 8 & 9 & 10 & 11 & 12 & 13 & 14 
& $S_i$\w{999} & $\dot S_i$\w{999} & $C'_i$\w{999} & $C_i$\w{999} & $\dot C_i$\w{999} & $S'_i$\w{999} \\
\w{9999}  & \w{99999} & \w{-999} & \w{-999} & \w{-999} & \w{-999} & \w{-999} & \w{-999} & \w{-999} & \w{-999} & \w{-999} & \w{-999} & \w{-999} & \w{-999} & \w{-999} 
& \w{+9999}$_{''}$\w{999999} & \w{+99}$_{''}$\w{999999} & \w{+99}$_{''}$\w{999999}
& \w{+9999}$_{''}$\w{999999} & \w{+99}$_{''}$\w{999999} & \w{+99}$_{''}$\w{999999} \\

 391 &   0 &   0 &   0 &   0 &   0 &   0 &   0 &   0 &   0 &  -1 &   0 &   0 &   4 &   2 &   -0.0000013 &    0.0000000 &    0.0000000 &    0.0000006 &    0.0000000 &    0.0000000 \\
 392 &   0 &   0 &   0 &   0 &   0 &   0 &   0 &   0 &   0 &  -1 &  -2 &   2 &   2 &   1 &   -0.0000014 &    0.0000000 &    0.0000000 &    0.0000008 &    0.0000000 &    0.0000000 \\
 393 &   0 &   0 &   0 &   0 &   0 &   0 &   0 &   0 &   0 &  -1 &   0 &   0 &  -2 &   2 &    0.0000013 &    0.0000000 &    0.0000000 &   -0.0000005 &    0.0000000 &    0.0000000 \\
 394 &   0 &   0 &   0 &   0 &   0 &   0 &   0 &   0 &   0 &   1 &   0 &  -2 &  -2 &   1 &   -0.0000017 &    0.0000000 &    0.0000000 &    0.0000009 &    0.0000000 &    0.0000000 \\
 395 &   0 &   0 &   0 &   0 &   0 &   0 &   0 &   0 &   0 &   0 &   0 &  -2 &  -2 &   1 &   -0.0000012 &    0.0000000 &    0.0000000 &    0.0000006 &    0.0000000 &    0.0000000 \\
\end{tabular}
\newpage

\begin{tabular}{rrrrrrrrrrrrrrrrrrrrr}
Term & \multicolumn{14}{c}{Fundamental Argument Multipliers $M_{i,j}$}& \multicolumn{3}{c}{$\Delta\psi$ Coefficients}& \multicolumn{3}{c}{$\Delta\epsilon$ Coefficients} \\
$i$ & \qquad$j$= 1 & 2 & 3 & 4 & 5 & 6 & 7 & 8 & 9 & 10 & 11 & 12 & 13 & 14 
& $S_i$\w{999} & $\dot S_i$\w{999} & $C'_i$\w{999} & $C_i$\w{999} & $\dot C_i$\w{999} & $S'_i$\w{999} \\
\w{9999}  & \w{99999} & \w{-999} & \w{-999} & \w{-999} & \w{-999} & \w{-999} & \w{-999} & \w{-999} & \w{-999} & \w{-999} & \w{-999} & \w{-999} & \w{-999} & \w{-999} 
& \w{+9999}$_{''}$\w{999999} & \w{+99}$_{''}$\w{999999} & \w{+99}$_{''}$\w{999999}
& \w{+9999}$_{''}$\w{999999} & \w{+99}$_{''}$\w{999999} & \w{+99}$_{''}$\w{999999} \\

 481 &   0 &   0 &   0 &   0 &   0 &   0 &   0 &   0 &   0 &  -2 &  -2 &   0 &   2 &   1 &   -0.0000003 &    0.0000000 &    0.0000000 &    0.0000001 &    0.0000000 &    0.0000000 \\
 482 &   0 &   0 &   0 &   0 &   0 &   0 &   0 &   0 &   0 &  -2 &   0 &  -2 &   4 &   0 &    0.0000000 &    0.0000000 &    0.0000000 &    0.0000000 &    0.0000000 &    0.0000000 \\
 483 &   0 &   0 &   0 &   0 &   0 &   0 &   0 &   0 &   0 &   1 &   2 &   2 &  -4 &   1 &   -0.0000003 &    0.0000000 &    0.0000000 &    0.0000001 &    0.0000000 &    0.0000000 \\
 484 &   0 &   0 &   0 &   0 &   0 &   0 &   0 &   0 &   0 &   1 &   1 &   2 &  -4 &   2 &    0.0000007 &    0.0000000 &    0.0000000 &   -0.0000003 &    0.0000000 &    0.0000000 \\
 485 &   0 &   0 &   0 &   0 &   0 &   0 &   0 &   0 &   0 &  -1 &   2 &   2 &  -2 &   1 &   -0.0000004 &    0.0000000 &    0.0000000 &    0.0000002 &    0.0000000 &    0.0000000 \\
\end{tabular}
\newpage

\begin{tabular}{rrrrrrrrrrrrrrrrrrrrr}
Term & \multicolumn{14}{c}{Fundamental Argument Multipliers $M_{i,j}$}& \multicolumn{3}{c}{$\Delta\psi$ Coefficients}& \multicolumn{3}{c}{$\Delta\epsilon$ Coefficients} \\
$i$ & \qquad$j$= 1 & 2 & 3 & 4 & 5 & 6 & 7 & 8 & 9 & 10 & 11 & 12 & 13 & 14 
& $S_i$\w{999} & $\dot S_i$\w{999} & $C'_i$\w{999} & $C_i$\w{999} & $\dot C_i$\w{999} & $S'_i$\w{999} \\
\w{9999}  & \w{99999} & \w{-999} & \w{-999} & \w{-999} & \w{-999} & \w{-999} & \w{-999} & \w{-999} & \w{-999} & \w{-999} & \w{-999} & \w{-999} & \w{-999} & \w{-999} 
& \w{+9999}$_{''}$\w{999999} & \w{+99}$_{''}$\w{999999} & \w{+99}$_{''}$\w{999999}
& \w{+9999}$_{''}$\w{999999} & \w{+99}$_{''}$\w{999999} & \w{+99}$_{''}$\w{999999} \\

 571 &   0 &   0 &   0 &   0 &   0 &   0 &   0 &   0 &   0 &   2 &   2 &   2 &  -2 &   2 &    0.0000008 &    0.0000000 &    0.0000000 &   -0.0000003 &    0.0000000 &    0.0000000 \\
 572 &   0 &   0 &   0 &   0 &   0 &   0 &   0 &   0 &   0 &   2 &   0 &   4 &  -4 &   2 &   -0.0000004 &    0.0000000 &    0.0000000 &    0.0000002 &    0.0000000 &    0.0000000 \\
 573 &   0 &   0 &   0 &   0 &   0 &   0 &   0 &   0 &   0 &  -1 &  -2 &   0 &   4 &   0 &    0.0000011 &    0.0000000 &    0.0000000 &    0.0000000 &    0.0000000 &    0.0000000 \\
 574 &   0 &   0 &   0 &   0 &   0 &   0 &   0 &   0 &   0 &  -1 &  -3 &   2 &   2 &   2 &   -0.0000003 &    0.0000000 &    0.0000000 &    0.0000001 &    0.0000000 &    0.0000000 \\
 575 &   0 &   0 &   0 &   0 &   0 &   0 &   0 &   0 &   0 &  -3 &   0 &   2 &   4 &   2 &    0.0000003 &    0.0000000 &    0.0000000 &   -0.0000001 &    0.0000000 &    0.0000000 \\
\end{tabular}
\newpage

\begin{tabular}{rrrrrrrrrrrrrrrrrrrrr}
Term & \multicolumn{14}{c}{Fundamental Argument Multipliers $M_{i,j}$}& \multicolumn{3}{c}{$\Delta\psi$ Coefficients}& \multicolumn{3}{c}{$\Delta\epsilon$ Coefficients} \\
$i$ & \qquad$j$= 1 & 2 & 3 & 4 & 5 & 6 & 7 & 8 & 9 & 10 & 11 & 12 & 13 & 14 
& $S_i$\w{999} & $\dot S_i$\w{999} & $C'_i$\w{999} & $C_i$\w{999} & $\dot C_i$\w{999} & $S'_i$\w{999} \\
\w{9999}  & \w{99999} & \w{-999} & \w{-999} & \w{-999} & \w{-999} & \w{-999} & \w{-999} & \w{-999} & \w{-999} & \w{-999} & \w{-999} & \w{-999} & \w{-999} & \w{-999} 
& \w{+9999}$_{''}$\w{999999} & \w{+99}$_{''}$\w{999999} & \w{+99}$_{''}$\w{999999}
& \w{+9999}$_{''}$\w{999999} & \w{+99}$_{''}$\w{999999} & \w{+99}$_{''}$\w{999999} \\

 661 &   0 &   0 &   0 &   0 &   0 &   0 &   0 &   0 &   0 &   2 &  -2 &   2 &   2 &   2 &   -0.0000004 &    0.0000000 &    0.0000000 &    0.0000002 &    0.0000000 &    0.0000000 \\
 662 &   0 &   0 &   0 &   0 &   0 &   0 &   0 &   0 &   0 &   0 &   0 &   2 &   4 &   0 &    0.0000007 &    0.0000000 &    0.0000000 &    0.0000000 &    0.0000000 &    0.0000000 \\
 663 &   0 &   0 &   0 &   0 &   0 &   0 &   0 &   0 &   0 &   1 &   0 &   2 &   3 &   2 &    0.0000007 &    0.0000000 &    0.0000000 &   -0.0000003 &    0.0000000 &    0.0000000 \\
 664 &   0 &   0 &   0 &   0 &   0 &   0 &   0 &   0 &   0 &   4 &   0 &   0 &   2 &   0 &    0.0000004 &    0.0000000 &    0.0000000 &    0.0000000 &    0.0000000 &    0.0000000 \\
 665 &   0 &   0 &   0 &   0 &   0 &   0 &   0 &   0 &   0 &   2 &   0 &   2 &   2 &   0 &    0.0000011 &    0.0000000 &    0.0000000 &    0.0000000 &    0.0000000 &    0.0000000 \\
\end{tabular}
\newpage

\begin{tabular}{rrrrrrrrrrrrrrrrrrrrr}
Term & \multicolumn{14}{c}{Fundamental Argument Multipliers $M_{i,j}$}& \multicolumn{3}{c}{$\Delta\psi$ Coefficients}& \multicolumn{3}{c}{$\Delta\epsilon$ Coefficients} \\
$i$ & \qquad$j$= 1 & 2 & 3 & 4 & 5 & 6 & 7 & 8 & 9 & 10 & 11 & 12 & 13 & 14 
& $S_i$\w{999} & $\dot S_i$\w{999} & $C'_i$\w{999} & $C_i$\w{999} & $\dot C_i$\w{999} & $S'_i$\w{999} \\
\w{9999}  & \w{99999} & \w{-999} & \w{-999} & \w{-999} & \w{-999} & \w{-999} & \w{-999} & \w{-999} & \w{-999} & \w{-999} & \w{-999} & \w{-999} & \w{-999} & \w{-999} 
& \w{+9999}$_{''}$\w{999999} & \w{+99}$_{''}$\w{999999} & \w{+99}$_{''}$\w{999999}
& \w{+9999}$_{''}$\w{999999} & \w{+99}$_{''}$\w{999999} & \w{+99}$_{''}$\w{999999} \\

 750 &   0 &   0 &   1 &   0 &   0 &   1 &   0 &   0 &   0 &   0 &   0 &  -1 &   1 &   0 &   -0.0000042 & 0.\w{9999999} &    0.0000020 &    0.0000000 & 0.\w{9999999} &    0.0000000 \\
 751 &   0 &   5 &  -6 &   0 &   0 &   0 &   0 &   0 &   0 &   0 &   0 &  -2 &   2 &   0 &   -0.0000010 & 0.\w{9999999} &    0.0000233 &    0.0000000 & 0.\w{9999999} &    0.0000000 \\
 752 &   0 &   5 &  -7 &   0 &   0 &   0 &   0 &   0 &   0 &   0 &   0 &  -1 &   1 &   1 &   -0.0000003 & 0.\w{9999999} &    0.0000000 &    0.0000001 & 0.\w{9999999} &    0.0000000 \\
 753 &   0 &   6 &  -8 &   0 &   0 &   0 &   0 &   0 &   0 &  -2 &   0 &   0 &   2 &   0 &    0.0000078 & 0.\w{9999999} &   -0.0000018 &    0.0000000 & 0.\w{9999999} &    0.0000000 \\
 754 &   0 &  -6 &   7 &   0 &   0 &   0 &   0 &   0 &   0 &   2 &   0 &   1 &  -3 &   1 &    0.0000000 & 0.\w{9999999} &    0.0000003 &    0.0000000 & 0.\w{9999999} &    0.0000001 \\
\end{tabular}
\newpage

\begin{tabular}{rrrrrrrrrrrrrrrrrrrrr}
Term & \multicolumn{14}{c}{Fundamental Argument Multipliers $M_{i,j}$}& \multicolumn{3}{c}{$\Delta\psi$ Coefficients}& \multicolumn{3}{c}{$\Delta\epsilon$ Coefficients} \\
$i$ & \qquad$j$= 1 & 2 & 3 & 4 & 5 & 6 & 7 & 8 & 9 & 10 & 11 & 12 & 13 & 14 
& $S_i$\w{999} & $\dot S_i$\w{999} & $C'_i$\w{999} & $C_i$\w{999} & $\dot C_i$\w{999} & $S'_i$\w{999} \\
\w{9999}  & \w{99999} & \w{-999} & \w{-999} & \w{-999} & \w{-999} & \w{-999} & \w{-999} & \w{-999} & \w{-999} & \w{-999} & \w{-999} & \w{-999} & \w{-999} & \w{-999} 
& \w{+9999}$_{''}$\w{999999} & \w{+99}$_{''}$\w{999999} & \w{+99}$_{''}$\w{999999}
& \w{+9999}$_{''}$\w{999999} & \w{+99}$_{''}$\w{999999} & \w{+99}$_{''}$\w{999999} \\

 840 &   0 &   3 &  -6 &   0 &   0 &   0 &   0 &   0 &   0 &   0 &   0 &   1 &  -1 &   0 &    0.0000000 & 0.\w{9999999} &    0.0000004 &    0.0000000 & 0.\w{9999999} &    0.0000000 \\
 841 &   0 &  -3 &   5 &   0 &   0 &   0 &   0 &   0 &   0 &   0 &   0 &   0 &   0 &   1 &   -0.0000021 & 0.\w{9999999} &   -0.0000006 &    0.0000011 & 0.\w{9999999} &   -0.0000003 \\
 842 &   0 &  -3 &   4 &   0 &   0 &   0 &   0 &   0 &   0 &   0 &   0 &   1 &  -1 &   2 &    0.0000000 & 0.\w{9999999} &   -0.0000003 &    0.0000000 & 0.\w{9999999} &   -0.0000001 \\
 843 &   0 &   0 &  -2 &   4 &   0 &   0 &   0 &   0 &   0 &   0 &   0 &   0 &   0 &   1 &   -0.0000011 & 0.\w{9999999} &   -0.0000021 &    0.0000006 & 0.\w{9999999} &   -0.0000011 \\
 844 &   0 &  -5 &   6 &   0 &   0 &   0 &   0 &   0 &   0 &   0 &   0 &   2 &  -2 &   1 &   -0.0000018 & 0.\w{9999999} &   -0.0000436 &    0.0000009 & 0.\w{9999999} &   -0.0000233 \\
\end{tabular}
\newpage

\begin{tabular}{rrrrrrrrrrrrrrrrrrrrr}
Term & \multicolumn{14}{c}{Fundamental Argument Multipliers $M_{i,j}$}& \multicolumn{3}{c}{$\Delta\psi$ Coefficients}& \multicolumn{3}{c}{$\Delta\epsilon$ Coefficients} \\
$i$ & \qquad$j$= 1 & 2 & 3 & 4 & 5 & 6 & 7 & 8 & 9 & 10 & 11 & 12 & 13 & 14 
& $S_i$\w{999} & $\dot S_i$\w{999} & $C'_i$\w{999} & $C_i$\w{999} & $\dot C_i$\w{999} & $S'_i$\w{999} \\
\w{9999}  & \w{99999} & \w{-999} & \w{-999} & \w{-999} & \w{-999} & \w{-999} & \w{-999} & \w{-999} & \w{-999} & \w{-999} & \w{-999} & \w{-999} & \w{-999} & \w{-999} 
& \w{+9999}$_{''}$\w{999999} & \w{+99}$_{''}$\w{999999} & \w{+99}$_{''}$\w{999999}
& \w{+9999}$_{''}$\w{999999} & \w{+99}$_{''}$\w{999999} & \w{+99}$_{''}$\w{999999} \\

 930 &   0 &   0 &   3 &  -6 &   0 &   0 &   0 &   0 &   0 &   0 &   0 &   0 &   0 &   0 &    0.0000000 & 0.\w{9999999} &    0.0000031 &    0.0000000 & 0.\w{9999999} &    0.0000000 \\
 931 &   0 &   0 &  -3 &   6 &   0 &   0 &   0 &   0 &   1 &   0 &   0 &   0 &   0 &   0 &   -0.0000005 & 0.\w{9999999} &    0.0000000 &    0.0000003 & 0.\w{9999999} &    0.0000001 \\
 932 &   0 &   0 &  -4 &   6 &   0 &   0 &   0 &   0 &   0 &   0 &   0 &   1 &  -1 &   1 &    0.0000004 & 0.\w{9999999} &    0.0000000 &   -0.0000002 & 0.\w{9999999} &    0.0000000 \\
 933 &   0 &   0 &  -3 &   6 &   0 &   0 &   0 &   0 &   1 &   0 &   0 &   0 &   0 &   0 &   -0.0000004 & 0.\w{9999999} &    0.0000000 &    0.0000002 & 0.\w{9999999} &    0.0000000 \\
 934 &   0 &   0 &  -3 &   6 &   0 &   0 &   0 &   0 &   2 &   0 &   0 &   0 &   0 &   0 &   -0.0000024 & 0.\w{9999999} &   -0.0000013 &    0.0000010 & 0.\w{9999999} &   -0.0000006 \\
\end{tabular}
\newpage

\begin{tabular}{rrrrrrrrrrrrrrrrrrrrr}
Term & \multicolumn{14}{c}{Fundamental Argument Multipliers $M_{i,j}$}& \multicolumn{3}{c}{$\Delta\psi$ Coefficients}& \multicolumn{3}{c}{$\Delta\epsilon$ Coefficients} \\
$i$ & \qquad$j$= 1 & 2 & 3 & 4 & 5 & 6 & 7 & 8 & 9 & 10 & 11 & 12 & 13 & 14 
& $S_i$\w{999} & $\dot S_i$\w{999} & $C'_i$\w{999} & $C_i$\w{999} & $\dot C_i$\w{999} & $S'_i$\w{999} \\
\w{9999}  & \w{99999} & \w{-999} & \w{-999} & \w{-999} & \w{-999} & \w{-999} & \w{-999} & \w{-999} & \w{-999} & \w{-999} & \w{-999} & \w{-999} & \w{-999} & \w{-999} 
& \w{+9999}$_{''}$\w{999999} & \w{+99}$_{''}$\w{999999} & \w{+99}$_{''}$\w{999999}
& \w{+9999}$_{''}$\w{999999} & \w{+99}$_{''}$\w{999999} & \w{+99}$_{''}$\w{999999} \\

1020 &   0 &  -2 &   4 &   0 &   0 &   0 &   0 &   0 &   2 &   0 &   0 &   0 &   0 &   0 &   -0.0000458 & 0.\w{9999999} &    0.0000000 &    0.0000198 & 0.\w{9999999} &    0.0000000 \\
1021 &   0 &  -6 &   9 &   0 &   0 &   0 &   0 &   0 &   2 &   0 &   0 &   0 &   0 &   0 &    0.0000000 & 0.\w{9999999} &   -0.0000045 &    0.0000000 & 0.\w{9999999} &   -0.0000020 \\
1022 &   0 &  -6 &   9 &   0 &   0 &   0 &   0 &   0 &   1 &   0 &   0 &   0 &   0 &   0 &    0.0000009 & 0.\w{9999999} &    0.0000000 &   -0.0000005 & 0.\w{9999999} &    0.0000000 \\
1023 &   0 &   6 &  -9 &   0 &   0 &   0 &   0 &   0 &   0 &   0 &   0 &   0 &   0 &   0 &    0.0000000 & 0.\w{9999999} &   -0.0000003 &    0.0000000 & 0.\w{9999999} &    0.0000000 \\
1024 &   0 &   0 &   1 &   0 &  -2 &   0 &   0 &   0 &   0 &   0 &   0 &   0 &   0 &   1 &    0.0000000 & 0.\w{9999999} &   -0.0000004 &   -0.0000001 & 0.\w{9999999} &   -0.0000002 \\
\end{tabular}
\newpage

\begin{tabular}{rrrrrrrrrrrrrrrrrrrrr}
Term & \multicolumn{14}{c}{Fundamental Argument Multipliers $M_{i,j}$}& \multicolumn{3}{c}{$\Delta\psi$ Coefficients}& \multicolumn{3}{c}{$\Delta\epsilon$ Coefficients} \\
$i$ & \qquad$j$= 1 & 2 & 3 & 4 & 5 & 6 & 7 & 8 & 9 & 10 & 11 & 12 & 13 & 14 
& $S_i$\w{999} & $\dot S_i$\w{999} & $C'_i$\w{999} & $C_i$\w{999} & $\dot C_i$\w{999} & $S'_i$\w{999} \\
\w{9999}  & \w{99999} & \w{-999} & \w{-999} & \w{-999} & \w{-999} & \w{-999} & \w{-999} & \w{-999} & \w{-999} & \w{-999} & \w{-999} & \w{-999} & \w{-999} & \w{-999} 
& \w{+9999}$_{''}$\w{999999} & \w{+99}$_{''}$\w{999999} & \w{+99}$_{''}$\w{999999}
& \w{+9999}$_{''}$\w{999999} & \w{+99}$_{''}$\w{999999} & \w{+99}$_{''}$\w{999999} \\

1110 &   0 &   0 &   0 &   3 &   0 &   0 &   0 &   0 &   2 &   0 &   0 &   0 &   0 &   0 &   -0.0000018 & 0.\w{9999999} &   -0.0000010 &    0.0000008 & 0.\w{9999999} &   -0.0000004 \\
1111 &   0 &   1 &   0 &   0 &   0 &   0 &   0 &   0 &   0 &   0 &   0 &   0 &   0 &   0 &   -0.0000004 & 0.\w{9999999} &   -0.0000028 &    0.0000000 & 0.\w{9999999} &    0.0000000 \\
1112 &   0 &   1 &   0 &   0 &   0 &   0 &   0 &   0 &   2 &   0 &   0 &   0 &   0 &   0 &   -0.0000005 & 0.\w{9999999} &    0.0000006 &    0.0000002 & 0.\w{9999999} &    0.0000003 \\
1113 &   0 &  -9 &  13 &   0 &   0 &   0 &   0 &   0 &   2 &   0 &   0 &   0 &   0 &   0 &   -0.0000003 & 0.\w{9999999} &    0.0000000 &    0.0000001 & 0.\w{9999999} &    0.0000000 \\
1114 &   0 &   0 &  -1 &   5 &   0 &   0 &   0 &   0 &   2 &   0 &   0 &   0 &   0 &   0 &   -0.0000005 & 0.\w{9999999} &   -0.0000009 &    0.0000002 & 0.\w{9999999} &   -0.0000004 \\
\end{tabular}
\newpage

\begin{tabular}{rrrrrrrrrrrrrrrrrrrrr}
Term & \multicolumn{14}{c}{Fundamental Argument Multipliers $M_{i,j}$}& \multicolumn{3}{c}{$\Delta\psi$ Coefficients}& \multicolumn{3}{c}{$\Delta\epsilon$ Coefficients} \\
$i$ & \qquad$j$= 1 & 2 & 3 & 4 & 5 & 6 & 7 & 8 & 9 & 10 & 11 & 12 & 13 & 14 
& $S_i$\w{999} & $\dot S_i$\w{999} & $C'_i$\w{999} & $C_i$\w{999} & $\dot C_i$\w{999} & $S'_i$\w{999} \\
\w{9999}  & \w{99999} & \w{-999} & \w{-999} & \w{-999} & \w{-999} & \w{-999} & \w{-999} & \w{-999} & \w{-999} & \w{-999} & \w{-999} & \w{-999} & \w{-999} & \w{-999} 
& \w{+9999}$_{''}$\w{999999} & \w{+99}$_{''}$\w{999999} & \w{+99}$_{''}$\w{999999}
& \w{+9999}$_{''}$\w{999999} & \w{+99}$_{''}$\w{999999} & \w{+99}$_{''}$\w{999999} \\

1200 &   0 &   0 &  -3 &   0 &   5 &   0 &   0 &   0 &   2 &   0 &   0 &   0 &   0 &   0 &    0.0000003 & 0.\w{9999999} &    0.0000000 &   -0.0000001 & 0.\w{9999999} &    0.0000000 \\
1201 &   0 &   1 &   1 &   0 &   0 &   0 &   0 &   0 &   0 &   0 &   0 &   0 &   0 &   0 &   -0.0000004 & 0.\w{9999999} &    0.0000000 &    0.0000000 & 0.\w{9999999} &    0.0000000 \\
1202 &   0 &   1 &   1 &   0 &   0 &   0 &   0 &   0 &   1 &   0 &   0 &   0 &   0 &   0 &    0.0000005 & 0.\w{9999999} &   -0.0000023 &   -0.0000003 & 0.\w{9999999} &   -0.0000012 \\
1203 &   0 &   1 &   1 &   0 &   0 &   0 &   0 &   0 &   2 &   0 &   0 &   0 &   0 &   0 &   -0.0000339 & 0.\w{9999999} &    0.0000000 &    0.0000147 & 0.\w{9999999} &    0.0000000 \\
1204 &   0 &  -9 &  12 &   0 &   0 &   0 &   0 &   0 &   2 &   0 &   0 &   0 &   0 &   0 &    0.0000000 & 0.\w{9999999} &   -0.0000010 &    0.0000000 & 0.\w{9999999} &   -0.0000005 \\
\end{tabular}
\newpage

\begin{tabular}{rrrrrrrrrrrrrrrrrrrrr}
Term & \multicolumn{14}{c}{Fundamental Argument Multipliers $M_{i,j}$}& \multicolumn{3}{c}{$\Delta\psi$ Coefficients}& \multicolumn{3}{c}{$\Delta\epsilon$ Coefficients} \\
$i$ & \qquad$j$= 1 & 2 & 3 & 4 & 5 & 6 & 7 & 8 & 9 & 10 & 11 & 12 & 13 & 14 
& $S_i$\w{999} & $\dot S_i$\w{999} & $C'_i$\w{999} & $C_i$\w{999} & $\dot C_i$\w{999} & $S'_i$\w{999} \\
\w{9999}  & \w{99999} & \w{-999} & \w{-999} & \w{-999} & \w{-999} & \w{-999} & \w{-999} & \w{-999} & \w{-999} & \w{-999} & \w{-999} & \w{-999} & \w{-999} & \w{-999} 
& \w{+9999}$_{''}$\w{999999} & \w{+99}$_{''}$\w{999999} & \w{+99}$_{''}$\w{999999}
& \w{+9999}$_{''}$\w{999999} & \w{+99}$_{''}$\w{999999} & \w{+99}$_{''}$\w{999999} \\

1290 &   0 &   6 &  -4 &   0 &   0 &   0 &   0 &   0 &   1 &   0 &   0 &   0 &   0 &   0 &    0.0000000 & 0.\w{9999999} &   -0.0000003 &    0.0000000 & 0.\w{9999999} &   -0.0000002 \\
1291 &   0 &   0 &   6 &   0 &   0 &   0 &   0 &   0 &   2 &   0 &   0 &   0 &   0 &   0 &    0.0000002 & 0.\w{9999999} &    0.0000009 &    0.0000003 & 0.\w{9999999} &    0.0000004 \\
1292 &   0 &   0 &   6 &   0 &   0 &   0 &   0 &   0 &   0 &   0 &   0 &   0 &   0 &   0 &    0.0000000 & 0.\w{9999999} &    0.0000000 &   -0.0000004 & 0.\w{9999999} &    0.0000000 \\
1293 &   0 &   0 &   6 &   0 &   0 &   0 &   0 &   0 &   0 &   0 &   0 &   0 &   0 &   0 &    0.0000008 & 0.\w{9999999} &    0.0000000 &    0.0000000 & 0.\w{9999999} &    0.0000000 \\
1294 &   0 &   0 &   6 &   0 &   0 &   0 &   0 &   0 &   1 &   0 &   0 &   0 &   0 &   0 &    0.0000000 & 0.\w{9999999} &    0.0000004 &    0.0000000 & 0.\w{9999999} &    0.0000002 \\
\end{tabular}

\normalsize 

\chapter*{Errata \& Updates}
\markboth{}{}
\addcontentsline{toc}{chapter}{Errata \& Updates}

\setlength{\parskip}{2ex}
\setlength{\parindent}{0em}

Errata in this \pubname\ and updates to it are given at \citet{url-thispub}.   Please consult the web page at this URL periodically.

\end{document}